\documentclass[acmsmall,screen,authordraft,nonacm,review=false,timestamp=false]{acmart}

\usepackage[final]{changes}
\usepackage{cleveref}

\usepackage{amsmath}
\usepackage{amsfonts}
\usepackage{tabularx}
\usepackage{arydshln}

\usepackage{multirow}
\usepackage{array}

\usepackage{wasysym}
\usepackage{dcolumn} 

\newcolumntype{d}[1]{D{.}{.}{#1}}

\usepackage{subcaption}

\usepackage{placeins}

\newcommand{\sAI}{sAI}
\newcommand{\sAIactive}{sAI active}
\newcommand{\sAIinactive}{sAI inactive}
\newcommand{\sstatus}{system status}
\newcommand{\SStatus}{System Status}

\AtBeginDocument{%
  }

\setcopyright{acmcopyright}
\copyrightyear{2018}
\acmYear{2018}
\acmDOI{XXXXXXX.XXXXXXX}
\acmConference[Conference acronym 'XX]{Make sure to enter the correct
  conference title from your rights confirmation email}{June 03--05,
  2018}{Woodstock, NY}
\acmPrice{15.00}
\acmISBN{978-1-4503-XXXX-X/18/06}

\begin{teaserfigure}
 \includegraphics[width=\linewidth]{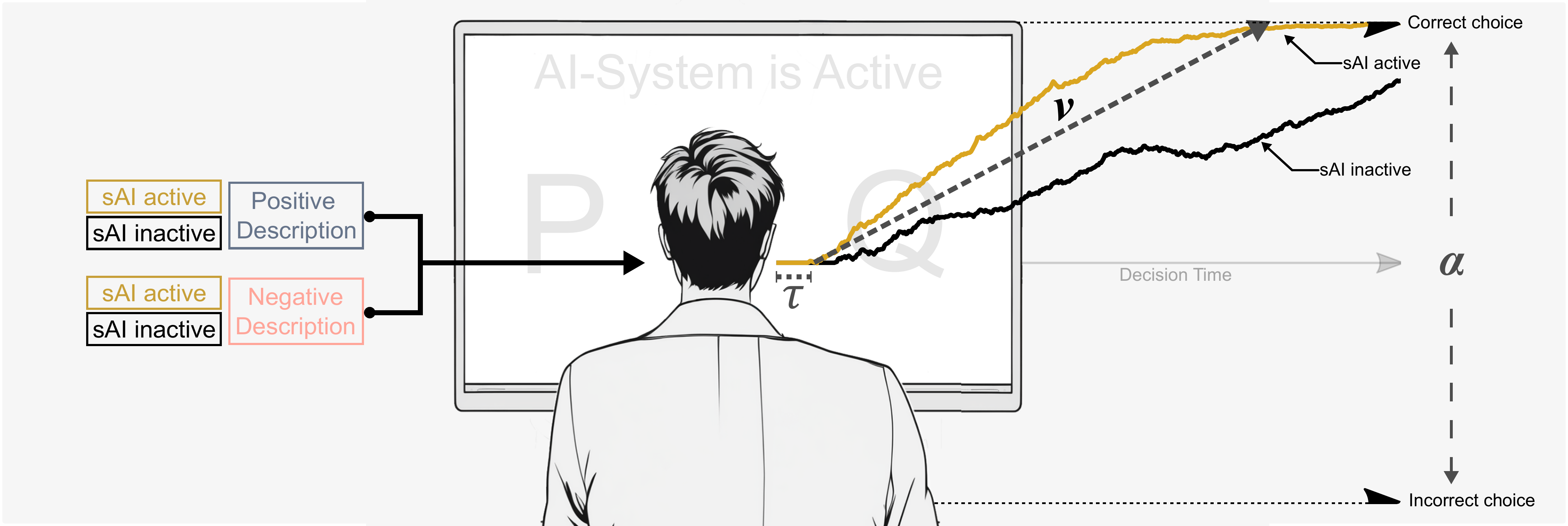}
 \caption{Schematic representation of the drift-diffusion process of decision-making with an increased drift-rate $\nu$ and a decreased non-decision time $\tau$, for the \replaced{sham-AI active}{sham-AI}  (sAI active) condition as compared to the \replaced{sham-AI inactive}{no-AI} (sAI inactive) condition. When using a sham AI, participants accumulate information faster.}
 \label{fig:teaser}
\end{teaserfigure}

\begin{document}

\title[The Placebo Effect Is Robust to Negative Descriptions of AI]{"AI enhances our performance, I have no doubt this one will do the same": The Placebo Effect Is Robust to Negative Descriptions of AI}


\settopmatter{authorsperrow=1}

\author{Agnes M. Kloft}
\orcid{0009-0008-8024-2398}
\affiliation{%
  \institution{Aalto University}
  \city{Espoo}
  \postcode{02150}
  \country{Finland}}
\email{agnes.kloft@aalto.fi}

\authornote{Shared first authorship: Both authors contributed equally to the paper}

\author{Robin Welsch}
\orcid{0000-0003-3890-1990}
\affiliation{%
  \institution{Aalto University}
  \city{Espoo}
  \postcode{02150}
  \country{Finland}}
\email{robin.welsch@aalto.fi}
\authornotemark[1]
\author{Thomas Kosch}
\orcid{}
\affiliation{%
  \institution{HU Berlin}
  \city{Berlin}
  \postcode{02150}
  \country{Germany}}

\author{Steeven Villa}
\orcid{}
\affiliation{%
  \institution{LMU Munich}
  \city{Munich}
  \postcode{80337}
  \country{Germany}}
\email{villa@posthci.com}

\renewcommand{\shortauthors}{Kloft et al.}

\begin{abstract}
Heightened AI expectations facilitate performance in human-AI interactions through placebo effects. While lowering expectations to control for placebo effects is advisable, overly negative expectations could induce nocebo effects. In a letter discrimination task, we informed participants that an AI would either increase or decrease their performance by adapting the interface, when in reality, no AI was present in any condition. A Bayesian analysis showed that participants had high expectations and performed descriptively better irrespective of the AI description when a sham-AI was present. Using cognitive modeling, we could trace this advantage back to participants gathering more information. A replication study verified that negative AI descriptions do not alter expectations, suggesting that performance expectations with AI are biased and robust to negative verbal descriptions. We discuss the impact of user expectations on AI interactions and evaluation.
\end{abstract}



\maketitle

\section{Introduction}

\replaced{Expectations regarding}{Beliefs about} Artificial intelligence (AI) fundamentally affect how we use this technology. 
The placebo effect of AI in Human-Computer Interaction (HCI) \cite{kosch2022placebo}, inspired by medical research ~\cite{stewart2004placebo,lasagna1954study, montgomery1996mechanisms,lasagna1954study, beecher1955powerful}, documents that a sham-AI \added{(\sAI){}} system can bring real subjective benefits accompanied by changes in \added{decision-making} and physiology \cite{kosch2022placebo,villa2023placebo}. \citet{kosch2022placebo} argued that much like in the medical context, user expectations about AI technology significantly influence study outcomes and thus undermine scientific evaluation if they are left uncontrolled. \added{The idea of controlling user expectations about novel technologies (such as AI) in human-centered studies has been discussed in the past~\cite{doi:10.1177/1745691613491271}, proposing to control results that originate from participant beliefs~\cite{https://doi.org/10.1111/j.1540-5915.2010.00292.x} rather than from an active system. Thus, user expectations play a critical role in assessing AI systems, regardless of a functional system in user studies.}

Prior research on placebo effects in HCI has been reported in gaming contexts, where fake power-up elements that make no difference to gameplay~\cite{Denisova2019} and sham descriptions of AI adaptation increase game immersion~\cite{denisova2015placebo}. In social media, sham control settings for a news feed can result in higher user satisfaction~\cite{vaccaro2018illusion}. ~\citet{kosch2022placebo} showed that \replaced{expecting}{a belief in receiving} benefits from using an adaptive AI can improve \added{subjective} performance. 
\citet{villa2023placebo} could show that high expectations regarding \replaced{sAI}{sham-AI}-based augmentation systems increase risky decision-making and affect information processing. Thus, AI technology can induce placebo effects that alter \added{subjective performance and decision-making, therefore experiences,} through heightened positive expectations. \added{Note that these placebo studies, where a control condition is compared to a placebo, must be distinguished from placebo-controlled studies, where an effective treatment is compared to a placebo condition. While the former is often used to understand the placebo effect scientifically, the latter presents a technique to control for it (see \citet{kosch2022placebo} for taxonomy in HCI). Note also that while conceptually similar to Wizard of Oz paradigms that are often employed in AI research (see \citet{SCHOONDERWOERD2022WOZ,dahlback1993wizard}), where an experimenter controls a computer to emulate an intelligent system, in placebo studies of AI the system is not functional.}

There are three major shortcomings in the placebo literature in HCI for AI. First, direct effects on a behavioral level are yet to be found~\cite{kosch2022placebo,villa2023placebo}. Second, it is unclear whether nocebo effects \replaced{(}{,} low expectations impairing behavior \replaced{)}{,} are equally influential as positive expectations based on verbal descriptions in HCI. \deleted{Third, we lack a behavioral marker for effectively designing adaptive AI interfaces that enhance decision-making amidst placebo responses.} \added{Third, we lack scientific studies that show how AI expectations affect interaction, and thus, study outcomes.}

This paper investigates the antecedents and consequences of AI's placebo effect in HCI. \added{In detail, we examine how descriptions can impact decision-making by raising or lowering expectations, thus using expectations as a mediator between descriptions and placebo or nocebo effects.} In an experimental study ($N$ = 66), we examined the influence of negative and positive verbal AI descriptions. We analyzed the impact of expectations on decision-making in a letter discrimination task, with \replaced{and}{or} without a \replaced{sAI}{sham-AI} system. 


First, in line with \citet{kosch2022placebo, villa2023placebo}, we found a subjective placebo effect: participants \replaced{upheld positive expectations for the sAI system's effectiveness}{retained belief in the sham-AI system's efficacy} post-interaction. Second, we observed a main effect at the behavioral level. Utilizing a Bayesian cognitive model of decision-making revealed that participants gathered information faster and altered their response style, giving us granular insights which aspects of interaction are affected by the placebo effect.
Third, contrary to previous work ~\cite{Denisova2019,kosch2022placebo,villa2023placebo}, we found no effect of verbal descriptions. Participants were biased, expecting increased performance with AI, irrespective of the verbal descriptions (AI performance bias). We replicate this bias in an online study ($N$ = 95). 

Our results resonate with recent calls in HCI to control for placebo effects in evaluating AI systems \cite{villa2023placebo, kosch2022placebo} and the power of AI narratives \cite{cave2019hopes,cave2019scary,sartori_minding_2023,KING2006740}. We add \deleted{to the literature}an AI performance bias \added{to the literature}, which makes the AI's placebo effect robust to manipulations of verbal system descriptions. \replaced{We describe which aspects of interaction are affected by the placebo effect}{We provide a behavioral marker for the placebo response to AI}, utilizing a cognitive model of decision-making \deleted{to reduce bias in AI-driven user interfaces}. We also discuss how, in a human-centered design process, the evaluation of AI must be done with user expectations in mind.

\section{Related Work}
\subsection{Expectations and the placebo effect of AI}
People hold expectations with regard to AI. Survey findings show that fears about AI's disruptive impact outweigh excitement in the British public \cite{cave2019hopes,cave2019scary}. This aligns with Sartori et al.'s report on the prevalence of \replaced{‘}{'}AI anxiety\replaced{’}{'} over perceived benefits \cite{sartori_minding_2023}. Interestingly, \replaced{it appears that the prevalence of concerns may also be influenced by narratives. For instance, science fiction portrayals have been suggested to contribute to the observed imbalance}{this is not driven by actual technology but by narratives, as even science fiction portrayals contribute to the imbalance} \cite{hermann2020beware}. The narratives about AI can differ among stakeholders and change over time \cite{bory2019deep}. Indeed, national policies in countries like China, Germany, the USA, and France underscore AI's disruptive potential \cite{bareis2022talking}, and these narratives are widely impactful, affecting usage \cite{bory2019deep,KING2006740}. Prior studies have explored key areas like transparency expectations \cite{meurisch_exploring_2020,park_human-AI_2021,purcell_fears_2023}, human-AI relationships \cite{zhang_ideal_2021}, \added{trust, \mbox{\cite{ueno2022TrustAI,liu2021TrustAI,vodrahalli2022TrustAI,ferrario2019TrustAI}}} and autonomy \cite{meurisch_exploring_2020}, forming the basis for AI interface design. However, there is a gap in understanding expectations of human-AI interaction outcomes, such as task performance with AI support \cite{kosch2022placebo}. \added{To address this gap, it is important to understand how exactly user expectations influence the outcome in human-AI interaction and to investigate how narratives play a role in this.}


The placebo effect \added{relies on expectations} \cite{Rickels1970Pills, Beckham1989ImprovementAE, lasagna1954study, Hrbjartsson2001IsTP, Diederich2008ThePT, Price2008ACR} \added{and} is not confined to medical contexts but also penetrates performance contexts like sports \cite{beedie2007positive}. Here, an inert substance given to athletes can improve but also deteriorate performance \cite{hurst2020placebo}. While placebo effects of AI in HCI and their effect on performance have recently been studied \cite{kosch2022placebo,villa2023placebo,denisova2015placebo,vaccaro2018illusion}, there is very little knowledge on nocebo effects.  
In HCI, a nocebo effect would negatively affect both performance and subjective metrics, like usability or user experience \cite{kosch2022placebo}. For example, \citet{halbhuber2022better} manipulated latency descriptions in gaming, showing that negative expectations reduced performance and user experience. \added{In human-AI interaction, }\citet{ragot2020} found that AI-generated art labeled as such was rated less favorably than if labeled as human-made. Thus, although first studies indicate the possibility of nocebo effects brought upon by technological artifacts, empirical studies directly leveling or even implementing negative expectations for AI are scarce \added{likewise it is unclear whether system descriptions as put forth by \mbox{\citet{kosch2022placebo}} determine AI expectations which bring placebo effects or whether general biases as in \mbox{\citet{ragot2020}} determine placebo effects. While the former could be addressed in a study context, the latter could only be adressed within a societal discourse \mbox{\cite{cave2019hopes,cave2019scary}}. \textbf{Therefore, it is critical to study how descriptions of AI influence placebo effects in HCI evaluation.}}

\subsection{Decision-making with AI}  
Decision-making, a process shaped by expectations and perceptions, involves selecting from a range of options \cite{SHAFIR199311}. The Drift Diffusion Model (DDM) serves as a cognitive framework for understanding this process, describing it as evidence accumulation until a decision boundary (a correct vs. an incorrect answer) is reached \cite{nunez2017attention,lerche2018speed,ratcliff_perceptual_2010,ratcliff_rouder_2000_dm_2C_Letteridentification}.
In its most basic form, the DDM models reaction times based on correct and incorrect responses in a random walk process toward a decision boundary, see \Cref{fig:teaser}. For a binary decision task with equal probability, we can assume three parameters. A drift-rate $\nu$, indicating the speed of gathering information, a boundary separation  $\alpha$, reflecting a decision-strategy, and a non-decision time $\tau$ parameter, reflecting motor preparation and perceptual processes unrelated to decision-making \cite{lerche2018speed}. 
\replaced{
 This model has been successful in predicting decision-making under uncertainty and in different cognitive tasks \mbox{\cite{ratcliff_perceptual_2010,ratcliff_rouder_2000_dm_2C_Letteridentification}}. Indeed, recent research argues that computational cognitive models like the DDM are central for interaction (see \citet{Oulas2022}). In line with this, the DDM has been applied to pedestrian crossing \cite{zgonnikov2022should}, moving target selection \mbox{\citet{lee2018moving}}, interactions with robots \mbox{\cite{huang2020human}} or teleoperations \mbox{\cite{corredor2016decision}}.}{The DDM's utility extends to HCI applications, including AI-enhanced interfaces, in various contexts like pedestrian crossing, social media, robot interactions or gaming\mbox{\cite{zgonnikov2022should, huang2020human,lee2018moving,corredor2016decision,chiossi2023short}}.}
Recent studies indicate that even sham adaptive AIs can influence user performance and risk-taking in decision-making \cite{kosch2022placebo,villa2023placebo}. However, the cognitive mechanisms behind these effects remain unclear.
 Applying the DDM could potentially shed light on the cognitive basis of the placebo effect for adaptive AI systems. Considering previous studies \added{by} \citet{kosch2022placebo} \added{and} \citet{villa2023placebo}, it appears plausible that the decision criterion may be affected by the implementation of positive expectations \added{(placebo)} improving performance (more liberal decision-making with decreased $\alpha$) and negative expectations (\added{nocebo), resulting in an} enlargement of $\alpha$\deleted{)}. \textbf{Consequently, users may make rapid, less accurate decisions when aided by an adaptive AI interface or slower, more accurate decisions when the AI system potentially hampers their performance.}

\section{Research model \deleted{and hypotheses}}

We conducted a mixed-design lab study with one between- and one within-subject factor, each with two levels. Two groups (between-subject) with different system descriptions, referred to as \textsc{Description} ("the AI system \replaced{worsens}{is worsening} performance and increases stress," referred to as \textsc{negative verbal description} condition vs. "the AI system \replaced{enhances}{is enhancing} performance and decreases stress," referred to as \textsc{positive verbal description} condition) were investigated. The within-subject factor for each group was the \added{s}AI' \replaced{\textsc{\sstatus{}}}{systems' \textsc{Status}} (\replaced{\textsc{\sAIactive{}}}{\textsc{sham-AI}} condition vs. \replaced{\textsc{\sAIinactive{}}}{\textsc{no-AI}}). The \added{\textsc{ORDER}}\footnote{\added{\textsc{order} was treated as a within-subject factor in the statistical analysis} \Cref{workloadandeda}\added{, addressing the question, "Is this condition from the first or the second half of the experiment."}} of conditions\added{ in \textsc{\sstatus{}}} was counterbalanced across participants \added{in both \textsc{description} groups}.

\deleted{
We address the following research questions and hypotheses with this design:}

\section{Method}
In the following, we motivate and document our methodological choices in realizing the study. The analysis with all associated measures can be found at \href{https://osf.io/8q7t6/?view_only=3c67187606a84cd990aa4c65ec3d16ec}{osf.io}. \added{The study was pre-registered, and the pre-registration details can be accessed at:} \href{https://aspredicted.org/blind.php?x=SX8_3BG}{aspredicted.org}. \added{Deviations from the pre-registration can be found in \Cref{tab:Deviations}.}

\subsection{\added{Verbal Description}}

\replaced{The study introduction varied in its verbal \textsc{description} between two groups. Participants in the \textsc{negative verbal description} group were informed that the system had previously "decreased task performance" and resulted in an "elevation in stress" among users. Moreover, they were informed that the system was new and untested, thus making it \textbf{"unreliable"} and \textbf{"risky"} for use in real-world scenarios. In contrast, the participants in the \textsc{positive verbal description} group were informed that the system had previously "enhanced task performance" while "reducing stress." They were also informed that the system was "cutting-edge," \textbf{"reliable"} and \textbf{"safe"} to use in real-world scenarios (\added{see \Cref{descriptions} for full descriptions}).
}{
For the \textsc{sham-AI} condition, participants were informed that the AI system was continuously adapting the task difficulty based on their task performance and stress levels, monitored through electrodermal activity via electrodes. They were advised that recognizing these pace adjustments might take some time.
In contrast, in the \textsc{\added{\sAIinactive{}}} condition, participants were informed that the AI system was not active and that the task pace was random.
}
\label{sec:sysDesc}
\replaced{
We informed all participants that they would be testing an AI system under two conditions: once with the AI' \textsc{\sstatus{}} set to active (\textsc{\sAIactive{}} condition) and once inactive (\textsc{\sAIinactive{}} condition). For the \textsc{\sAIactive{}} condition, participants were informed that the AI system was continuously adapting the task difficulty based on their task performance and stress levels, monitored through electrodermal activity via electrodes (see \Cref{status:AIActive}).
In contrast, in the \textsc{\sAIinactive{}} condition, participants were informed that the AI system was not active and that the task pace was random (see \Cref{status:AIInactive}).}{Participants in the \textsc{negative verbal description} group were informed that the system had previously "decreased task performance" and resulted in an "elevation in stress" among users. Moreover, they were informed that the system was new and untested, thus making it \textbf{"unreliable"} and \textbf{"risky"} for use in real-world scenarios. In contrast, the participants in the \textsc{positive verbal description} group were informed that the system had previously "enhanced task performance" while "reducing stress." They were also informed that the system was "cutting-edge," \textbf{"reliable"} and \textbf{"safe"} to use in real-world scenarios.}

\subsection{Measures} \label{Measures}

\subsubsection{Letter discrimination task}

\begin{figure} [!htbp]
    \centering
    \includegraphics[width = 14.5cm]{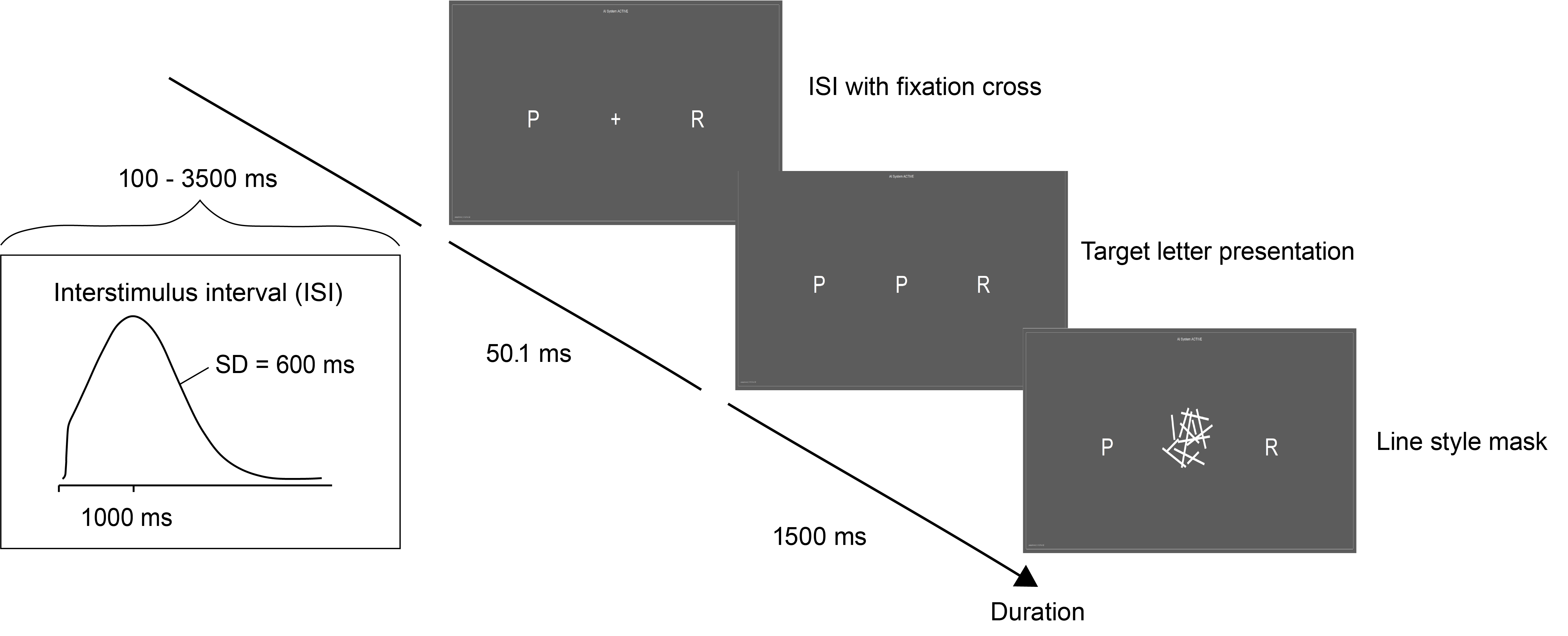}
    \caption{Trial sequence during the letter discrimination task. The duration of the ISI followed a Gaussian distribution (\textit{M} = 1000 ms, \textit{SD} = 600 ms). Key responses (left or right arrow) were logged during the presentation of the mask.}
    \label{trial_sequence}
\end{figure}

Two-alternative forced choice tasks, such as letter discrimination tasks, model simple decision-making and its underlying cognitive processes \cite{thapar_2003_Letter_discrimination_task,ratcliff_rouder_2000_dm_2C_Letteridentification, lerche_2017_trials_dm, Voss_2015_ddm_cognitiveprocesses}. 
In the task, participants must identify which of two letters, displayed on either side of a central target letter, matches the target. We used four letter pairs (E/F, P/R, C/G, Q/O), selected from \citet{ratcliff_rouder_2000_dm_2C_Letteridentification}. Each trial \added{consisted of a three-component trial sequence, which }began with a fixation cross centrally displayed between the letters for a variable time (interstimulus interval, ISI), facilitating perception of the system's adaptability similar to \cite{villa2023placebo}. 
Then, one of the letters was shown for 50.1 ms in the center of the screen \cite{thapar_2003_Letter_discrimination_task}. After this, a randomly sketched line mask rotated by $x\cdot360$ degrees (x $\in$ [0,1\replaced{[}{]}) and mirrored (vertically and/or horizontally, or neither) was shown in place of the target letter for 1500 ms; see \Cref{trial_sequence}. 
During the line mask presentation, the participants responded by pressing the left or the right arrow key (either index or middle finger). The first key press response during mask presentation time was recorded. The only critical change made to the task of \citet{thapar_2003_Letter_discrimination_task} was the randomly varying ISI. This was done to allow participants to track potential changes related to adaptation and should not affect task performance.

\FloatBarrier 

Each participant underwent 400 trials derived from \replaced{two}{2} Blocks $\times$ 100 trials of one random letter pair $\times$ \replaced{two}{2}\replaced{ \textsc{\sstatus} conditions (\textsc{\sAIactive} vs. \textsc{\sAIinactive}). }{\textsc{Status} (\textsc{sham-AI} vs. \textsc{no-AI}} \added{The order of the \textsc{\sstatus{}} conditions was counterbalanced across the participants in each \textsc{Description} group. The duration of each trial varied based on the randomized duration of the ISI in the trial sequence, which followed a Gaussian distribution with a mean of 1000 ms and a standard deviation of 600 ms. The shortest trial lasted 1650.1 ms, the longest lasted 5050.1 ms, and the median trial duration was 2550.1 ms. The overall median task duration for all 400 trials was approximately 17 minutes.} After each block, participants were offered a short break.

\subsubsection{Questionnaires} \label{Questionnaires}

\paragraph{Assessment of expectations}

We measured user expectations of performance and how they persisted after the interaction.
For overall performance expectations (judgments prior to interaction), we used four questions: A seven-point Likert item (1: Strongly disagree to 7: Strongly agree), \added{ indicating the expected overall performance (}\textit{I think I will perform better in the task with the AI system than in the task without the AI system.}\added{)}, \added{a slider item 
 from zero (slower) to 100 (faster)}\added{ as an indicator for the subjectively estimated task speed (}\deleted{on a slider (slower 0 to 100 faster)}\textit{I will be [slower/faster] in the task with the AI system active than in the task with the AI system inactive.}\added{)}, and two \replaced{open text}{slider} questions \replaced{(allowed response range: 0 to 100)}{from zero to 200} asking participants the expected number of correct letter discriminations in each condition (\textit{Out of 200 actions, how many do you expect to get correct [with/without] the AI system active?}). 
To evaluate judgments of performance following the interaction, identical questions phrased in the past tense were assessed.
An additional questionnaire adapted from \citet{villa2023placebo} was termed "System evaluation" and implemented to assess the participant's judgment of performance after the interaction, see \Cref{tab:customnscales}. 

\paragraph{Task load}
To \replaced{measure}{test H1 and H2 with respect to} workload, we implemented the NASA-TLX ~\citep{HART1988139}, a well-established questionnaire \cite{tosch2023}, with six dimensions: mental demand, physical demand, temporal demand, performance, effort, and frustration. Participants rated each dimension on a scale of 1 to 20, with higher scores indicating higher task loads. We calculated the raw score by summing up the item scores (Raw-TLX, \cite{hart_nasa-task_2006}).

\paragraph{Additional Questionnaires}
We assess user experience using the UEQ-S \cite{Schrepp_2017_UEQS} (8 item pairs; Likert scale from -3 to +3) with its two dimensions, pragmatic quality and hedonic quality. For measuring Usability, we used an adapted version of the System Usability Scale (SUS) \cite{Brooke_SUS_1996}, changing "system" to "AI system," adding the synonym \textit{awkward} for \textit{cumbersome} \cite{Finstad_SUS_non-native-speaker}, and computed the \replaced{SUS score by summing the score contributions of each item and multiplying the sum by a factor of 2.5 in line with \citet{Brooke_SUS_1996}.}{weighted sum score as an index of usability}

\subsubsection{\replaced{Electrodermal activity}{EDA} recording and pre-processing}

\added{Skin conductance, reflecting physiological arousal, was measured as an indicator for cognitive workload} \added{\cite{tosch2023}} \replaced{f}{F}ollowing the framework for \added{Electrodermal Activity (EDA) research in HCI} 
\cite{Babaei_2021_critique_EDA}\replaced{.}{,} EDA was recorded using standard Ag/AgCl electrodes (24 mm surface diameter) placed on the distal surfaces of the proximal phalanges of the index and middle fingers of the participant's non-dominant hand. Before testing, participants washed their hands with soap and cleaned the areas where the electrodes were placed using a 70\% alcohol wipe.
For data acquisition, we used the BITalino biomedical toolkit~\cite{guerreiro2013bitalino} to acquire the EDA signals via Bluetooth connection.
The \textit{OpenSignals (r)evolution} Python API Version 1.2.6\footnote{\url{https://github.com/BITalinoWorld/revolution-python-api\#bitalino-revolution-python-api/}} was set at \added{a} sampling rate of 100\,Hz. Time-series data were recorded using the Lab Streaming Layer (LSL)\footnote{\url{https://github.com/labstreaminglayer/}}. 
For offline data preprocessing, we used \added{the} Neurokit toolbox~\cite{makowski2021neurokit2}. After non-negative deconvolution analysis, we derived one metric of physiological arousal\replaced{:}{,} the mean tonic SCL in each block.

\subsection{Participants}
\label{sec:p1}
Participants were recruited through print advertisements in the [anonymized] area. Eligibility criteria included: no background in computer science, age above 18, self-reported normal or corrected-to-normal vision, no silver allergy, and no use of medication or history of epilepsy or other cognitive/motor impairments. The participants received 20 [anonymized] gift vouchers as compensation for their participation. The study was approved by an ethics committee (Grant Nr. <removed>).

We tested 66 participants in our study\footnote{Deviation from pre-registration see \Cref{tab:Deviations}}, excluding one for insufficient English proficiency \added{and one for careless responding (i.e. responding consistently with the maximum on a scale)}. Our final sample size consisted of \replaced{64 participants ($N$ = 64,  \mars = 24, \female = 40, zero non-binary or did not disclose) with an average age of 27.64 years ($SD$ = 6.49; min = 18; max = 49) reporting an average technical competence of 4.80 ($SD$ = 1.25)}{65 participants ($N$ = 65,  \mars = 25, \female = 40, zero non-binary or did not disclose) with an average age of 27.54 years ($SD$ = 6.49) reporting an average technical competence of 4.82 ($SD$ = 1.25)} on a 1 (low) - 7 (high) Likert \replaced{item}{scale}. To ensure that the two samples \added{(\textsc{description}: $n$\textsubscript{positive} = 31, $n$\textsubscript{negative} = 33)} are comparable, we checked their AI literacy using the Meta AI Literacy Scale\footnote{We only implemented items of factors loading onto the dimension "AI Literacy"} (MAILS) \cite{Carolus_2023_MAILS}, the Checklist for Trust between People and Automation (TiA) \cite{jian_2000_Trust_Automated_Systems} and the Subjective Information Processing Awareness Scale (SIPAS) \cite{Schrills_Franke_2021_SIPAS, Schrills_2021_SIPAS_validation, Schrills_Franke_2023_SIPAS_validity}. We indeed found no differences as a function of \textsc{Description} (see \Cref{table:MAILSTiASIPAS}).

\subsection{Procedure}
After consenting in line with the Declaration of Helsinki, the Bitalino device's electrodes were attached, and the device was activated and secured. The experimental program appeared on the screen. We then collected data on age, profession, handedness, caffeine or medication use, experimenter familiarity, and technical competence.

\begin{figure} [!htbp]
    \centering
    \includegraphics[width=1\textwidth]{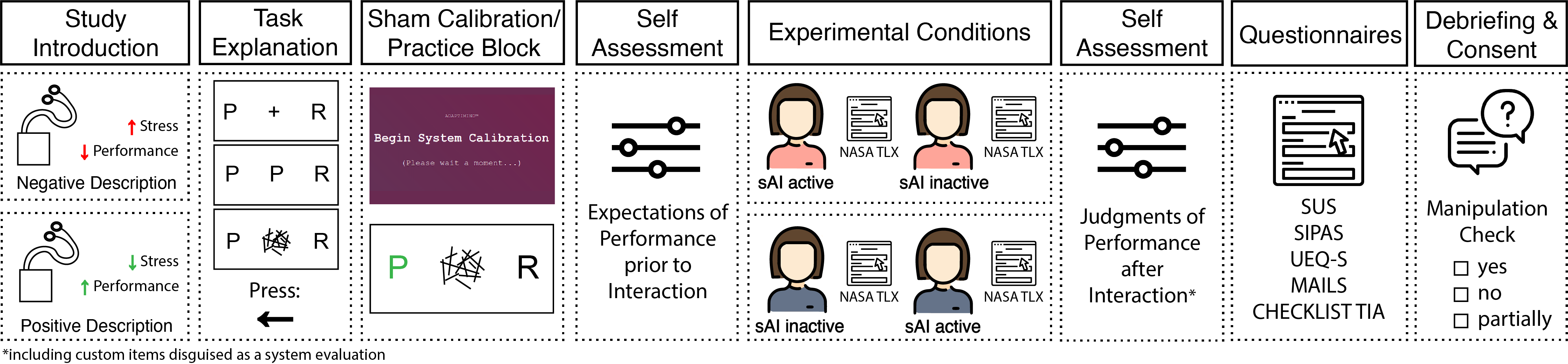}
    \caption{Study Procedure: In this mixed-design study examining the induction of placebo and nocebo effects, participants were divided into two groups (\textsc{Description}), with each group receiving altered system descriptions \added{(\textsc{negative}: AI decreased task performance and increased stress in users/ \textsc{positive}: AI increased task performance and decreased stress in users)}. Participants in each group performed a letter discrimination task under two conditions (\replaced{\textsc{\sstatus}}{\textsc{Status}}): in the \replaced{\textsc{sham-AI (\sAI){} active}}{\textsc{sham-AI}} condition, they were informed \replaced{that an AI system was active and adjusting the task pace based on their measured stress responses}{the task pace was adjusted by an AI system based on their measured stress responses}; in the \replaced{\textsc{\sAIinactive{}}}{\textsc{no-AI}} condition, they were told \added{that the AI system was inactive and} adjustments in task pace were random. \added{The \textsc{order} of \textsc{\sstatus{}} alternated within each  \textsc{description} group.} Before and after interacting with the \replaced{\sAI}{sham-AI} system, expectations on performance with and without the \replaced{sAI}{sham-AI} system set as active were assessed. After the tasks and before debriefing, additional questionnaires assessing i.e., user experience and AI literacy were implemented. Ultimately, we revealed the manipulation and assessed the participants' belief in the manipulation.}
    \label{fig:study_procedure}
    
\end{figure}

Participants read an introductory text explaining the AI system and apparatus (see \Cref{fig:study_procedure}). Depending on the \textsc{Description} assignment, the text included a positive or negative verbal description (\Cref{sec:sysDesc}) before interacting with the \replaced{sAI}{sham-AI}. This was followed by a survey asking for information on the system being evaluated, see \citet{villa2023placebo}.

\FloatBarrier %

Before the task, participants completed 50 practice trials with visual feedback labeled as calibration. We then assessed their performance expectations with and without \replaced{the AI system set to active}{AI}. \added{Next, participants performed the task, starting with either the \textsc{\sAIactive{}} or \textsc{\sAIinactive{}} condition}\footnote{\added{During the entire task, information was displayed on the screen indicating that the AI' \textsc{\sstatus{}} was set to either active or inactive.}}\added{, depending on the assigned \textsc{order}.} \deleted{Participants then performed the task in either the \textsc{sham-AI} or \textsc{no-AI} condition, based on group assignment.} Task load was evaluated post-condition using TLX \cite{HART1988139}. After both conditions, the AI system was further assessed (\Cref{Questionnaires}). Participants were then debriefed, re-consented, and their belief in the manipulation checked (\Cref{manipulationcheck}) before thanking and compensating them. \added{The entire experiment lasted approximately 70 minutes.}

\subsection{Bayesian Data Analysis and Inference}

We adopted a Bayesian approach, utilizing Bayesian linear mixed models\footnote{For a guide on Bayesian techniques, see \cite{schad2021toward,van2021bayesian,dix2022bayesian, burkner2017brms,kay2016researcher}\deleted{\mbox{\cite{URBANIAK2022107371,ACKERMANS2019248,GUERONSELA2023107572}} }}.
 \added{For parameter estimation, we used brms \mbox{\cite{burkner2017brms}}, a wrapper for the STAN-sampler \mbox{\cite{carpenter2017stan}} executed in R \mbox{\cite{R-base}}. Two Hamilton-Monte-Carlo chains were computed, each with 8,000-40,000 iterations and a 20\% warm-up. Trace plots of the Markov-chain Monte-Carlo permutations were inspected for divergent transitions and autocorrelation, and we checked for local convergence. All Rubin-Gelman statistics~\mbox{\cite{gelman1992inference}} were well below 1.1 and the Effective Sampling Size was over 1000. Model specifications and their non-informative priors alongside all estimated parameters can be found in \Cref{modelsparam}}.

We then analyzed the posterior of the model. To investigate a parameter's distinguishability from zero, we utilized $p_b$, which resembles the classical $p$-value but quantifies the effect's likelihood of being zero or opposite \cite{hoijtink2018testing,shi2020reconnecting}. Effects with $p_b$ $\leq$ 2.5\% were deemed distinguishable. We also calculated the 95\% High-Density Interval (HDI) for each model parameter. For population-level effects in simple regression models, we set priors for regression parameters to one standard deviation of the outcome variable.
All binary factors were effect coded (\textsc{Time} (pre/post): 1, -1; \replaced{\textsc{\sstatus{}}}{\textsc{status}} (\replaced{\textsc{\sAIactive{}}}{\textsc{sham-AI}}/\replaced{\textsc{\sAIinactive{}}}{\textsc{no-AI}}): 1, -1; \textsc{Description} (negative/positive): 1, -1); \added{\textsc{Order} (first condition/ second condition): 1,-1}.

\subsection{\added{Apparatus}}
The experiment was carried out using Chromium on a Linux (Ubuntu 22.04.2 LTS) laptop (Dell Latitude 7310) with an i5 (Intel Core i5-1031U) processor and 16GB of RAM. A separate monitor (HP E27uG4) displayed the paradigm with a screen size of 27 inches (2160px*1440px) and a refresh rate of 60\,Hz. The monitor's position was adjusted according to the participant's eye level. Screen distance was roughly 60\,cm (23,6\,inch). 
We built a custom experiment that ran locally using JavaScript using the lab.js library version 20.2.4 \cite{henninger_2021_labjs}.

 \begin{figure}[tbh]
     \centering
     \includegraphics[width=0.45\textwidth]{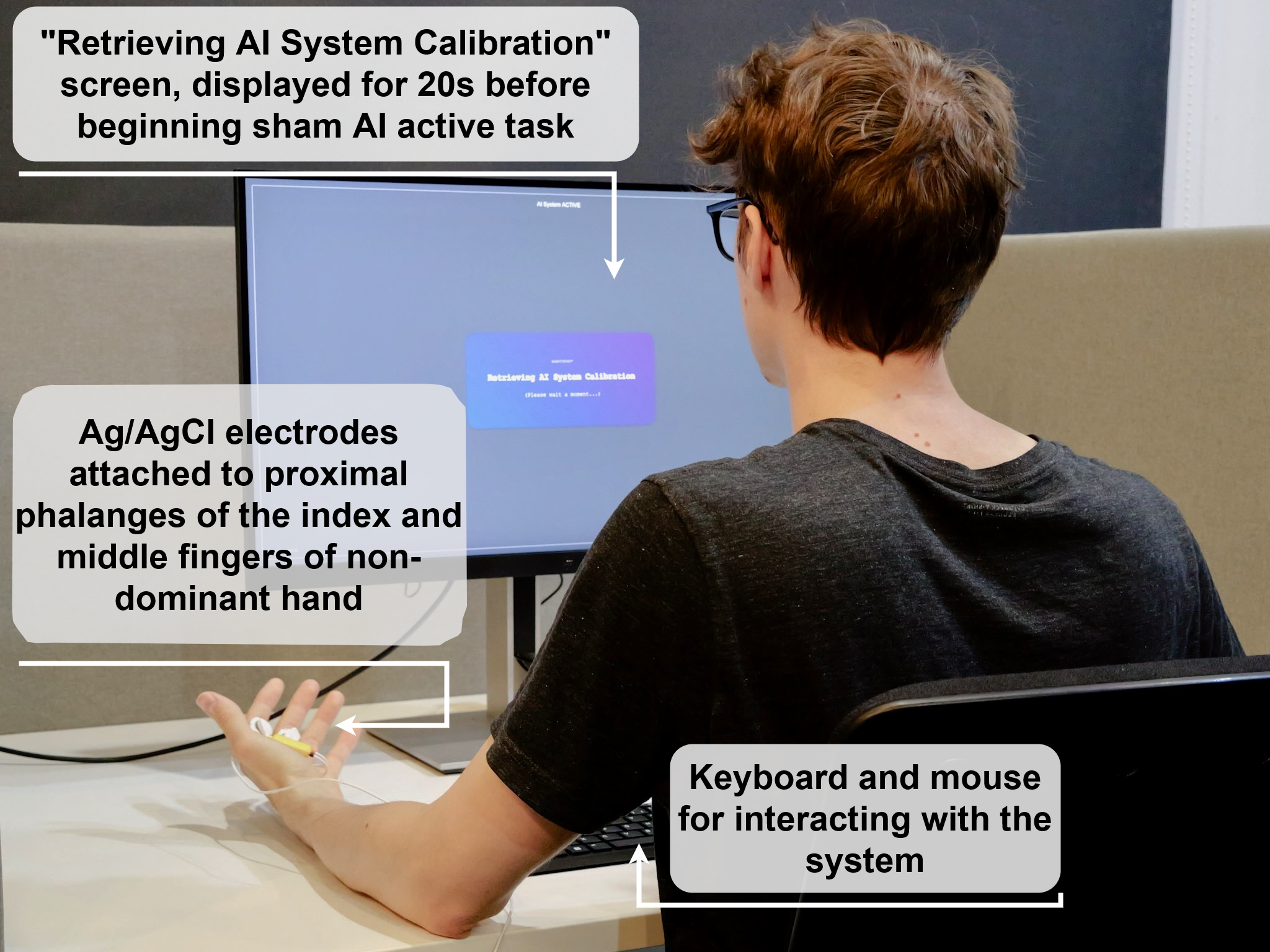}
     \caption{The participants interacted with the system with their dominant hand using a keyboard and a mouse.}
     \label{fig:setup}
     \FloatBarrier
 \end{figure}
\FloatBarrier

\section{\added{Research Questions and Hypotheses}}

We address the following research questions and hypotheses\deleted{ with this design}:

\begin{table}[h]
\centering
\renewcommand{\arraystretch}{1.2}
\caption{Research Questions and Hypotheses}
\label{tab:ResearchQuestionsHypotheses}
\resizebox{\textwidth}{!}{
\begin{tabular}{p{4.9cm}p{9.5cm}}
\multicolumn{1}{c}{Research Question} & \multicolumn{1}{c}{Hypotheses} \\ \toprule
\multirow{2}{5cm}{\textbf{RQ1}: Do subjective ratings on performance and mental workload differ between negative and positive verbal descriptions (nocebo/placebo)?} & \vspace{-11px} \textbf{H1}: Lower subjective performance (\textbf{H1.1}) and higher mental workload (\textbf{H1.2}) in \replaced{\sAIactive{}}{sham-AI} with a negative description (nocebo) compared to \replaced{\sAIinactive{}}{no-AI}. \\ 
 & \textbf{H2}: Higher subjective performance (\textbf{H2.1}) and lower mental workload (\textbf{H2.2}) in \replaced{\sAIactive{}}{\textsc{sham-AI}} with a positive description (placebo) compared to \replaced{\sAIinactive{}}{no-AI}. \\ \\
\multirow{2}{5cm}{\textbf{RQ2}: Do verbal descriptions of a \replaced{\sAI{}}{sham-AI} affect decision-making (e.g., in a letter discrimination task)?} & \vspace{-11px} \textbf{H3}: More conservative speed-accuracy trade-off in \replaced{\sAIactive{}}{sham-AI} with a negative description (nocebo) compared to \replaced{\sAIinactive{}}{no-AI}. \\ 
 & \textbf{H4}: More liberal speed-accuracy trade-off in \replaced{\sAIactive{}}{sham-AI} with a positive description (placebo) compared to \replaced{\sAIinactive{}}{no-AI}. \\ \\
\multirow{2}{5cm}{\textbf{RQ3}: Do verbal descriptions of a \replaced{\sAI{}}{sham-AI} affect physiological indicators of cognition when compared to \replaced{no implementation of a \sAI{}}{no-AI}?} & \vspace{-11px} \textbf{H5}: Higher levels of physiological arousal (measured by mean tonic skin conductance level (SCL)) in \replaced{\sAIactive{}}{sham-AI} with a negative description (nocebo) compared to \replaced{\sAIinactive{}}{no-AI}. \\ 
 & \textbf{H6}: Lower levels of physiological arousal (measured by mean tonic SCL) in \replaced{\sAIactive{}}{sham-AI} with a positive description (placebo) compared to \replaced{\sAIinactive{}}{no-AI}. \\ \bottomrule
\end{tabular}}
\end{table}

\section{Results}

\subsection{Manipulation Check}\label{manipulationcheck}
To the question \textit{Did you believe that an AI system was implemented to adapt task pace?} with possible answers being \textit{Yes, No} or \textit{Partially}, 
 \added{10 of 64 (15.62\%; 6 of 33 negative description; 4 of 31 positive description) responded with "no" and did not believe in the system's capabilities. 27 out of 64 participants (42.19\%; 13 for positive description, 14 for negative description) participants reported some suspicion of the system's functionality. Thus, 27 of 64 participants fully believed in the implemented system.}

\subsection{Performance Expectations and Judgments of Performance}
\subsubsection{Subjective overall performance}
\label{Subjective overall performance}
To analyze expected overall performance, we centered the values by subtracting four points of the \replaced{Likert item}{scale} so that 0 indicates not favoring any condition and modeled overall performance estimates as a function of \textsc{Time} and \textsc{Description}\footnote{Gaussian link-function with default priors.}. \added{Overall, participants were positive about the \sAI{}, $Intercept$ = 0.51 [0.25, 0.77], $p_b$ = 0.00\%. However, participants showed no difference in subjective performance before and after interaction with the \sAI{} ($\tilde{b}_\text{Time}$ = 0.19 [-0.03, 0.42], $p_b$ = 6.70\%). 
There was no main effect of \textsc{Description} ($\tilde{b}_\text{Description}$= -0.23 [-0.50, 0.03], $p_b$ = 4.66\%) 
 and no interaction effects ($\tilde{b}_\text{Time $\times$ Description}$= 0.16 [-0.06, 0.38], $p_b$ = 7.99\% 
), see also \Cref{fig:expect}A.}

\subsubsection{Subjective estimated task speed}
\label{Subjective estimated task speed}
\added{We computed a similar model to investigate \added{the }participants' expected task speed by subtracting 50 points so that zero indicates a neutral response.} \Cref{fig:expect}B \added{shows the average expected speed across all conditions being positive, $Intercept$ = 8.54 [5.41, 11.78], $p_b$ = 0.00\%). The participants believed to be faster with the \sAIactive{} before interacting with the system ($M$ = 62.47, $SD$ = 17.31) than after ($M$ = 54.56, $SD$ = 18.01). This difference ($d_z$ = 0.30) could be distinguished from zero, $\tilde{b}_\text{Time}$ =  3.96 [0.92,  7.03] , $p_b$ = 0.50\%.
We found no differences for \textsc{Description} $\tilde{b}_\text{Description}$ = -1.31 [-4.52,  1.88], $p_b$ = 21.09\%
or interaction effects, \textsc{Description $\times$ Time} $\tilde{b}_\text{Description $\times$ Time}$ = -1.16 [-4.17,  1.93], $p_b$ = 22.46\%.}

\subsubsection{Subjective estimated number of correct responses}
\label{Subjective estimated number correct}
\begin{figure}[tbh]
    \centering
    \includegraphics[width=1\textwidth]{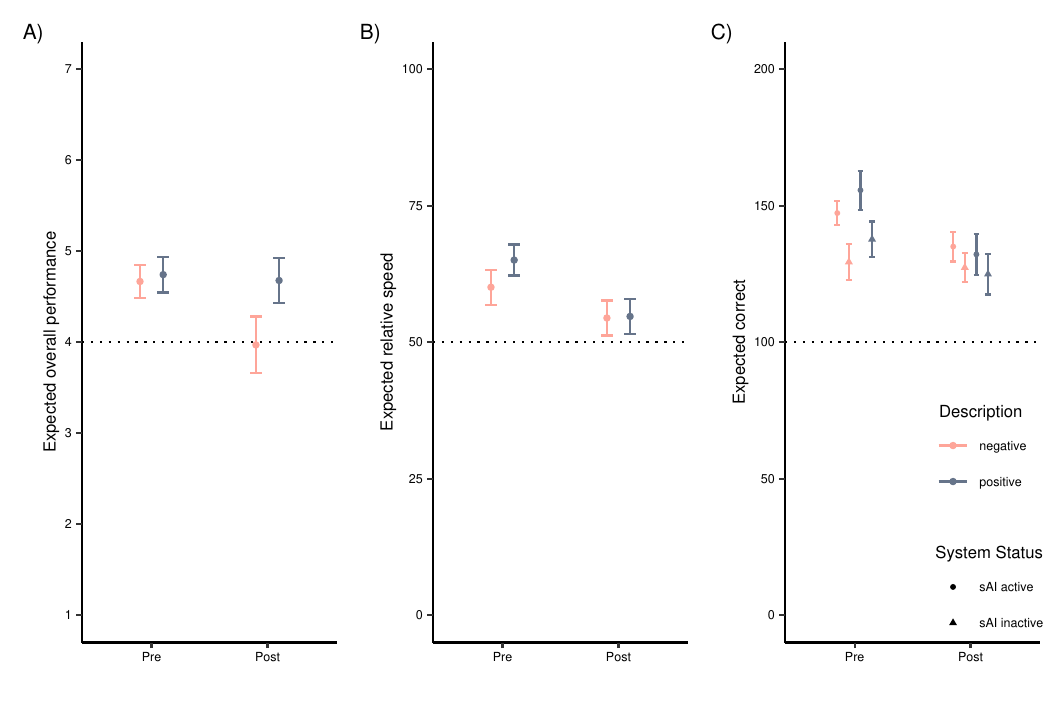}
    \caption{A: Mean expected performance as a function of \textsc{Time} and \textsc{Description}. B: Mean expected relative speed as a function of \textsc{Time} and \textsc{Description}. C: Mean \added{expected }correct responses before and after interacting with the \replaced{\sAI}{sham-AI} system as a function of \textsc{Time}, \added{\textsc{\SStatus{}}} and \textsc{Description}. Error bars denote +-1 standard error of the mean.}
    \label{fig:expect}
\end{figure}

\added{We expanded the statistical model to consider \textsc{\sstatus{}} for estimated points (no transformation) in each condition; see also \Cref{fig:expect}C. Participants indicated that in the \textsc{\sAIactive{}} condition ($M$ = 142.46, $SD$ = 35.87), they would score more points than in the \textsc{\sAIinactive{}} condition ($M$ = 129.77, $SD$ = 36.56). This difference was not zero $\tilde{b}_\text{\SStatus{}}$ =  6.33  [ 3.23, 9.26] , $p_b$ = 0.00\%, $d_z$ = 0.53. Participants believed to score more points before performing the task ($M$ = 142.36, $SD$ = 36.16) than after ($M$ = 129.88, $SD$ = 33.31 $d_z$ = 0.47), $\tilde{b}_\text{Time}$ =  6.31 [3.31, 9.33], $p_b$ = 0.00\%, resembling \citet{kosch2022placebo}.
We found no distinguishable effects for \textsc{Description} $\tilde{b}_\text{Description}$ = -1.35 [-8.78, 5.78], $p_b$ = 35.51\%,
or any interaction effects $p_b$ > 4.07\%, see also \Cref{fig:expect}C.}

Therefore, participants were biased toward a superior performance with AI even when given a negative verbal description of the system. We refer to this as \textsc{AI performance bias}.

\subsection{Performance data}
\replaced{We excluded 6 out of 64 participants (9.38\%) only from the behavioral data analysis as they did not comply with our task (percent correct <60\% in one of the conditions or very large number of misses >35\%). 
We deleted the first trial in each trial block along with too-short responses by filtering reaction times (RT) under 150 ms (519 out of 23084; 2.25\%)}{We excluded 6 out of 65 participants (9.23\%) from the behavioral data analysis as they did not comply with our task (percent correct <60\% in one of the conditions or very large number of misses >35\%). We deleted the first trial in each block along with too short responses by filtering reaction times (RT) under 150 ms (522 out of 23484; 2.22\%)}\footnote{Deviation from pre-registration see \Cref{tab:Deviations}}\replaced{and missed responses with RT > 1499 ms (32 out of 22565; 0.14\%).} {and missed responses with RT > 1499 ms (32 out of 22930; 0.14\%).}

\begin{table}[h]
\centering
\caption{\added{Mean percent correct and reaction time (RT) for both correct and incorrect trials as a function of \textsc{\added{\SStatus{}}} and \textsc{Description}}}
\label{table:combined_results}
\resizebox{\textwidth}{!}{
\begin{tabular}{lcccccc}
\toprule
 & \multicolumn{3}{c}{\replaced{\sAIactive{}}{Sham-AI}} & \multicolumn{3}{c}{\replaced{\sAIinactive{}}{No-AI}} \\
\cmidrule(lr){2-4} \cmidrule(lr){5-7}
 Description & Correct \% & Correct RT & Incorrect RT & Correct \% & Correct RT & Incorrect RT \\
\midrule
Negative  & 91.11 (8.86) & 586.80 (75.61) & 719.16 (168.01) & 90.35 (6.68) & 596.56 (43.71) & 739.32 (156.84) \\
Positive & 88.87 (9.61) & 599.44 (96.61) & 723.33 (147.19) & 88.21 (10.76) & 603.19 (95.69) & 722.31 (161.11)\\
\bottomrule
\end{tabular}}
\end{table}

\begin{figure}[tbh]
    \centering
    \includegraphics[width=1\textwidth]{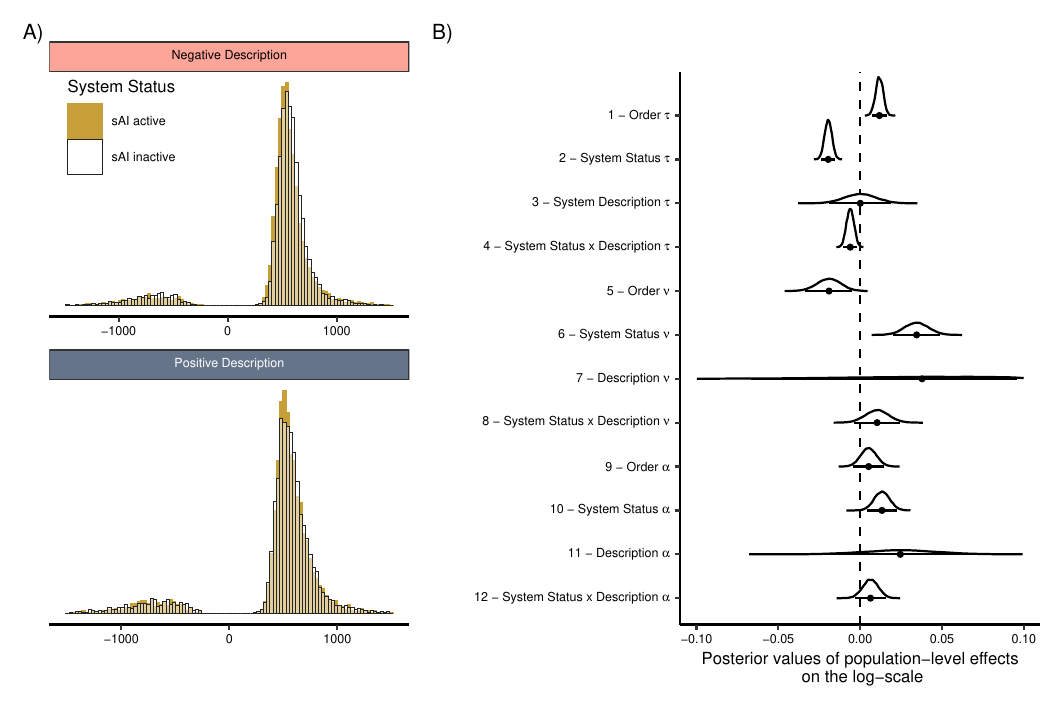}
    \caption{A: Reaction time distribution as a function of \textsc{\replaced{\SStatus{}}{status}} (\added{\sAIactive{}} vs. \added{\sAIinactive{}}) and \textsc{Description} (incorrect trials are multiplied with -1). B: Posterior density plot for the parameter values for all population-level parameters 95\% High-Density Interval (HDI). If the HDI does not cross the midline $p_b$  will be <2.5\%.}
    \label{fig:DDM}
\end{figure}

\added{To explore our interventions, we computed two separate regression models with varying intercepts for each participant and, \textsc{Order} (first vs second experimental block), \textsc{\SStatus{}} and \textsc{Description} as population-level effects for RT (Gaussian-link function) and correctness of response (Bernoulli-link function). For RT, we found an effect for \textsc{\SStatus{}}, $\tilde{b}_\text{\SStatus{}}$=  -4.17 ms  [-6.14,  -2.17], $p_b$ = 0.00\%. Participants reacted on average faster in the \added{\textsc{\sAIactive{}}} condition ($M$ = 604 ms, $SD$ = 92 ms) to stimuli as compared to the \added{\textsc{\sAIinactive{}}} condition ($M$ = 611 ms, $SD$ = 79 ms; Cohen's $d_z$ = 0.12) \Cref{fig:DDM}A. We also found that participants increased their response speed from the first to the second experimental condition; we found an effect for \textsc{Order}, $\tilde{b}_\text{Order}$=  11.05 ms  [9.06,  13.06], $p_b$ = 0.00\% (First condition: $M$ =619 ms, $SD$ = 91 ms; Second condition: $M$ = 596 ms, $SD$ = 79 ms;  Cohen's $d_z$ = 0.39). There was no effect of \textsc{Description}, $\tilde{b}_\text{Description}$=  -4.76 ms [-26.43,  16.43], $p_b$ = 32.80\%. For the correctness of responses, we found the same pattern of results. Participants were more likely to respond correctly in the \added{\textsc{\sAIactive{}}}} ($M$ = 90.07\%, $SD$ = 9.20\%) condition as compared to the \added{\textsc{\sAIinactive{}}} condition ($M$ = 89.35\%, $SD$ = 8.80\%; $\tilde{b}_\text{\SStatus{}}$=  -0.05 [0.00,  0.09], $p_b$ = 2.05\%; Odds = 0.95) and improved in accuracy along the course if the experiment (\textsc{Order}), $\tilde{b}_\text{Order}$=  -0.05 [-0.09,  0.00], $p_b$ = 1.37\%, Odds = 1.05 (First condition: $M$ = 89.29\%, $SD$ = 9.75\%; Second condition: $M$ = 90.13\%, $SD$ = 8.18\%). There was no effect of \textsc{Description}, $\tilde{b}_\text{Description}$=   0.10 [-0.12,  0.33], $p_b$ = 19.00\%.

\added{Be reminded that in the DDM, we model  RTs based on correct and incorrect responses by fitting data to a model that represents decision-making as the noisy accumulation of information ($\nu$ denoting the average rate of accumulation), for one choice or the other, until a threshold is reached ($\alpha$; boundary separation). A starting point from which the accumulation process starts and a parameter  $\tau$ denoting non-decision time is added to the model. For a visual representation, see} \Cref{fig:teaser}.

We computed the DDM to test H3 \& H4  on the reaction time data\footnote{Deviation from pre-registration see \Cref{tab:Deviations}}, see \Cref{fig:DDM}A. A  hierarchical form of this model was built accounting for inter-subject variability with a varying intercept and a population-level effect for each \textsc{\replaced{\SStatus{}}{status}} and an interaction term for \textsc{Description} for each $\tau$, $\nu$ and $\alpha$.

\begin{figure}[tbh]
    \centering
    \includegraphics[width=1\textwidth]{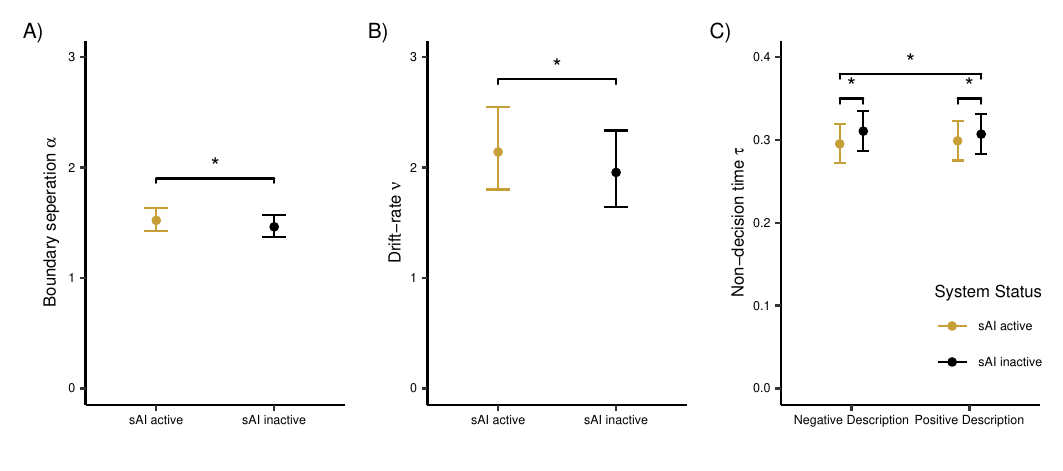}
    \caption{ A: Estimates with High-Density Interval (HDI) 95\% for  boundary separation, $\alpha$, as a function of \textsc{\replaced{\SStatus{}}{status}}. B: Estimates with HDI 95\% for the drift rate $\nu$ as a function of \textsc{\replaced{\SStatus{}}{status}}. C: Estimates of non-decision time $\tau$ with HDI 95\% as a function of \textsc{\replaced{\SStatus{}}{status}} and \textsc{Description}.}
    \label{fig:wienerest}
\end{figure}

We inspected the parameters of the model, see \Cref{fig:DDM}B, for differences in \textsc{\replaced{\SStatus{}}{status}} \added{\Cref{fig:DDM}B-10} and \textsc{Description} \added{\Cref{fig:DDM}B-11} for boundary separation $\alpha$\replaced{. S}{,s}ee \Cref{fig:wienerest}A, to see whether the difference in reaction time \added{and percent correct} comes from a change in the participant's strategy, e.g., prioritizing speed over the accuracy. We found that in the \added{\sAIactive{}} condition, participants had a slightly larger boundary separation, $\alpha$, making them slightly more conservative as compared to the \added{\sAIinactive{}} condition \added{(Figures \ref{fig:DDM}B-10 and \ref{fig:wienerest}A)}. However, we also found that $\nu$ (drift rate), see \Cref{fig:wienerest}B, was higher for \added{\sAIactive{}} as compared to \added{\sAIinactive{}}. Thus, information accumulation was relatively faster in the \added{\sAIactive{}} condition, see \Cref{fig:DDM}B\added{-6}. \added{With a relatively faster accumulation of information, $\nu$, and more conservative boundaries, $\alpha$, in the \added{\sAIactive{}} condition as compared to \added{\sAIinactive{}}, we can explain the differences between conditions for the singular analysis of RT and the correctness of trials (for a schematic representation of this difference for \textsc{\replaced{\SStatus{}}{status}}, see \Cref{fig:teaser}). Note that when we inspected the posterior distribution for each participant, as well as their RT difference as a function of \textsc{\replaced{\SStatus{}}{status}}, we found that the effect did not vary as a function of post-experimental belief in the system, see \cref{manipulationcheck}. Therefore, the model seems to hold for all participants, irrespective of their beliefs after debriefing. }

Similarly, $\tau$ was also affected by \textsc{\replaced{\SStatus{}}{status}} with an interaction with \textsc{Description} qualifying the effect, see \Cref{tab:model-outputs}. Looking at \Cref{fig:DDM}A and \Cref{fig:wienerest}C, we can see that the group with the negative description had a slightly earlier onset in RT. For all parameter values, see \Cref{tab:model-outputs} and for the mathematical formulation and priors, see \Cref{BRMS_DDM}. \added{To contextualize the effect size on RT, we also predicted the RT from the model, averaged across conditions, and calculated Cohen's $d_z$ for \textsc{order}, at 1.21, and for \textsc{\replaced{\SStatus{}}{status}}, at 0.74. }

\subsection{Workload and physiological arousal}
\label{workloadandeda}

Investigating H1.2, H2.2, H5 and H6, we computed a regression model for the NASA-TLX raw data with \textsc{\replaced{\SStatus{}}{status}}, \textsc{Description}, their interaction and \textsc{Order} as predictors found. We found no differences for \textsc{\replaced{\SStatus{}}{status}} $\tilde{b}_\text{\replaced{\SStatus{}}{status}}$ = -0.08[-2.04, 1.90], $p_b$ = 46.82\%, \textsc{Description} $\tilde{b}_\text{Description}$ = 0.83 [3.72, 5.42], $p_b$ = 35.91\%, their interaction effects $\tilde{b}_\text{\replaced{\SStatus{}}{status} $\times$ Description}$ = 0.43 [-1.54, 2.39], $p_b$ = 33.28\% or \textsc{Order}, $\tilde{b}_\text{Order}$ = 1.45 [-0.52,  3.42], $p_b$ = 7.28\%. \deleted{The EDA data largely resembled the TLX data}For EDA,\footnote{same predictor formula; 6 participants excluded due to poor signal quality}, there was no effect of the \textsc{\replaced{\SStatus{}}{status}}, $\tilde{b}_\text{\replaced{\SStatus{}}{status}}$ =  -0.20  [-0.53,  0.13], $p_b$ = 11.88\%, 
no effect of \textsc{Description}  $\tilde{b}_\text{Description}$ = 0.25 [-0.10,  0.59], $p_b$ = 7.37\%,  
 no interaction effect, $\tilde{b}_\text{Description $\times$ \replaced{\SStatus{}}{status}}$ = 0.07 [-0.24,  0.38], $p_b$ = 32.70\% or \textsc{Order} effect, $\tilde{b}_\text{Order}$ = 0.20 [-0.03,  0.42], $p_b$ = 4.20\%.

\subsection{Usability and User Experience}

Except for \textit{The AI system made the task easier} \added{(item 2)}, which was viewed more favorably with a positive description, there were no significant differences in \textsc{Description} (\Cref{tab:customnscales}). Participants slightly disagreed with \textit{The task was easy} \added{(item 1)} and were slightly negative about \textit{The AI system improved my cognitive abilities} \added{(item 7)}. Yet, similar to \citet{kosch2022placebo,villa2023placebo}, they agreed that the AI has future potential.

For UEQ-S scales, we found an overall positive user experience, with no group differences on hedonic or pragmatic attributes, with both having positive values indicating a positive user experience. SUS ratings indicated that the system was rated average in terms of usability unaffected by \textsc{Description}.

\begin{table}[tbh]
\centering
\renewcommand{\arraystretch}{1.1}
\small
\caption{\replaced{The customized system evaluation was answered on nine 7-point Likert items (1: strongly disagree; 7: strongly agree)}{Custom items for System evaluation were answered on a 7-point Likert scale (1 - strongly disagree; 7 - strongly agree)}. We estimate the difference towards a neutral value and compare the samples across \textsc{Description}. A neutral value for the custom items was 4, for the UEQ-S scales, \added{it} was set to zero. We fitted a robust regression model for each comparison. For the \added{adapted} SUS, the expected average is 68. Distinguishable effects from a neutral value (expected) or for each \textsc{Description} are marked with *. We used a studentized link-function with priors scaled to one $SD$}
\resizebox{\textwidth}{!}{
\begin{tabular}{lcccccc}
\toprule
Item/scale & $M_{neg}$ ($SD$) & $M_{pos}$ ($SD$) &   $\tilde\Delta_\text{expected}$ [HDI 95\%]  &  $p_b$ & $\tilde{b}_\text{Description}$  [HDI 95\%] &  $p_b$\\
\midrule
System evaluation &  &  &  &  &  &  \\
\textit{\hspace{1mm} 1 - The task was easy.}                             & 3.24 (1.62) & 3.61 (1.82)& -0.60 [-1.03,-0.16]& 0.45\%*& -0.17  [-0.63,  0.25]& 21.58\%\\
\textit{\hspace{1mm}The AI system}                             &  &  &  &  &  &  \\
\textit{\hspace{1mm} 2 - made the task easier.}            & 3.36 (1.78) & 4.19 (1.42)& -0.24  [-0.66,  0.16]& 12.54\%& -0.41  [-0.82, 0.00]& 2.48\%*\\

\textit{\hspace{1mm} 3 - made the task more enjoyable.}    & 3.33 (1.83) & 4.03 (1.43)& -0.32  [-0.74, 0.10]& 6.39\%& -0.35  [-0.76, 0.07]& 5.15\%\\

\textit{\hspace{1mm} 4 - made me more self-confident.}     & 3.39 (1.49) & 3.97 (1.72)& -0.33  [-0.74, 0.08]& 5.87\%& -0.27  [-0.68, 0.14]& 9.35\%\\

\textit{\hspace{1mm} 5 - made me more efficient.}          & 3.87 (1.55) & 4.16 (1.27)& 0.02  [-0.33, 0.39]& 24.80\%& -0.12  [-0.48, 0.23]& 44.47\%\\

\textit{\hspace{1mm} 6 - improved my performance.}         & 3.96 (1.55) & 4.19 (1.33)& 0.09  [-0.27, 0.45]& 31.37\%& -0.11  [-0.46, 0.27]& 28.33\%\\

\textit{\hspace{1mm} 7 - improved my cognitive abilities.} & 3.55 (1.39) & 3.77 (1.34 )& -0.35  [-0.68, -0.01]& 2.06\%*& -0.09 [-0.43,  0.25] & 29.47\%\\

\textit{\hspace{1mm} 8 - has a lot of potential for} & \multirow{2}{*}{5.03 (1.15)} & \multirow{2}{*}{5.42 (1.12)}& \multirow{2}{*}{1.23  [0.93, 1.51]}& \multirow{2}{*}{0.00\%*}  & \multirow{2}{*}{-0.20  [-0.49, 0.09]}& \multirow{2}{*}{8.98\%}\\ 

\textit{\hspace{4mm} future development.} &  &  &  &  &  &  \\

UEQ-S-Pragmatic       & 0.54 (0.93) & 0.85 (1.19) & 0.73  [0.46, 0.99] & 0.00\%* & -0.17 [-0.43, 0.10] & 10.79\%\\
UEQ-S-Hedonic         & 0.73 (1.03) & 0.80 (1.32) & 0.78 [0.49, 1.08] & 0.00\%*  & -0.04 [-0.33, 0.26] & 40.80\%\\
SUS\added{-Adapted}  & 64.62 (13.86) & 67.74 (17.43) & -1.41 [-5.39, 2.47] & 23.55\% & -1.71 [-5.69, 2.18] & 19.28\%\\

\bottomrule
\end{tabular}}
\label{tab:customnscales}
\\[1ex]  
\small
\raggedright
\textit{Note: In this figure, "\textit{neg}" denotes a negative system description, while "\textit{pos}" represents a positive one.}
\end{table}

\section{Replication Study: Positive expectations for negative descriptions}

To confirm the \textsc{AI performance bias}, we conducted an \added{additional} online replication study with \added{the} negative system description\deleted{s}. We replicated the first part of the previous study\replaced{ using the previous negative verbal description and subjective questions to assess expectations and judgments in the previous study}{\mbox{\footnote{We used the same questions to assess expectations and judgments in the previous study}}}. \added{Subsequently, we replaced the adaptation description, which initially referred to utilizing real-time EDA analysis to measure stress responses, with the use of computer vision technology to analyze facial expressions in real-time, as per \citet{kosch2022placebo} (no data was recorded).} \replaced{To address potential concerns about participants not fully comprehending the instructions,}{As one could argue that participants did not comprehend the instructions,} we set up the experiment to enforce comprehension of verbal descriptions. \replaced{Based on this, the participants were divided into two groups. Both groups read the negative system description. However, one group was asked to complete a comprehension check (\textsc{Comprehension}), ensuring they fully understood the negative description, before being able to continue to the next part of the study. In the \textsc{no-comprehension} group, participants were not bound by the same requirement, allowing for variations in their engagement with the negative description. This decision was made to facilitate a comparison between participants in the comprehension group, where individuals were required to fully understand the text, indicating a predicted decline in performance, and the no-comprehension group. While some may not have read it thoroughly, others may have held pre-existing expectations. This contrast allows a nuanced exploration of how differing levels of understanding might influence participants' responses.}{Participants read the negative system description and were asked to complete a comprehension check (\textsc{Comprehension}).}

\added{We recruited 95 participants via prolific. Five participants had to be excluded due to incomplete data, e.g., missing responses in demographics, or too short or incomprehensible responses to open questions, leaving 90 participants (Age: $M$ =30.69, $SD$ = 9.17, Min = 18, Max = 65) for analysis. The first group ($N_{\mbox{\scriptsize {No-Comprehension}}} = 44$) completed the check and got no feedback on correctness, while the second group ($N_{\mbox{\scriptsize {Comprehension}}} = 46$) had to answer all questions correctly (coded no: -1/yes: 1) to continue with the study}\deleted{. The system adaptation description was based on real-time video of their facial expressions, as per \mbox{\citet{kosch2022placebo}} (no data was recorded). The data was z-standardized for modeling; regression weights, thus, show deviation from the mean, akin to Cohen's D}. \added{After the check the participants gave their assessment of how they expected to perform with the AI system. }Finally, participants explained their point choices in an open text field. \added{The study took, on average, about 10 minutes to complete. Participants were compensated at £13.48/hr, resulting in a payment of £2.25 for a 10 minute-survey.}


\subsection{Quantitative results}


\begin{table}

\caption{\replaced{Summary statistics for performance expectations as a function of \textsc{Comprehension}}{Summary statistics for performance expectations as a function of \textsc{Comprehension}}}
\centering
\label{quantvaltable}
\resizebox{\textwidth}{!}{
\begin{tabular}[t]{lccc}
\toprule
 & \multicolumn{2}{c}{Comprehension} \\
Performance expectations with \added{\sAIactive{}} when compared to \added{\sAIinactive{}} & No & Yes \\
\midrule
Overall performance*  & 5.11* (1.13) & 4.59* (0.98) \\
Task speed*  & 70.05* (18.48) & 60.70* (17.19) \\
\parbox[t]{8cm}{\raggedright Difference in the number of correct responses} & -20.16* (29.71) & -2.72 (39.41) \\
\bottomrule
\end{tabular}}
\\[1ex]  
\small
\raggedright
\textit{Note: Differences between groups are highlighted in the variable with a *. Means that are distinguishably more positive than their neutral value (4 for overall performance, 50 for task speed and r zero for Difference in the number for correct responses) are marked with *.}
\end{table}

\Cref{quantvaltable} \added{shows all means of the subjective performance expectations for each group. \textsc{Comprehension} had an effect on overall performance, $\tilde{b}_\text{Comprehension}$=  -0.26  [-0.48, -0.04] , $p_b$ = 1.05\% and expected task speed, $\tilde{b}_\text{Comprehension}$ = -4.60 [-8.34, -0.86], $p_b$ = 0.82\%. For estimated correct, we added \textsc{\replaced{\SStatus{}}{status}} to the model, $\tilde{b}_\text{Comprehension}$ = -0.51  [-7.80,  6.74], $p_b$ = 44.50\%, \textsc{Comprehension} had no effect. However,  a difference for \textsc{\replaced{\SStatus{}}{status}} emerged, $\tilde{b}_\text{\replaced{\SStatus{}}{status}}$ = 5.71  [2.00,  9.39] , $p_b$ = 0.15\% and an interaction effect $\tilde{b}_\text{\replaced{\SStatus{}}{status} $\times$ Comprehension}$  = -4.36  [-8.08, -0.69], $p_b$ = 1.03\%.} Participants in the group without the enforced comprehension check estimated to answer more accurately with the \replaced{\sAIactive{}}{system active} than without $p_b\text{diff}$ = 0.00\%, while in the comprehension check group, this difference was not present, $p_b\text{diff}$ = 30.40\%. Most importantly, participants were optimistic with regard to overall performance and expected speed, irrespective of \textsc{Comprehension}. Only for \replaced{the difference in the number of}{$\Delta n$ for} expected correct \replaced{responses}{respinses}, we find that the \textsc{Comprehension} leveled participants to neutral expectations.

\subsection{Qualitative results}
\replaced{After participants estimated their subjective performance, they were further prompted to elaborate on the rationale behind their responses. To gain deeper insights into participants' perceptions and expectations regarding their performance with AI, a qualitative analysis was performed. The focus was on revising statements made by participants regarding their expectations of performing better or worse with an AI system when informed of a potential performance decline (\textsc{negative description}). This qualitative exploration aimed to uncover nuanced reasons underlying the participants' convictions about performance with AI and the perceived speed advantage or disadvantage. \\ The analysis involved clustering statements based on the participants' subjective assessments of their expected speed and overall performance on the Likert items. Two researchers independently performed a qualitative analysis of the statements, grouping them according to their semantic meaning. Afterward, a consensus was reached, identifying five distinct categories: AI Trust, AI Assistance, Uncertainty, Neutral, Self-Awareness, and AI Antagonism. \\ \mbox{\Cref{tab:Statement categorization}} provides a detailed breakdown of the distribution of statements across these categories, revealing predominant themes. Notably, the majority of statements (out of 180) primarily align with AI Trust (27 statements), AI Assistance (64 statements), and Uncertainty (44 statements). AI Trust reflects the participants' positive expectations and trust in the capabilities of AI systems as powerful tools that ensure an advantage. AI Assistance describes the perception of AI as a helpful assistant that facilitates task completion. Uncertainty portrays the participants' uncertainty toward the AI system's influence on task completion. These prevalent themes indicate that the majority of participants expected a positive influence (AI Trust and AI Assistance) on task performance ($N_{\mbox{\scriptsize Statements}} = 41$) and speed ($N_{\mbox{\scriptsize Statements}} = 50$) from the AI system, with some expressing uncertainty instead of negative sentiment toward the AI system despite being informed of potential performance decline.}{Participants were asked to explain the reasoning behind their responses after the estimation of performance. We analyzed all statements clustered for expected speed and overall performance. Two researchers independently performed a reflexive qualitative analysis \mbox{\citet{Braun2019}}, of the statements, grouped them based on their semantic meaning, and finally agreed on a set of five categories, see \mbox{\Cref{tab:Statement categorization}}.}

\begin{table}[h]
\centering
\caption{Subjective influence of the AI system on expected performance and speed before task completion: Number of statements \added{and percentage} per category}
\resizebox{\textwidth}{!}{
\begin{tabular}{p{2cm} p{4cm} p{4.4cm} p{1.4cm} p{1.4cm}} 
\toprule
Category & Description & Statement Examples & \multicolumn{2}{l}{\added{Statement} Count\added{s} ($\%$)}\\
 &  &  &  Performance &  Speed\\
\midrule
AI Trust &\small{ Trust and positive \replaced{expectations toward the}{belief in} capabilities of AI systems in general. Seeing AI as a powerful tool that ensures an advantage.} &  \small{\textit{AI models enhance our performances, so I have no doubt that this one will do the same.}} (22P; P = 6) \small{\textit{Because I trust AI, IT IT FAST [sic] and quite reliable for most activities.} (11S; S = 68)} & 15 (\replaced{$16.67\%$}{$15.96\%$}) & 12 (\replaced{$13.33\%$}{$12.77\%$}) \\
AI Assistance &\small{ AI is a helpful assistant that will facilitate task completion.} &  \small{\textit{I think the AI will assist me as it will be programmed to do the task, and I am not.}} (35P; P = 6); \small{\textit{With the help of AI, I will be able to work fast because it will be assisting me rather than having to figure this out myself.} (37S; S = 85); \textit{My effort and AI combined we will produce better results.} (40S; S = 99)} & \replaced{26}{28} (\replaced{$28.89\%$}{$29.79\%$}) & \replaced{38}{40} (\replaced{$42.22\%$}{$42.55\%$}) \\
Uncertainty & \small{Uncertainty toward the AI's systems influence on task completion.} &  \small{\textit{I don't know what to expect, really, maybe I could do better or not.} (18P; P = 4 );} \small{\textit{[...] I do not know the effects of the AI system on my performance yet.} (56S; S = 51) }& \replaced{29}{31} (\replaced{$32.22\%$}{$32.98\%$}) & 15 (\replaced{$16.67\%$}{$15.96\%$}) \\
Neutral & \small{AI will neither  have a positive or negative influence. AI won't make a difference in the task.} &  \small{\textit{I don't think there will be a large effect either way.} (2P; P = 4);} \small{ \textit{Because AI shouldn't have an effect on how I respond.} (15S; S = 56}) & 7 (\replaced{$7.78\%$}{$7.45\%$}) & 9 (\replaced{$10.00\%$}{$9.57\%$}) \\
Self-Awareness & \small{Self-reliance, and confidence in individual abilities, regardless of AI assistance, emphasizing autonomy and individual skill. }&  \small{\textit{Because I do not depend on enhancement to complete my tasks.} (7P; P = 6);}   \small{\textit{Cause I am a bit smarter for now than the AI system.} (39S; S = 99)}, & 9 (\replaced{$10.00\%$}{$9.57\%$}) & \replaced{7}{8} (\replaced{$7.78\%$}{$8.51\%$}) \\
AI Antagonism & \small{Lack of trust in the AI system, believing it will hamper performance, and skepticism towards AI's usefulness. }&   \small{\textit{As far as I understand, the AI will confuse me more than be of any help.} (60P; P = 4)};  \small{\textit{The AI might distract me and make me a little slower.} (9S; S = 27)} & 4 (\replaced{$4.44\%$}{$4.26\%$}) & 9 (\replaced{$10.00\%$}{$10.64\%$}) \\
\bottomrule
\end{tabular}}
\\[1ex]  
\small
\raggedright
\textit{Note: The \added{participants'} statements \added{on the AI systems influence on their performance and speed} \replaced{were}{have been} grammatically corrected to ensure good readability. Any quotes that remain unchanged are marked with [sic]. Each quote is followed by parentheses indicating the statement item number and whether the statement is related to \added{the participants' assessments of their} expected performance (P) or speed (S). The number after the semicolon indicates the \added{participants' subjective assessments of their} expected performance on a Likert \replaced{item}{Scale} ranging from 1 (strongly disagree) to 7 (strongly agree)\replaced{. Similarly, for expected speed, participants provided scores on a scale ranging from 1 (slower) to 100 (faster).}{, or expected speed, ranging from 1 (slower) to 100 (faster).}}
\label{tab:Statement categorization}
\end{table}
\FloatBarrier

\section{Discussion}
In this study, we set out to implement negative expectations and study the nocebo effect of AI (RQ1). However, we found that the placebo effect of AI in HCI \cite{kosch2022placebo} is robust to the manipulation of expectations by \added{a} negative verbal description (contrary to H1.1 and H1.2). Even when we told participants that the AI system would make their task harder and more stressful, they still believed it would improve their performance. This was the same for those who \replaced{read}{heard} positive descriptions of the AI (rendering H1, H3 \& H5 void). We refer to this expectation of high performance as \textsc{AI performance bias}. We replicate this bias in a dedicated online study. 

We found that heightened expectations (supporting H2.1.) carry over to the way participants make decisions (RQ2). Participants in the \added{\sAIactive{}} condition responded slightly faster and more accurately when informed they were interacting with an adaptive AI system. Using the DDM model to analyze decision-making, we found that just believing an AI is involved can make participants gather information more quickly, respond more conservatively, and make them more alert (partial support for H4). We found no effects on workload or physiological arousal (no support for H2.1, H6).

\subsection{\added{Beyond demand characteristics and system descriptions}}

\added{Critics may argue that placebo effects in AI are not genuine and stem from demand characteristics, which often influence experimental studies and HCI evaluations \mbox{\cite{weber1972subject, dell2012yours}}. In our study, despite participants being primed to view the AI negatively, their improved performance and positive ratings contradicted these expectations, suggesting that demand characteristics cannot account for the AI's placebo effect. 
One could also assert that our system descriptions were not effective in producing expectations. We used similar positive and negative verbal descriptions as studies in sports science, e.g., \cite{beedie2007positive,hurst2020placebo}. Also, the manipulation of \textsc{\replaced{\SStatus{}}{status}} influenced participants both subjectively and behaviorally, irrespective of their post-experimental accounts of believing in the system's capabilities, see \cref{manipulationcheck}. Moreover, in study 2, participants who understood the negative AI description (comprehension check) adjusted their expectations accordingly. This indicates that while our negative portrayal had some impact, it was less influential than AI narratives, creating high expectations. Future research should further explore this by comparing a sham AI system with a non-AI system (e.g., controlled by a sham operator) or by screening for AI expectations a-priori and comparing the placebo response with a rather neutral and minimal AI description.}

\subsection{AI performance bias as an antecedent of the placebo effect of AI} 

It appears that the prevailing positive perceptions about AI are influential enough to overshadow context-specific negative verbal descriptions, \added{irrespective of reported belief after the experiment}. This could be due to participants bringing their daily experiences and narratives of AI into the evaluation, biasing both their subjective evaluations and behavior, see \Cref{tab:Statement categorization}. From a mental model perspective \cite{wilson1989mental}, performance-reducing AI assistance may not fit into a coherent representation of human-AI interaction. 
It follows that the placebo mechanism for AI interfaces presented in the HCI literature is invalid \cite{kosch2022placebo,villa2023placebo}, as they focus on verbal system descriptions producing a placebo effect of AI. 
Based on our qualitative data, we follow that the effect is not specific to verbal descriptions of the system but may arise out of the socio-technical context as a function of the user's mental model. 

The AI performance bias presents an intriguing contrast with \citet{sartori_minding_2023} findings on AI Anxiety. While individuals often express strong negative attitudes about AI replacing them in certain tasks, it appears that when humans and AI work together, even in a non-functional AI setting, joint performance is judged to be superior.
Past studies have demonstrated that task performance \replaced{in human-AI collaborations can surpass individual AI or human performance}{can surpass individual AI or human performance in human-AI collaborations} \cite{CognitiveChallenges}. However, our findings shed new light on these findings. The human-AI performance gain may not arise from the summation of individual capabilities but also involves an elevation in human performance influenced by performance expectations. 
This suggests that (HCI) designers may harness the advantages of human-AI collaboration when focusing on systems that leverage a symbiotic relationship rather than fully automated tasks. However, future studies should explore not only the context of collaboration similar to \citet{villa2023placebo} and \citet{kosch2022placebo} but also consider human-AI competition.

\subsection{The Impact of sham-AI on Decision-making}

\citet{villa2023placebo} explored the impact of the placebo effect on decision-making in risky situations. They found that individuals with high expectations of AI system support tended to take greater risks compared to those without AI assistance. This emphasizes how people's actions can be shaped by the narrative surrounding AI systems. In our study, we extended this research by investigating how positive and negative verbal descriptions affect decision-making processes. Our model showed that when people believed to have AI support, they gathered information faster than when not supported by AI. Yet, the type of narrative (positive or negative) did not have an impact on parameters in the DDM and, thus, the underlying decision-making process. Prior research indicates that a participant's confidence can substantially influence the drift rate in a DDM \cite{Lee2021EvidenceOC, Liu2022DriftDM}. Therefore, it is possible that our findings can be explained by the participants feeling more confident when using the AI system. Also, we find a slightly more conservative decision boundary, with participants gathering more information until making a decision when supported with \replaced{sAI}{sham-AI}. With AI support, participants might prioritize accuracy (a strategy that can be experimentally induced \cite{starns2010effects}), which also improves their overall performance. Lastly, \replaced{sAI}{sham-AI} also shortened participants' non-decision time, indicating they were in a more prepared state when making decisions, especially for negative descriptions. Note, however, that while some proponents associate a reduced non-decision time with better attention, as argued by \citet{nunez2017attention}, or disinhibition \cite{schuch2016task}, others have developed models without this parameter \cite{van2020dstarm}, as it is sensitive to contaminants. 
Thus, our computational model shows that the belief in using AI influenced participants' decision-making processes when interacting with a computing system.

\subsection{Limitations \& Implications}
The study presents multiple limitations. First, by applying the social-affective perspective from \citet{atlas2021social} to our findings, it is evident that we did not account for the influence of emotions. 
While fostering a comfortable and friendly environment is commonly recommended in HCI evaluations \cite{rubin2008handbook,lazar2016human}, prior research \cite{geers2019testing} has indicated that positive emotions can counteract the nocebo response in pain experiments. Positive affect could explain why we observed no nocebo effects. \added{Analysis of EDA and TLX data over time showed that participants, at the very least, were not strained by the task.} \deleted{It is worth noting, however, that our study could not induce negative expectations to begin with, as confirmed by our validation study.} Nonetheless, future research should take into account the impact of emotions during tests, perhaps by deliberately altering them, as suggested by \citet{geers2019testing}. 

\added{It is worth noting that in addition to positive affect possibly accounting for the absence of nocebo effects, the fact that only around 17\% of participants didn't fully believe in the AI system's capabilities could also serve as an explanation. However, this percentage is lower than the number of participants who either fully believed in or had some level of suspicion towards the system. Yet, the effect was present in most nonbelievers nevertheless (see \autoref{ap:individual-react}).}

In line with \citet{berkel2023}, we highlight two major domains of implications of our work. 
First, methodologically,  given that a drift rate in the DDM can be estimated fast \cite{Voss_2015_ddm_cognitiveprocesses}, the DDM could be used to compute a robust behavioral indicator of a placebo response for an AI interface. 
Second, it is crucial for the HCI community to understand that technology narratives can significantly bias AI performance expectations to the point where even negative descriptions cannot mitigate their influence on evaluation and interaction.
For instance, positive expectations (placebo) may lead to overconfidence regarding the attributes of the system, such as its usability or user experience \cite{kosch2022placebo}. \added{Our findings demonstrate that individuals tended to be overly confident about their performance. This could potentially mislead those evaluating the technology, fundamentally undermining the principles of human-centered design.
One could argue that our behavioral effects are small and, thus, the placebo effect of AI is negligible to human-centered design. We will outline why this is unproblematic. First, while our behavioral effects were small ($d_z = 0.12$), and arguably they become larger when controlling for the speed-accuracy trade-off, effects on subjective measures were medium-sized ($d_z = 0.53$). Second, we used minimal intervention by only describing a sham AI system. A more severe intervention, including more placebo characteristics (for an overview, see \cite{Price2008ACR}) may yield more substantial effects. In the context of a user study, a false-positive due to placebo could have severe consequences (for a discussion, see \citet{kosch2022placebo}). \citet{RN526770} argue that small effects in studies with minimal interventions are particularly meaningful, much like \citet{gotz2022small} that posit how small effects are essential to progress in science. Third, placebo/nocebo interventions in sports contexts are also tied to small effects (\cite{hurst2020placebo} $d_{placebo}$ = .36, $d_{nocebo}$ = .37). Note also that studies on aging populations with similar tasks only find medium effects \cite{ratcliff_perceptual_2010}. Given the medium-sized subjective effects that align with our small behavioral results, we deem our results meaningful for applied contexts.}

\subsection{\added{Strategies to Mitigate the Placebo Effect of AI Technologies}}
\added{
Building on previous studies demonstrating a placebo effect in HCI \mbox{\cite{kosch2022placebo, villa2023placebo}}, our research investigated the impact of positive or negative descriptions of AI in eliciting a placebo or nocebo effect. Contrary to our hypotheses, we were unable to induce a nocebo effect (negative descriptions leading to the expectation of a poorer performance) with AI technology. Even when AI is framed negatively, people expect it to be effective and improve performance. Based on these findings, we propose strategies for mitigating the potential influence of prior expectations when evaluating AI technologies, which should be investigated in future research:
}


\begin{enumerate}

\item \added{\textbf{Monitor Decision-Making Processes:} Observe changes in participants' judgments or behaviors in response to negative/positive information about the system, utilizing subjective, behavioral, and psychophysiological measures} \cite{villa2023placebo}.
\item \added{\textbf{Minimize AI Disclosure:} Refrain from informing users about the AI's involvement to avoid biasing their experiences and thus control for contextual placebo-related information} \citep{wager2015neuroscience}.
\item \added{\textbf{Transparent AI Disclosure When Necessary:} If AI disclosure is unavoidable, clearly communicate its limitations and development status to encourage critical evaluation based on performance rather than expectations.}
\item \added{\textbf{Incorporate Sham Conditions:} Use a non-functional AI (sham) condition alongside the functional AI in experiments to differentiate the AI's actual effect from user expectations.}
\item \added{\textbf{Evaluate Expectation Narratives:} Conduct interviews to understand user anticipations and perceptions regarding specific technologies to see how pre-existing expectations influence the study outcome.}
    
\end{enumerate}

\section{Conclusion}
We found that even when we told participants to expect poor performance from a fake AI system, they still performed better and responded faster, showing a robust placebo effect. Contrary to previous work, this indicates that the placebo effect of AI is not easily negated by negative verbal descriptions, which raises questions about current methods for controlling for expectations in HCI studies. Additionally, the belief in having AI assistance facilitated decision-making processes, even when the narrative about AI was negative, thereby emphasizing that the influence of AI goes beyond simple narratives. This highlights the complexity and impact of AI narratives and suggests the need for a more nuanced approach in both research and practical user evaluation of AI.



\bibliographystyle{ACM-Reference-Format}
\bibliography{main}


\begin{thebibliography}{92}


\ifx \showCODEN    \undefined \def \showCODEN     #1{\unskip}     \fi
\ifx \showDOI      \undefined \def \showDOI       #1{#1}\fi
\ifx \showISBNx    \undefined \def \showISBNx     #1{\unskip}     \fi
\ifx \showISBNxiii \undefined \def \showISBNxiii  #1{\unskip}     \fi
\ifx \showISSN     \undefined \def \showISSN      #1{\unskip}     \fi
\ifx \showLCCN     \undefined \def \showLCCN      #1{\unskip}     \fi
\ifx \shownote     \undefined \def \shownote      #1{#1}          \fi
\ifx \showarticletitle \undefined \def \showarticletitle #1{#1}   \fi
\ifx \showURL      \undefined \def \showURL       {\relax}        \fi
\providecommand\bibfield[2]{#2}
\providecommand\bibinfo[2]{#2}
\providecommand\natexlab[1]{#1}
\providecommand\showeprint[2][]{arXiv:#2}

\bibitem[Atlas(2021)]%
        {atlas2021social}
\bibfield{author}{\bibinfo{person}{Lauren~Y Atlas}.}
  \bibinfo{year}{2021}\natexlab{}.
\newblock \showarticletitle{A social affective neuroscience lens on placebo
  analgesia}.
\newblock \bibinfo{journal}{\emph{Trends in Cognitive Sciences}}
  \bibinfo{volume}{25}, \bibinfo{number}{11} (\bibinfo{date}{Nov.}
  \bibinfo{year}{2021}), \bibinfo{pages}{992--1005}.
\newblock
\urldef\tempurl%
\url{https://doi.org/10.1016/j.tics.2021.07.016}
\showDOI{\tempurl}


\bibitem[Babaei et~al\mbox{.}(2021)]%
        {Babaei_2021_critique_EDA}
\bibfield{author}{\bibinfo{person}{Ebrahim Babaei}, \bibinfo{person}{Benjamin
  Tag}, \bibinfo{person}{Tilman Dingler}, {and} \bibinfo{person}{Eduardo
  Velloso}.} \bibinfo{year}{2021}\natexlab{}.
\newblock \showarticletitle{A Critique of Electrodermal Activity Practices at
  CHI}. In \bibinfo{booktitle}{\emph{Proceedings of the 2021 CHI Conference on
  Human Factors in Computing Systems}} (Yokohama, Japan)
  \emph{(\bibinfo{series}{Chi '21})}. \bibinfo{publisher}{Association for
  Computing Machinery}, \bibinfo{address}{New York, NY, USA}, Article
  \bibinfo{articleno}{177}, \bibinfo{numpages}{14}~pages.
\newblock
\showISBNx{9781450380966}
\urldef\tempurl%
\url{https://doi.org/10.1145/3411764.3445370}
\showDOI{\tempurl}


\bibitem[Bareis and Katzenbach(2022)]%
        {bareis2022talking}
\bibfield{author}{\bibinfo{person}{Jascha Bareis} {and}
  \bibinfo{person}{Christian Katzenbach}.} \bibinfo{year}{2022}\natexlab{}.
\newblock \showarticletitle{Talking AI into being: The narratives and
  imaginaries of national AI strategies and their performative politics}.
\newblock \bibinfo{journal}{\emph{Science, Technology, \& Human Values}}
  \bibinfo{volume}{47}, \bibinfo{number}{5} (\bibinfo{date}{May}
  \bibinfo{year}{2022}), \bibinfo{pages}{855--881}.
\newblock
\urldef\tempurl%
\url{https://doi.org/10.1177/01622439211030007}
\showDOI{\tempurl}


\bibitem[Beckham(1989)]%
        {Beckham1989ImprovementAE}
\bibfield{author}{\bibinfo{person}{Ernest~Edward Beckham}.}
  \bibinfo{year}{1989}\natexlab{}.
\newblock \showarticletitle{Improvement after evaluation in psychotherapy of
  depression: evidence of a placebo effect?}
\newblock \bibinfo{journal}{\emph{Journal of clinical psychology}}
  \bibinfo{volume}{45}, \bibinfo{number}{6} (\bibinfo{date}{Nov.}
  \bibinfo{year}{1989}), \bibinfo{pages}{945--950}.
\newblock
\urldef\tempurl%
\url{https://doi.org/10.1002/1097-4679(198911)45:6<945::aid-jclp2270450620>3.0.co;2-2}
\showDOI{\tempurl}


\bibitem[Beecher(1955)]%
        {beecher1955powerful}
\bibfield{author}{\bibinfo{person}{Henry~K. Beecher}.}
  \bibinfo{year}{1955}\natexlab{}.
\newblock \showarticletitle{The powerful placebo}.
\newblock \bibinfo{journal}{\emph{Journal of the American Medical Association}}
  \bibinfo{volume}{159}, \bibinfo{number}{17} (\bibinfo{date}{Dec.}
  \bibinfo{year}{1955}), \bibinfo{pages}{1602--1606}.
\newblock
\showISSN{0002-9955}
\urldef\tempurl%
\url{https://doi.org/10.1001/jama.1955.02960340022006}
\showDOI{\tempurl}
\showeprint{https://jamanetwork.com/journals/jama/articlepdf/303530/jama\_159\_17\_006.pdf}


\bibitem[Beedie et~al\mbox{.}(2007)]%
        {beedie2007positive}
\bibfield{author}{\bibinfo{person}{Christopher~J Beedie},
  \bibinfo{person}{Damian~A Coleman}, {and} \bibinfo{person}{Abigail~J Foad}.}
  \bibinfo{year}{2007}\natexlab{}.
\newblock \showarticletitle{Positive and negative placebo effects resulting
  from the deceptive administration of an ergogenic aid}.
\newblock \bibinfo{journal}{\emph{International journal of sport nutrition and
  exercise metabolism}} \bibinfo{volume}{17}, \bibinfo{number}{3}
  (\bibinfo{date}{June} \bibinfo{year}{2007}), \bibinfo{pages}{259--269}.
\newblock
\urldef\tempurl%
\url{https://doi.org/10.1123/ijsnem.17.3.259}
\showDOI{\tempurl}


\bibitem[Boot et~al\mbox{.}(2013)]%
        {doi:10.1177/1745691613491271}
\bibfield{author}{\bibinfo{person}{Walter~R. Boot}, \bibinfo{person}{Daniel~J.
  Simons}, \bibinfo{person}{Cary Stothart}, {and} \bibinfo{person}{Cassie
  Stutts}.} \bibinfo{year}{2013}\natexlab{}.
\newblock \showarticletitle{The Pervasive Problem With Placebos in Psychology:
  Why Active Control Groups Are Not Sufficient to Rule Out Placebo Effects}.
\newblock \bibinfo{journal}{\emph{Perspectives on Psychological Science}}
  \bibinfo{volume}{8}, \bibinfo{number}{4} (\bibinfo{year}{2013}),
  \bibinfo{pages}{445--454}.
\newblock
\urldef\tempurl%
\url{https://doi.org/10.1177/1745691613491271}
\showDOI{\tempurl}


\bibitem[Bory(2019)]%
        {bory2019deep}
\bibfield{author}{\bibinfo{person}{Paolo Bory}.}
  \bibinfo{year}{2019}\natexlab{}.
\newblock \showarticletitle{Deep new: The shifting narratives of artificial
  intelligence from Deep Blue to AlphaGo}.
\newblock \bibinfo{journal}{\emph{Convergence}} \bibinfo{volume}{25},
  \bibinfo{number}{4} (\bibinfo{date}{Feb.} \bibinfo{year}{2019}),
  \bibinfo{pages}{627--642}.
\newblock
\urldef\tempurl%
\url{https://doi.org/10.1177/1354856519829679}
\showDOI{\tempurl}


\bibitem[Brooke(1996)]%
        {Brooke_SUS_1996}
\bibfield{author}{\bibinfo{person}{John Brooke}.}
  \bibinfo{year}{1996}\natexlab{}.
\newblock \showarticletitle{SUS: A quick and dirty usability scale}.
\newblock \bibinfo{journal}{\emph{Usability Evaluation in Industry}}
  \bibinfo{volume}{189} (\bibinfo{date}{Jan.} \bibinfo{year}{1996}).
\newblock


\bibitem[B{\"u}rkner(2017)]%
        {burkner2017brms}
\bibfield{author}{\bibinfo{person}{Paul-Christian B{\"u}rkner}.}
  \bibinfo{year}{2017}\natexlab{}.
\newblock \showarticletitle{brms: An R package for Bayesian multilevel models
  using Stan}.
\newblock \bibinfo{journal}{\emph{Journal of statistical software}}
  \bibinfo{volume}{80}, \bibinfo{number}{1} (\bibinfo{date}{Aug.}
  \bibinfo{year}{2017}), \bibinfo{pages}{1--28}.
\newblock
\urldef\tempurl%
\url{https://doi.org/10.18637/jss.v080.i01}
\showDOI{\tempurl}


\bibitem[Carolus et~al\mbox{.}(2023)]%
        {Carolus_2023_MAILS}
\bibfield{author}{\bibinfo{person}{Astrid Carolus}, \bibinfo{person}{Martin
  Koch}, \bibinfo{person}{Samantha Straka}, \bibinfo{person}{Marc Latoschik},
  {and} \bibinfo{person}{Carolin Wienrich}.} \bibinfo{year}{2023}\natexlab{}.
\newblock \bibinfo{title}{MAILS -- Meta AI Literacy Scale: Development and
  Testing of an AI Literacy Questionnaire Based on Well-Founded Competency
  Models and Psychological Change- and Meta-Competencies}.
  (\bibinfo{date}{Feb.} \bibinfo{year}{2023}).
\newblock
\showeprint[arxiv]{2302.09319}~[cs.AI]


\bibitem[Carpenter et~al\mbox{.}(2017)]%
        {carpenter2017stan}
\bibfield{author}{\bibinfo{person}{Bob Carpenter}, \bibinfo{person}{Andrew
  Gelman}, \bibinfo{person}{Matthew~D. Hoffman}, \bibinfo{person}{Daniel Lee},
  \bibinfo{person}{Ben Goodrich}, \bibinfo{person}{Michael Betancourt},
  \bibinfo{person}{Marcus Brubaker}, \bibinfo{person}{Jiqiang Guo},
  \bibinfo{person}{Peter Li}, {and} \bibinfo{person}{Allen Riddell}.}
  \bibinfo{year}{2017}\natexlab{}.
\newblock \showarticletitle{Stan: A Probabilistic Programming Language}.
\newblock \bibinfo{journal}{\emph{Journal of Statistical Software}}
  \bibinfo{volume}{76}, \bibinfo{number}{1} (\bibinfo{date}{Jan.}
  \bibinfo{year}{2017}), \bibinfo{pages}{1–32}.
\newblock
\urldef\tempurl%
\url{https://doi.org/10.18637/jss.v076.i01}
\showDOI{\tempurl}


\bibitem[Cave et~al\mbox{.}(2019)]%
        {cave2019scary}
\bibfield{author}{\bibinfo{person}{Stephen Cave}, \bibinfo{person}{Kate
  Coughlan}, {and} \bibinfo{person}{Kanta Dihal}.}
  \bibinfo{year}{2019}\natexlab{}.
\newblock \showarticletitle{"Scary Robots": Examining Public Responses to AI}.
  In \bibinfo{booktitle}{\emph{Proceedings of the 2019 AAAI/ACM Conference on
  AI, Ethics, and Society}} (Honolulu, HI, USA) \emph{(\bibinfo{series}{Aies
  '19})}. \bibinfo{publisher}{Association for Computing Machinery},
  \bibinfo{address}{New York, NY, USA}, \bibinfo{pages}{331–337}.
\newblock
\showISBNx{9781450363242}
\urldef\tempurl%
\url{https://doi.org/10.1145/3306618.3314232}
\showDOI{\tempurl}


\bibitem[Cave and Dihal(2019)]%
        {cave2019hopes}
\bibfield{author}{\bibinfo{person}{Stephen Cave} {and} \bibinfo{person}{Kanta
  Dihal}.} \bibinfo{year}{2019}\natexlab{}.
\newblock \showarticletitle{Hopes and fears for intelligent machines in fiction
  and reality}.
\newblock \bibinfo{journal}{\emph{Nature Machine Intelligence}}
  \bibinfo{volume}{1} (\bibinfo{date}{Feb.} \bibinfo{year}{2019}),
  \bibinfo{pages}{74--78}.
\newblock
\urldef\tempurl%
\url{https://doi.org/10.1038/s42256-019-0020-9}
\showDOI{\tempurl}


\bibitem[Corredor et~al\mbox{.}(2017)]%
        {corredor2016decision}
\bibfield{author}{\bibinfo{person}{Javier Corredor}, \bibinfo{person}{Jorge
  Sofrony}, {and} \bibinfo{person}{Angelika Peer}.}
  \bibinfo{year}{2017}\natexlab{}.
\newblock \showarticletitle{Decision-Making Model for Adaptive Impedance
  Control of Teleoperation Systems}.
\newblock \bibinfo{journal}{\emph{Institute of Electrical and Electronics
  Engineers (IEEE) Transactions on Haptics}} \bibinfo{volume}{10},
  \bibinfo{number}{1} (\bibinfo{date}{Jan.} \bibinfo{year}{2017}),
  \bibinfo{pages}{5--16}.
\newblock
\urldef\tempurl%
\url{https://doi.org/10.1109/toh.2016.2581807}
\showDOI{\tempurl}


\bibitem[Dahlbäck et~al\mbox{.}(1993)]%
        {dahlback1993wizard}
\bibfield{author}{\bibinfo{person}{Nils Dahlbäck}, \bibinfo{person}{Arne
  Jönsson}, {and} \bibinfo{person}{Lars Ahrenberg}.}
  \bibinfo{year}{1993}\natexlab{}.
\newblock \bibinfo{title}{Wizard of Oz studies}.
\newblock
\newblock


\bibitem[Dell et~al\mbox{.}(2012)]%
        {dell2012yours}
\bibfield{author}{\bibinfo{person}{Nicola Dell}, \bibinfo{person}{Vidya
  Vaidyanathan}, \bibinfo{person}{Indrani Medhi}, \bibinfo{person}{Edward
  Cutrell}, {and} \bibinfo{person}{William Thies}.}
  \bibinfo{year}{2012}\natexlab{}.
\newblock \showarticletitle{" Yours is better!" participant response bias in
  HCI}. In \bibinfo{booktitle}{\emph{Proceedings of the sigchi conference on
  human factors in computing systems}}. \bibinfo{pages}{1321--1330}.
\newblock


\bibitem[Denisova and Cairns(2015)]%
        {denisova2015placebo}
\bibfield{author}{\bibinfo{person}{Alena Denisova} {and} \bibinfo{person}{Paul
  Cairns}.} \bibinfo{year}{2015}\natexlab{}.
\newblock \showarticletitle{The Placebo Effect in Digital Games: Phantom
  Perception of Adaptive Artificial Intelligence}. In
  \bibinfo{booktitle}{\emph{Proceedings of the 2015 Annual Symposium on
  Computer-Human Interaction in Play}} (London, United Kingdom)
  \emph{(\bibinfo{series}{Chi Play '15})}. \bibinfo{publisher}{Association for
  Computing Machinery}, \bibinfo{address}{New York, NY, USA},
  \bibinfo{pages}{23–33}.
\newblock
\showISBNx{9781450334662}
\urldef\tempurl%
\url{https://doi.org/10.1145/2793107.2793109}
\showDOI{\tempurl}


\bibitem[Denisova and Cook(2019)]%
        {Denisova2019}
\bibfield{author}{\bibinfo{person}{Alena Denisova} {and}
  \bibinfo{person}{Eliott Cook}.} \bibinfo{year}{2019}\natexlab{}.
\newblock \showarticletitle{Power-Ups in Digital Games: The Rewarding Effect of
  Phantom Game Elements on Player Experience}. In
  \bibinfo{booktitle}{\emph{Proceedings of the Annual Symposium on
  Computer-Human Interaction in Play}} (Barcelona, Spain)
  \emph{(\bibinfo{series}{Chi Play '19})}. \bibinfo{publisher}{Association for
  Computing Machinery}, \bibinfo{address}{New York, NY, USA},
  \bibinfo{pages}{161–168}.
\newblock
\showISBNx{9781450366885}
\urldef\tempurl%
\url{https://doi.org/10.1145/3311350.3347173}
\showDOI{\tempurl}


\bibitem[Diederich and Goetz(2008)]%
        {Diederich2008ThePT}
\bibfield{author}{\bibinfo{person}{Nico~J. Diederich} {and}
  \bibinfo{person}{Christopher~G. Goetz}.} \bibinfo{year}{2008}\natexlab{}.
\newblock \showarticletitle{The placebo treatments in neurosciences}.
\newblock \bibinfo{journal}{\emph{Neurology}} \bibinfo{volume}{71},
  \bibinfo{number}{9} (\bibinfo{date}{Aug.} \bibinfo{year}{2008}),
  \bibinfo{pages}{677--684}.
\newblock
\showISSN{0028-3878}
\urldef\tempurl%
\url{https://doi.org/10.1212/01.wnl.0000324635.49971.3d}
\showDOI{\tempurl}


\bibitem[Dix(2022)]%
        {dix2022bayesian}
\bibfield{author}{\bibinfo{person}{Alan Dix}.} \bibinfo{year}{2022}\natexlab{}.
\newblock \showarticletitle{Bayesian statistics}.
\newblock In \bibinfo{booktitle}{\emph{Bayesian Methods for Interaction and
  Design}}, \bibfield{editor}{\bibinfo{person}{John~H. Williamson},
  \bibinfo{person}{Antti Oulasvirta}, \bibinfo{person}{Per~Ola Kristensson},
  {and} \bibinfo{person}{Nikola Banovic}} (Eds.). \bibinfo{publisher}{Cambridge
  University Press}, \bibinfo{pages}{81–114}.
\newblock
\urldef\tempurl%
\url{https://doi.org/10.1017/9781108874830.004}
\showDOI{\tempurl}


\bibitem[Ferrario et~al\mbox{.}(2019)]%
        {ferrario2019TrustAI}
\bibfield{author}{\bibinfo{person}{Andrea Ferrario}, \bibinfo{person}{Michele
  Loi}, {and} \bibinfo{person}{Eleonora Viganò}.}
  \bibinfo{year}{2019}\natexlab{}.
\newblock \showarticletitle{In AI We Trust Incrementally: a Multi-layer Model
  of Trust to Analyze Human-Artificial Intelligence Interactions}.
\newblock \bibinfo{journal}{\emph{Philosophy and Technology}}
  \bibinfo{volume}{33} (\bibinfo{year}{2019}), \bibinfo{pages}{523--539}.
\newblock


\bibitem[Finstad(2006)]%
        {Finstad_SUS_non-native-speaker}
\bibfield{author}{\bibinfo{person}{Kraig Finstad}.}
  \bibinfo{year}{2006}\natexlab{}.
\newblock \showarticletitle{The System Usability Scale and Non-Native English
  Speakers}.
\newblock \bibinfo{journal}{\emph{Journal of User Experience}}
  \bibinfo{volume}{1}, \bibinfo{number}{4} (\bibinfo{date}{Aug.}
  \bibinfo{year}{2006}), \bibinfo{pages}{185–188}.
\newblock


\bibitem[F\"{u}gener et~al\mbox{.}(2022)]%
        {CognitiveChallenges}
\bibfield{author}{\bibinfo{person}{Andreas F\"{u}gener},
  \bibinfo{person}{J\"{o}rn Grahl}, \bibinfo{person}{Alok Gupta}, {and}
  \bibinfo{person}{Wolfgang Ketter}.} \bibinfo{year}{2022}\natexlab{}.
\newblock \showarticletitle{Cognitive Challenges in Human–Artificial
  Intelligence Collaboration: Investigating the Path Toward Productive
  Delegation}.
\newblock \bibinfo{journal}{\emph{Information Systems Research}}
  \bibinfo{volume}{33}, \bibinfo{number}{2} (\bibinfo{date}{June}
  \bibinfo{year}{2022}), \bibinfo{pages}{678--696}.
\newblock
\urldef\tempurl%
\url{https://doi.org/10.1287/isre.2021.1079}
\showDOI{\tempurl}
\showeprint{https://doi.org/10.1287/isre.2021.1079}


\bibitem[Geers et~al\mbox{.}(2019)]%
        {geers2019testing}
\bibfield{author}{\bibinfo{person}{Andrew~L Geers}, \bibinfo{person}{Shane
  Close}, \bibinfo{person}{Fawn~C Caplandies}, \bibinfo{person}{Charles~L
  Vogel}, \bibinfo{person}{Ashley~B Murray}, \bibinfo{person}{Yopina Pertiwi},
  \bibinfo{person}{Ian~M Handley}, {and} \bibinfo{person}{Lene Vase}.}
  \bibinfo{year}{2019}\natexlab{}.
\newblock \showarticletitle{Testing a positive-affect induction to reduce
  verbally induced nocebo hyperalgesia in an experimental pain paradigm}.
\newblock \bibinfo{journal}{\emph{Pain}} \bibinfo{volume}{160},
  \bibinfo{number}{10} (\bibinfo{date}{Oct.} \bibinfo{year}{2019}),
  \bibinfo{pages}{2290--2297}.
\newblock
\urldef\tempurl%
\url{https://doi.org/10.1097/j.pain.0000000000001618}
\showDOI{\tempurl}


\bibitem[Gelman et~al\mbox{.}(1992)]%
        {gelman1992inference}
\bibfield{author}{\bibinfo{person}{Andrew Gelman}, \bibinfo{person}{Donald~B
  Rubin}, {et~al\mbox{.}}} \bibinfo{year}{1992}\natexlab{}.
\newblock \showarticletitle{Inference from iterative simulation using multiple
  sequences}.
\newblock \bibinfo{journal}{\emph{Statistical science}} \bibinfo{volume}{7},
  \bibinfo{number}{4} (\bibinfo{year}{1992}), \bibinfo{pages}{457--472}.
\newblock


\bibitem[G{\"o}tz et~al\mbox{.}(2022)]%
        {gotz2022small}
\bibfield{author}{\bibinfo{person}{Friedrich~M G{\"o}tz},
  \bibinfo{person}{Samuel~D Gosling}, {and} \bibinfo{person}{Peter~J
  Rentfrow}.} \bibinfo{year}{2022}\natexlab{}.
\newblock \showarticletitle{Small effects: The indispensable foundation for a
  cumulative psychological science}.
\newblock \bibinfo{journal}{\emph{Perspectives on Psychological Science}}
  \bibinfo{volume}{17}, \bibinfo{number}{1} (\bibinfo{year}{2022}),
  \bibinfo{pages}{205--215}.
\newblock


\bibitem[Guerreiro et~al\mbox{.}(2013)]%
        {guerreiro2013bitalino}
\bibfield{author}{\bibinfo{person}{José Guerreiro}, \bibinfo{person}{Raúl
  Martins}, \bibinfo{person}{Hugo Silva}, \bibinfo{person}{André Lourenço},
  {and} \bibinfo{person}{Ana Fred}.} \bibinfo{year}{2013}\natexlab{}.
\newblock \showarticletitle{BITalino - A Multimodal Platform for Physiological
  Computing}. In \bibinfo{booktitle}{\emph{Proceedings of the 10th
  International Conference on Informatics in Control, Automation and Robotics -
  Volume 1: ICINCO}}. \bibinfo{pages}{500--506}.
\newblock
\showISBNx{978-989-8565-70-9}
\urldef\tempurl%
\url{https://doi.org/10.5220/0004594105000506}
\showDOI{\tempurl}


\bibitem[Halbhuber et~al\mbox{.}(2022)]%
        {halbhuber2022better}
\bibfield{author}{\bibinfo{person}{David Halbhuber},
  \bibinfo{person}{Maximilian Schlenczek}, \bibinfo{person}{Johanna Bogon},
  {and} \bibinfo{person}{Niels Henze}.} \bibinfo{year}{2022}\natexlab{}.
\newblock \showarticletitle{Better Be Quiet about It! The Effects of Phantom
  Latency on Experienced First-Person Shooter Players}. In
  \bibinfo{booktitle}{\emph{Proceedings of the 21st International Conference on
  Mobile and Ubiquitous Multimedia}} (Lisbon, Portugal)
  \emph{(\bibinfo{series}{Mum '22})}. \bibinfo{publisher}{Association for
  Computing Machinery}, \bibinfo{address}{New York, NY, USA},
  \bibinfo{pages}{172–181}.
\newblock
\showISBNx{9781450398206}
\urldef\tempurl%
\url{https://doi.org/10.1145/3568444.3568448}
\showDOI{\tempurl}


\bibitem[Hart(2006)]%
        {hart_nasa-task_2006}
\bibfield{author}{\bibinfo{person}{Sandra~G. Hart}.}
  \bibinfo{year}{2006}\natexlab{}.
\newblock \showarticletitle{Nasa-Task Load Index (NASA-TLX); 20 Years Later}.
\newblock \bibinfo{journal}{\emph{Proceedings of the Human Factors and
  Ergonomics Society Annual Meeting}} \bibinfo{volume}{50}, \bibinfo{number}{9}
  (\bibinfo{date}{Oct.} \bibinfo{year}{2006}), \bibinfo{pages}{904--908}.
\newblock
\urldef\tempurl%
\url{https://doi.org/10.1177/154193120605000909}
\showDOI{\tempurl}


\bibitem[Hart and Staveland(1988)]%
        {HART1988139}
\bibfield{author}{\bibinfo{person}{Sandra~G. Hart} {and}
  \bibinfo{person}{Lowell~E. Staveland}.} \bibinfo{year}{1988}\natexlab{}.
\newblock \showarticletitle{Development of NASA-TLX (Task Load Index): Results
  of Empirical and Theoretical Research}.
\newblock In \bibinfo{booktitle}{\emph{Human Mental Workload}},
  \bibfield{editor}{\bibinfo{person}{Peter~A. Hancock} {and}
  \bibinfo{person}{Najmedin Meshkati}} (Eds.). \bibinfo{series}{Advances in
  Psychology}, Vol.~\bibinfo{volume}{52}. \bibinfo{publisher}{North-Holland},
  \bibinfo{pages}{139--183}.
\newblock
\showISSN{0166-4115}
\urldef\tempurl%
\url{https://doi.org/10.1016/S0166-4115(08)62386-9}
\showDOI{\tempurl}


\bibitem[Henninger et~al\mbox{.}(2021)]%
        {henninger_2021_labjs}
\bibfield{author}{\bibinfo{person}{Felix Henninger}, \bibinfo{person}{Yury
  Shevchenko}, \bibinfo{person}{Ulf Mertens}, \bibinfo{person}{Pascal~J.
  Kieslich}, {and} \bibinfo{person}{Benjamin~E. Hilbig}.}
  \bibinfo{year}{2021}\natexlab{}.
\newblock \bibinfo{booktitle}{\emph{lab.js: A free, open, online experiment
  builder}}.
\newblock
\urldef\tempurl%
\url{https://doi.org/10.5281/zenodo.5233003}
\showDOI{\tempurl}


\bibitem[Hermann(2020)]%
        {hermann2020beware}
\bibfield{author}{\bibinfo{person}{Isabella Hermann}.}
  \bibinfo{year}{2020}\natexlab{}.
\newblock \showarticletitle{Beware of fictional AI narratives}.
\newblock \bibinfo{journal}{\emph{Nature Machine Intelligence}}
  \bibinfo{volume}{2}, \bibinfo{number}{11} (\bibinfo{date}{Oct.}
  \bibinfo{year}{2020}), \bibinfo{pages}{654--654}.
\newblock
\urldef\tempurl%
\url{https://doi.org/10.1038/s42256-020-00256-0}
\showDOI{\tempurl}


\bibitem[Hoijtink and van~de Schoot(2018)]%
        {hoijtink2018testing}
\bibfield{author}{\bibinfo{person}{Herbert Hoijtink} {and}
  \bibinfo{person}{Rens van~de Schoot}.} \bibinfo{year}{2018}\natexlab{}.
\newblock \showarticletitle{Testing small variance priors using prior-posterior
  predictive p values.}
\newblock \bibinfo{journal}{\emph{Psychological Methods}} \bibinfo{volume}{23},
  \bibinfo{number}{3} (\bibinfo{year}{2018}), \bibinfo{pages}{561--569}.
\newblock
\urldef\tempurl%
\url{https://doi.org/10.1037/met0000131}
\showDOI{\tempurl}


\bibitem[Hr{\~o}bjartsson and G{\o}tzsche(2001)]%
        {Hrbjartsson2001IsTP}
\bibfield{author}{\bibinfo{person}{Asbj{\o}rn Hr{\~o}bjartsson} {and}
  \bibinfo{person}{Peter~Christian G{\o}tzsche}.}
  \bibinfo{year}{2001}\natexlab{}.
\newblock \showarticletitle{Is the placebo powerless? An analysis of clinical
  trials comparing placebo with no treatment.}
\newblock \bibinfo{journal}{\emph{The New England journal of medicine}}
  \bibinfo{volume}{344 21} (\bibinfo{date}{May} \bibinfo{year}{2001}),
  \bibinfo{pages}{1594--602}.
\newblock
\urldef\tempurl%
\url{https://doi.org/10.1056/nejm200105243442106}
\showDOI{\tempurl}


\bibitem[Huang et~al\mbox{.}(2020)]%
        {huang2020human}
\bibfield{author}{\bibinfo{person}{Jie Huang}, \bibinfo{person}{Wenhua Wu},
  \bibinfo{person}{Zhenyi Zhang}, {and} \bibinfo{person}{Yutao Chen}.}
  \bibinfo{year}{2020}\natexlab{}.
\newblock \showarticletitle{A human decision-making behavior model for
  human-robot interaction in multi-robot systems}.
\newblock \bibinfo{journal}{\emph{Institute of Electrical and Electronics
  Engineers (IEEE) Access}}  \bibinfo{volume}{8} (\bibinfo{date}{Nov.}
  \bibinfo{year}{2020}), \bibinfo{pages}{197853--197862}.
\newblock
\urldef\tempurl%
\url{https://doi.org/10.1109/access.2020.3035348}
\showDOI{\tempurl}


\bibitem[Hurst et~al\mbox{.}(2020)]%
        {hurst2020placebo}
\bibfield{author}{\bibinfo{person}{Philip Hurst}, \bibinfo{person}{Lieke
  Schipof-Godart}, \bibinfo{person}{Attila Szabo}, \bibinfo{person}{John
  Raglin}, \bibinfo{person}{Florentina Hettinga}, \bibinfo{person}{Bart
  Roelands}, \bibinfo{person}{Andrew Lane}, \bibinfo{person}{Abby Foad},
  \bibinfo{person}{Damian Coleman}, {and} \bibinfo{person}{Chris Beedie}.}
  \bibinfo{year}{2020}\natexlab{}.
\newblock \showarticletitle{The Placebo and Nocebo effect on sports
  performance: A systematic review}.
\newblock \bibinfo{journal}{\emph{European Journal of Sport Science}}
  \bibinfo{volume}{20}, \bibinfo{number}{3} (\bibinfo{date}{Aug.}
  \bibinfo{year}{2020}), \bibinfo{pages}{279--292}.
\newblock
\urldef\tempurl%
\url{https://doi.org/10.1080/17461391.2019.1655098}
\showDOI{\tempurl}


\bibitem[Jian et~al\mbox{.}(2000)]%
        {jian_2000_Trust_Automated_Systems}
\bibfield{author}{\bibinfo{person}{Jiun-Yin Jian}, \bibinfo{person}{Ann~M.
  Bisantz}, {and} \bibinfo{person}{Colin~G. Drury}.}
  \bibinfo{year}{2000}\natexlab{}.
\newblock \showarticletitle{Foundations for an Empirically Determined Scale of
  Trust in Automated Systems}.
\newblock \bibinfo{journal}{\emph{International Journal of Cognitive
  Ergonomics}} \bibinfo{volume}{4}, \bibinfo{number}{1} (\bibinfo{date}{March}
  \bibinfo{year}{2000}), \bibinfo{pages}{53--71}.
\newblock
\showISSN{1088-6362}
\urldef\tempurl%
\url{https://doi.org/10.1207/S15327566IJCE0401_04}
\showDOI{\tempurl}


\bibitem[Kay et~al\mbox{.}(2016)]%
        {kay2016researcher}
\bibfield{author}{\bibinfo{person}{Matthew Kay}, \bibinfo{person}{Gregory~L.
  Nelson}, {and} \bibinfo{person}{Eric~B. Hekler}.}
  \bibinfo{year}{2016}\natexlab{}.
\newblock \showarticletitle{Researcher-Centered Design of Statistics: Why
  Bayesian Statistics Better Fit the Culture and Incentives of HCI}. In
  \bibinfo{booktitle}{\emph{Proceedings of the 2016 CHI Conference on Human
  Factors in Computing Systems}} (San Jose, California, USA)
  \emph{(\bibinfo{series}{Chi '16})}. \bibinfo{publisher}{Association for
  Computing Machinery}, \bibinfo{address}{New York, NY, USA},
  \bibinfo{pages}{4521–4532}.
\newblock
\showISBNx{9781450333627}
\urldef\tempurl%
\url{https://doi.org/10.1145/2858036.2858465}
\showDOI{\tempurl}


\bibitem[King and He(2006)]%
        {KING2006740}
\bibfield{author}{\bibinfo{person}{William~R. King} {and} \bibinfo{person}{Jun
  He}.} \bibinfo{year}{2006}\natexlab{}.
\newblock \showarticletitle{A meta-analysis of the technology acceptance
  model}.
\newblock \bibinfo{journal}{\emph{Information \& Management}}
  \bibinfo{volume}{43}, \bibinfo{number}{6} (\bibinfo{date}{Sept.}
  \bibinfo{year}{2006}), \bibinfo{pages}{740--755}.
\newblock
\showISSN{0378-7206}
\urldef\tempurl%
\url{https://doi.org/10.1016/j.im.2006.05.003}
\showDOI{\tempurl}


\bibitem[Kosch et~al\mbox{.}(2023a)]%
        {tosch2023}
\bibfield{author}{\bibinfo{person}{Thomas Kosch}, \bibinfo{person}{Jakob
  Karolus}, \bibinfo{person}{Johannes Zagermann}, \bibinfo{person}{Harald
  Reiterer}, \bibinfo{person}{Albrecht Schmidt}, {and}
  \bibinfo{person}{Pawe\l{}~W. Wo\'{z}niak}.} \bibinfo{year}{2023}\natexlab{a}.
\newblock \showarticletitle{A Survey on Measuring Cognitive Workload in
  Human-Computer Interaction}.
\newblock \bibinfo{journal}{\emph{ACM Comput. Surv.}} \bibinfo{volume}{55},
  \bibinfo{number}{13s}, Article \bibinfo{articleno}{283} (\bibinfo{date}{July}
  \bibinfo{year}{2023}), \bibinfo{numpages}{39}~pages.
\newblock
\showISSN{0360-0300}
\urldef\tempurl%
\url{https://doi.org/10.1145/3582272}
\showDOI{\tempurl}


\bibitem[Kosch et~al\mbox{.}(2023b)]%
        {kosch2022placebo}
\bibfield{author}{\bibinfo{person}{Thomas Kosch}, \bibinfo{person}{Robin
  Welsch}, \bibinfo{person}{Lewis Chuang}, {and} \bibinfo{person}{Albrecht
  Schmidt}.} \bibinfo{year}{2023}\natexlab{b}.
\newblock \showarticletitle{The Placebo Effect of Artificial Intelligence in
  Human–Computer Interaction}.
\newblock \bibinfo{journal}{\emph{ACM Trans. Comput.-Hum. Interact.}}
  \bibinfo{volume}{29}, \bibinfo{number}{6}, Article \bibinfo{articleno}{56}
  (\bibinfo{date}{Jan.} \bibinfo{year}{2023}), \bibinfo{numpages}{32}~pages.
\newblock
\showISSN{1073-0516}
\urldef\tempurl%
\url{https://doi.org/10.1145/3529225}
\showDOI{\tempurl}


\bibitem[Lasagna et~al\mbox{.}(1954)]%
        {lasagna1954study}
\bibfield{author}{\bibinfo{person}{Louis Lasagna}, \bibinfo{person}{Frederick
  Mosteller}, \bibinfo{person}{John~M. {von Felsinger}}, {and}
  \bibinfo{person}{Henry~K. Beecher}.} \bibinfo{year}{1954}\natexlab{}.
\newblock \showarticletitle{A study of the placebo response}.
\newblock \bibinfo{journal}{\emph{The American Journal of Medicine}}
  \bibinfo{volume}{16}, \bibinfo{number}{6} (\bibinfo{date}{June}
  \bibinfo{year}{1954}), \bibinfo{pages}{770--779}.
\newblock
\showISSN{0002-9343}
\urldef\tempurl%
\url{https://doi.org/10.1016/0002-9343(54)90441-6}
\showDOI{\tempurl}


\bibitem[Lazar et~al\mbox{.}(2016)]%
        {lazar2016human}
\bibfield{author}{\bibinfo{person}{Jonathan Lazar}, \bibinfo{person}{Julio
  Abascal}, \bibinfo{person}{Simone Barbosa}, \bibinfo{person}{Jeremy
  Barksdale}, \bibinfo{person}{Batya Friedman}, \bibinfo{person}{Jens
  Grossklags}, \bibinfo{person}{Jan Gulliksen}, \bibinfo{person}{Jeff Johnson},
  \bibinfo{person}{Tom McEwan}, \bibinfo{person}{Lo\"{\i}c
  Mart\'{\i}nez-Normand}, \bibinfo{person}{Wibke Michalk},
  \bibinfo{person}{Janice Tsai}, \bibinfo{person}{Gerrit van~der Veer},
  \bibinfo{person}{Hans von Axelson}, \bibinfo{person}{Ake Walldius},
  \bibinfo{person}{Gill Whitney}, \bibinfo{person}{Marco Winckler},
  \bibinfo{person}{Volker Wulf}, \bibinfo{person}{Elizabeth~F. Churchill},
  \bibinfo{person}{Lorrie Cranor}, \bibinfo{person}{Janet Davis},
  \bibinfo{person}{Alan Hedge}, \bibinfo{person}{Harry Hochheiser},
  \bibinfo{person}{Juan~Pablo Hourcade}, \bibinfo{person}{Clayton Lewis},
  \bibinfo{person}{Lisa Nathan}, \bibinfo{person}{Fabio Paterno},
  \bibinfo{person}{Blake Reid}, \bibinfo{person}{Whitney Quesenbery},
  \bibinfo{person}{Ted Selker}, {and} \bibinfo{person}{Brian Wentz}.}
  \bibinfo{year}{2016}\natexlab{}.
\newblock \showarticletitle{Human–Computer Interaction and International
  Public Policymaking: A Framework for Understanding and Taking Future
  Actions}.
\newblock \bibinfo{journal}{\emph{Foundations and Trends in Human-Computer
  Interaction}} \bibinfo{volume}{9}, \bibinfo{number}{2} (\bibinfo{date}{May}
  \bibinfo{year}{2016}), \bibinfo{pages}{69–149}.
\newblock
\showISSN{1551-3955}
\urldef\tempurl%
\url{https://doi.org/10.1561/1100000062}
\showDOI{\tempurl}


\bibitem[Lee et~al\mbox{.}(2018)]%
        {lee2018moving}
\bibfield{author}{\bibinfo{person}{Byungjoo Lee}, \bibinfo{person}{Sunjun Kim},
  \bibinfo{person}{Antti Oulasvirta}, \bibinfo{person}{Jong-In Lee}, {and}
  \bibinfo{person}{Eunji Park}.} \bibinfo{year}{2018}\natexlab{}.
\newblock \showarticletitle{Moving Target Selection: A Cue Integration Model}.
  In \bibinfo{booktitle}{\emph{Proceedings of the 2018 CHI Conference on Human
  Factors in Computing Systems}} (Montreal QC, Canada)
  \emph{(\bibinfo{series}{Chi '18})}. \bibinfo{publisher}{Association for
  Computing Machinery}, \bibinfo{address}{New York, NY, USA},
  \bibinfo{pages}{1–12}.
\newblock
\showISBNx{9781450356206}
\urldef\tempurl%
\url{https://doi.org/10.1145/3173574.3173804}
\showDOI{\tempurl}


\bibitem[Lee et~al\mbox{.}(2021)]%
        {Lee2021EvidenceOC}
\bibfield{author}{\bibinfo{person}{Douglas~G. Lee}, \bibinfo{person}{Jean
  Daunizeau}, {and} \bibinfo{person}{Giovanni Pezzulo}.}
  \bibinfo{year}{2021}\natexlab{}.
\newblock \showarticletitle{Evidence or Confidence: What Is Really Monitored
  during a Decision?}
\newblock \bibinfo{journal}{\emph{Psychonomic Bulletin \& Review}}
  (\bibinfo{date}{March} \bibinfo{year}{2021}), \bibinfo{pages}{1--20}.
\newblock
\urldef\tempurl%
\url{https://doi.org/10.3758/s13423-023-02255-9}
\showDOI{\tempurl}


\bibitem[Lerche and Voss(2018)]%
        {lerche2018speed}
\bibfield{author}{\bibinfo{person}{Veronika Lerche} {and}
  \bibinfo{person}{Andreas Voss}.} \bibinfo{year}{2018}\natexlab{}.
\newblock \showarticletitle{Speed–accuracy manipulations and diffusion
  modeling: Lack of discriminant validity of the manipulation or of the
  parameter estimates?}
\newblock \bibinfo{journal}{\emph{Behavior Research Methods}}
  \bibinfo{volume}{50} (\bibinfo{date}{March} \bibinfo{year}{2018}),
  \bibinfo{pages}{2568--2585}.
\newblock
\urldef\tempurl%
\url{https://doi.org/10.3758/s13428-018-1034-7}
\showDOI{\tempurl}


\bibitem[Lerche et~al\mbox{.}(2017)]%
        {lerche_2017_trials_dm}
\bibfield{author}{\bibinfo{person}{Veronika Lerche}, \bibinfo{person}{Andreas
  Voss}, {and} \bibinfo{person}{Markus Nagler}.}
  \bibinfo{year}{2017}\natexlab{}.
\newblock \showarticletitle{How many trials are required for parameter
  estimation in diffusion modeling? A comparison of different optimization
  criteria}.
\newblock \bibinfo{journal}{\emph{Behavior Research Methods}}
  \bibinfo{volume}{49} (\bibinfo{date}{April} \bibinfo{year}{2017}),
  \bibinfo{pages}{513--537}.
\newblock
\urldef\tempurl%
\url{https://doi.org/10.3758/s13428-016-0740-2}
\showDOI{\tempurl}


\bibitem[Liu(2021)]%
        {liu2021TrustAI}
\bibfield{author}{\bibinfo{person}{Bingjie Liu}.}
  \bibinfo{year}{2021}\natexlab{}.
\newblock \showarticletitle{{In AI We Trust? Effects of Agency Locus and
  Transparency on Uncertainty Reduction in Human–AI Interaction}}.
\newblock \bibinfo{journal}{\emph{Journal of Computer-Mediated Communication}}
  \bibinfo{volume}{26}, \bibinfo{number}{6} (\bibinfo{date}{09}
  \bibinfo{year}{2021}), \bibinfo{pages}{384--402}.
\newblock
\showISSN{1083-6101}
\urldef\tempurl%
\url{https://doi.org/10.1093/jcmc/zmab013}
\showDOI{\tempurl}
\showeprint{https://academic.oup.com/jcmc/article-pdf/26/6/384/41139653/zmab013.pdf}


\bibitem[Liu and Lourenco(2022)]%
        {Liu2022DriftDM}
\bibfield{author}{\bibinfo{person}{Yaxin Liu} {and} \bibinfo{person}{Stella~F.
  Lourenco}.} \bibinfo{year}{2022}\natexlab{}.
\newblock \showarticletitle{Drift diffusion modeling informs how affective
  factors affect visuospatial decision making}.
\newblock \bibinfo{journal}{\emph{Journal of Vision}}  \bibinfo{volume}{22}
  (\bibinfo{date}{Dec.} \bibinfo{year}{2022}), \bibinfo{pages}{3394}.
\newblock
\urldef\tempurl%
\url{https://doi.org/10.1167/jov.22.14.3394}
\showDOI{\tempurl}


\bibitem[Makowski et~al\mbox{.}(2021)]%
        {makowski2021neurokit2}
\bibfield{author}{\bibinfo{person}{Dominique Makowski}, \bibinfo{person}{Tam
  Pham}, \bibinfo{person}{Zen~J. Lau}, \bibinfo{person}{Jan~C. Brammer},
  \bibinfo{person}{François Lespinasse}, \bibinfo{person}{Hung Pham},
  \bibinfo{person}{Christopher Schölzel}, {and} \bibinfo{person}{S.~H.~Annabel
  Chen}.} \bibinfo{year}{2021}\natexlab{}.
\newblock \showarticletitle{NeuroKit2: A Python toolbox for neurophysiological
  signal processing}.
\newblock \bibinfo{journal}{\emph{Behavior Research Methods}}
  \bibinfo{volume}{53} (\bibinfo{date}{Feb.} \bibinfo{year}{2021}),
  \bibinfo{pages}{1689--1696}.
\newblock
\urldef\tempurl%
\url{https://doi.org/10.3758/s13428-020-01516-y}
\showDOI{\tempurl}


\bibitem[Meurisch et~al\mbox{.}(2020)]%
        {meurisch_exploring_2020}
\bibfield{author}{\bibinfo{person}{Christian Meurisch},
  \bibinfo{person}{Cristina~A. Mihale-Wilson}, \bibinfo{person}{Adrian
  Hawlitschek}, \bibinfo{person}{Florian Giger}, \bibinfo{person}{Florian
  M\"{u}ller}, \bibinfo{person}{Oliver Hinz}, {and} \bibinfo{person}{Max
  M\"{u}hlh\"{a}user}.} \bibinfo{year}{2020}\natexlab{}.
\newblock \showarticletitle{Exploring User Expectations of Proactive AI
  Systems}.
\newblock \bibinfo{journal}{\emph{Proc. ACM Interact. Mob. Wearable Ubiquitous
  Technol.}} \bibinfo{volume}{4}, \bibinfo{number}{4}, Article
  \bibinfo{articleno}{146} (\bibinfo{date}{Dec.} \bibinfo{year}{2020}),
  \bibinfo{numpages}{22}~pages.
\newblock
\urldef\tempurl%
\url{https://doi.org/10.1145/3432193}
\showDOI{\tempurl}


\bibitem[Montgomery and Kirsch(1996)]%
        {montgomery1996mechanisms}
\bibfield{author}{\bibinfo{person}{Guy Montgomery} {and}
  \bibinfo{person}{Irving Kirsch}.} \bibinfo{year}{1996}\natexlab{}.
\newblock \showarticletitle{Mechanisms of Placebo Pain Reduction: An Empirical
  Investigation}.
\newblock \bibinfo{journal}{\emph{Psychological Science}}  \bibinfo{volume}{7}
  (\bibinfo{date}{May} \bibinfo{year}{1996}), \bibinfo{pages}{174--176}.
\newblock
\urldef\tempurl%
\url{https://doi.org/10.1111/j.1467-9280.1996.tb00352.x}
\showDOI{\tempurl}


\bibitem[Nunez et~al\mbox{.}(2017)]%
        {nunez2017attention}
\bibfield{author}{\bibinfo{person}{Michael~D. Nunez}, \bibinfo{person}{Joachim
  Vandekerckhove}, {and} \bibinfo{person}{Ramesh Srinivasan}.}
  \bibinfo{year}{2017}\natexlab{}.
\newblock \showarticletitle{How attention influences perceptual decision
  making: Single-trial EEG correlates of drift-diffusion model parameters}.
\newblock \bibinfo{journal}{\emph{Journal of Mathematical Psychology}}
  \bibinfo{volume}{76} (\bibinfo{date}{Feb.} \bibinfo{year}{2017}),
  \bibinfo{pages}{117--130}.
\newblock
\urldef\tempurl%
\url{https://doi.org/10.1016/j.jmp.2016.03.003}
\showDOI{\tempurl}


\bibitem[Oulasvirta et~al\mbox{.}(2022)]%
        {Oulas2022}
\bibfield{author}{\bibinfo{person}{Antti Oulasvirta}, \bibinfo{person}{Jussi
  P.~P. Jokinen}, {and} \bibinfo{person}{Andrew Howes}.}
  \bibinfo{year}{2022}\natexlab{}.
\newblock \showarticletitle{Computational Rationality as a Theory of
  Interaction}. In \bibinfo{booktitle}{\emph{Proceedings of the 2022 CHI
  Conference on Human Factors in Computing Systems}} (New Orleans, LA, USA)
  \emph{(\bibinfo{series}{CHI '22})}. \bibinfo{publisher}{Association for
  Computing Machinery}, \bibinfo{address}{New York, NY, USA}, Article
  \bibinfo{articleno}{359}, \bibinfo{numpages}{14}~pages.
\newblock
\showISBNx{9781450391573}
\urldef\tempurl%
\url{https://doi.org/10.1145/3491102.3517739}
\showDOI{\tempurl}


\bibitem[Park et~al\mbox{.}(2021)]%
        {park_human-AI_2021}
\bibfield{author}{\bibinfo{person}{Hyanghee Park}, \bibinfo{person}{Daehwan
  Ahn}, \bibinfo{person}{Kartik Hosanagar}, {and} \bibinfo{person}{Joonhwan
  Lee}.} \bibinfo{year}{2021}\natexlab{}.
\newblock \showarticletitle{Human-AI Interaction in Human Resource Management:
  Understanding Why Employees Resist Algorithmic Evaluation at Workplaces and
  How to Mitigate Burdens}. In \bibinfo{booktitle}{\emph{Proceedings of the
  2021 CHI Conference on Human Factors in Computing Systems}} (Yokohama, Japan)
  \emph{(\bibinfo{series}{Chi '21})}. \bibinfo{publisher}{Association for
  Computing Machinery}, \bibinfo{address}{New York, NY, USA}, Article
  \bibinfo{articleno}{154}, \bibinfo{numpages}{15}~pages.
\newblock
\showISBNx{9781450380966}
\urldef\tempurl%
\url{https://doi.org/10.1145/3411764.3445304}
\showDOI{\tempurl}


\bibitem[Prentice and Miller(2016)]%
        {RN526770}
\bibfield{author}{\bibinfo{person}{Deborah~A. Prentice} {and}
  \bibinfo{person}{Dale~T. Miller}.} \bibinfo{year}{2016}\natexlab{}.
\newblock \bibinfo{title}{When small effects are impressive}.
\newblock , \bibinfo{numpages}{99-105}~pages.
\newblock
\showISBNx{1-4338-2091-9 (Hardcover); 1-4338-2140-0 (Digital (undefined
  format)); 1-4338-2092-7 (Paperback); 978-1-4338-2091-5 (Hardcover);
  978-1-4338-2140-0 (Digital (undefined format)); 978-1-4338-2092-2
  (Paperback)}
\urldef\tempurl%
\url{https://doi.org/10.1037/14805-006}
\showDOI{\tempurl}


\bibitem[Price et~al\mbox{.}(2008)]%
        {Price2008ACR}
\bibfield{author}{\bibinfo{person}{Donald~D. Price}, \bibinfo{person}{Damien~G.
  Finniss}, {and} \bibinfo{person}{Fabrizio Benedetti}.}
  \bibinfo{year}{2008}\natexlab{}.
\newblock \showarticletitle{A Comprehensive Review of the Placebo Effect:
  Recent Advances and Current Thought}.
\newblock \bibinfo{journal}{\emph{Annual Review of Psychology}}
  \bibinfo{volume}{59} (\bibinfo{date}{Jan.} \bibinfo{year}{2008}),
  \bibinfo{pages}{565--590}.
\newblock
\urldef\tempurl%
\url{https://doi.org/10.1146/annurev.psych.59.113006.095941}
\showDOI{\tempurl}


\bibitem[Purcell et~al\mbox{.}(2023)]%
        {purcell_fears_2023}
\bibfield{author}{\bibinfo{person}{Zoe~A. Purcell}, \bibinfo{person}{Mengchen
  Dong}, \bibinfo{person}{Anne-Marie Nussberger}, \bibinfo{person}{Nils
  Köbis}, {and} \bibinfo{person}{Maurice Jakesch}.}
  \bibinfo{year}{2023}\natexlab{}.
\newblock \bibinfo{title}{Fears about AI-mediated communication are grounded in
  different expectations for one's own versus others' use}.
\newblock
\newblock
\showeprint[arxiv]{2305.01670}~[cs.HC]


\bibitem[Ragot et~al\mbox{.}(2020)]%
        {ragot2020}
\bibfield{author}{\bibinfo{person}{Martin Ragot}, \bibinfo{person}{Nicolas
  Martin}, {and} \bibinfo{person}{Salom\'{e} Cojean}.}
  \bibinfo{year}{2020}\natexlab{}.
\newblock \showarticletitle{AI-Generated vs. Human Artworks. A Perception Bias
  Towards Artificial Intelligence?}. In \bibinfo{booktitle}{\emph{Extended
  Abstracts of the 2020 CHI Conference on Human Factors in Computing Systems}}
  (Honolulu, HI, USA) \emph{(\bibinfo{series}{Chi Ea '20})}.
  \bibinfo{publisher}{Association for Computing Machinery},
  \bibinfo{address}{New York, NY, USA}, \bibinfo{pages}{1–10}.
\newblock
\showISBNx{9781450368193}
\urldef\tempurl%
\url{https://doi.org/10.1145/3334480.3382892}
\showDOI{\tempurl}


\bibitem[Ratcliff and Rouder(2000)]%
        {ratcliff_rouder_2000_dm_2C_Letteridentification}
\bibfield{author}{\bibinfo{person}{Roger Ratcliff} {and}
  \bibinfo{person}{Jeffrey~N. Rouder}.} \bibinfo{year}{2000}\natexlab{}.
\newblock \showarticletitle{A diffusion model account of masking in two-choice
  letter identification.}
\newblock \bibinfo{journal}{\emph{Journal of experimental psychology. Human
  perception and performance}} \bibinfo{volume}{26}, \bibinfo{number}{1}
  (\bibinfo{date}{Feb.} \bibinfo{year}{2000}), \bibinfo{pages}{127--40}.
\newblock
\urldef\tempurl%
\url{https://doi.org/10.1037//0096-1523.26.1.127}
\showDOI{\tempurl}


\bibitem[Ratcliff and Smith(2010)]%
        {ratcliff_perceptual_2010}
\bibfield{author}{\bibinfo{person}{Roger Ratcliff} {and}
  \bibinfo{person}{Philip~L. Smith}.} \bibinfo{year}{2010}\natexlab{}.
\newblock \showarticletitle{Perceptual discrimination in static and dynamic
  noise: The temporal relation between perceptual encoding and decision
  making.}
\newblock \bibinfo{journal}{\emph{Journal of Experimental Psychology: General}}
  \bibinfo{volume}{139}, \bibinfo{number}{1} (\bibinfo{date}{Feb.}
  \bibinfo{year}{2010}), \bibinfo{pages}{70--94}.
\newblock
\showISSN{1939-2222, 0096-3445}
\urldef\tempurl%
\url{https://doi.org/10.1037/a0018128}
\showDOI{\tempurl}


\bibitem[Rickels et~al\mbox{.}(1970)]%
        {Rickels1970Pills}
\bibfield{author}{\bibinfo{person}{K. Rickels}, \bibinfo{person}{P.~T.
  Hesbacher}, \bibinfo{person}{C.~C. Weise}, \bibinfo{person}{B. Gray}, {and}
  \bibinfo{person}{H.~S. Feldman}.} \bibinfo{year}{1970}\natexlab{}.
\newblock \showarticletitle{Pills and improvement: A study of placebo response
  in psychoneurotic outpatients}.
\newblock \bibinfo{journal}{\emph{Psychopharmacologia}}  \bibinfo{volume}{16}
  (\bibinfo{date}{Jan.} \bibinfo{year}{1970}), \bibinfo{pages}{318--328}.
\newblock
\urldef\tempurl%
\url{https://doi.org/10.1007/bf00404738}
\showDOI{\tempurl}


\bibitem[Rubin and Chisnell(2008)]%
        {rubin2008handbook}
\bibfield{author}{\bibinfo{person}{Jeffrey Rubin} {and} \bibinfo{person}{Dana
  Chisnell}.} \bibinfo{year}{2008}\natexlab{}.
\newblock \bibinfo{booktitle}{\emph{Handbook of usability testing: How to plan,
  design, and conduct effective tests}}.
\newblock \bibinfo{publisher}{John Wiley \& Sons}.
\newblock


\bibitem[Sartori and Bocca(2023)]%
        {sartori_minding_2023}
\bibfield{author}{\bibinfo{person}{Laura Sartori} {and} \bibinfo{person}{Giulia
  Bocca}.} \bibinfo{year}{2023}\natexlab{}.
\newblock \showarticletitle{Minding the gap(s): public perceptions of AI and
  socio-technical imaginaries}.
\newblock \bibinfo{journal}{\emph{Ai \& Society}} \bibinfo{volume}{38},
  \bibinfo{number}{2} (\bibinfo{date}{April} \bibinfo{year}{2023}),
  \bibinfo{pages}{443--458}.
\newblock
\urldef\tempurl%
\url{https://doi.org/10.1007/s00146-022-01422-1}
\showDOI{\tempurl}


\bibitem[Schad et~al\mbox{.}(2021)]%
        {schad2021toward}
\bibfield{author}{\bibinfo{person}{Daniel~J Schad}, \bibinfo{person}{Michael
  Betancourt}, {and} \bibinfo{person}{Shravan Vasishth}.}
  \bibinfo{year}{2021}\natexlab{}.
\newblock \showarticletitle{Toward a principled Bayesian workflow in cognitive
  science.}
\newblock \bibinfo{journal}{\emph{Psychological methods}} \bibinfo{volume}{26},
  \bibinfo{number}{1} (\bibinfo{year}{2021}), \bibinfo{pages}{103}.
\newblock


\bibitem[Schoonderwoerd et~al\mbox{.}(2022)]%
        {SCHOONDERWOERD2022WOZ}
\bibfield{author}{\bibinfo{person}{Tjeerd~A.J. Schoonderwoerd},
  \bibinfo{person}{Emma~M. van Zoelen}, \bibinfo{person}{Karel van~den Bosch},
  {and} \bibinfo{person}{Mark~A. Neerincx}.} \bibinfo{year}{2022}\natexlab{}.
\newblock \showarticletitle{Design patterns for human-AI co-learning: A
  wizard-of-Oz evaluation in an urban-search-and-rescue task}.
\newblock \bibinfo{journal}{\emph{International Journal of Human-Computer
  Studies}}  \bibinfo{volume}{164} (\bibinfo{year}{2022}),
  \bibinfo{pages}{102831}.
\newblock
\showISSN{1071-5819}
\urldef\tempurl%
\url{https://doi.org/10.1016/j.ijhcs.2022.102831}
\showDOI{\tempurl}


\bibitem[Schrepp et~al\mbox{.}(2017)]%
        {Schrepp_2017_UEQS}
\bibfield{author}{\bibinfo{person}{Martin Schrepp}, \bibinfo{person}{Andreas
  Hinderks}, {and} \bibinfo{person}{J{\"o}rg Thomaschewski}.}
  \bibinfo{year}{2017}\natexlab{}.
\newblock \showarticletitle{Design and Evaluation of a Short Version of the
  User Experience Questionnaire (UEQ-S)}.
\newblock \bibinfo{journal}{\emph{International Journal of Interactive
  Multimedia and Artificial Intelligence}}  \bibinfo{volume}{4}
  (\bibinfo{year}{2017}), \bibinfo{pages}{103--108}.
\newblock
\urldef\tempurl%
\url{https://doi.org/10.9781/ijimai.2017.09.001}
\showDOI{\tempurl}


\bibitem[Schrills and Franke(2021)]%
        {Schrills_Franke_2021_SIPAS}
\bibfield{author}{\bibinfo{person}{Tim Schrills} {and} \bibinfo{person}{Thomas
  Franke}.} \bibinfo{year}{2021}\natexlab{}.
\newblock \showarticletitle{Subjective Information Processing Awareness Scale
  (SIPAS)}.
\newblock  (\bibinfo{date}{07} \bibinfo{year}{2021}).
\newblock


\bibitem[Schrills and Franke(2023)]%
        {Schrills_Franke_2023_SIPAS_validity}
\bibfield{author}{\bibinfo{person}{Tim Schrills} {and} \bibinfo{person}{Thomas
  Franke}.} \bibinfo{year}{2023}\natexlab{}.
\newblock \showarticletitle{How Do Users Experience Traceability of AI Systems?
  Examining Subjective Information Processing Awareness in Automated Insulin
  Delivery (AID) Systems}.
\newblock \bibinfo{journal}{\emph{ACM Trans. Interact. Intell. Syst.}}
  (\bibinfo{date}{March} \bibinfo{year}{2023}).
\newblock
\showISSN{2160-6455}
\urldef\tempurl%
\url{https://doi.org/10.1145/3588594}
\showURL{%
\tempurl}


\bibitem[Schrills et~al\mbox{.}(2021)]%
        {Schrills_2021_SIPAS_validation}
\bibfield{author}{\bibinfo{person}{Tim Schrills}, \bibinfo{person}{Mourad
  Zoubir}, \bibinfo{person}{Mona Bickel}, \bibinfo{person}{Susanne Kargl},
  {and} \bibinfo{person}{Thomas Franke}.} \bibinfo{year}{2021}\natexlab{}.
\newblock \showarticletitle{Are Users in the Loop? Development of the
  Subjective Information Processing Awareness Scale to Assess XAI}.
\newblock


\bibitem[Schuch(2016)]%
        {schuch2016task}
\bibfield{author}{\bibinfo{person}{Stefanie Schuch}.}
  \bibinfo{year}{2016}\natexlab{}.
\newblock \showarticletitle{Task inhibition and response inhibition in older
  vs. younger adults: A diffusion model analysis}.
\newblock \bibinfo{journal}{\emph{Frontiers in Psychology}}
  \bibinfo{volume}{7} (\bibinfo{year}{2016}), \bibinfo{pages}{1722}.
\newblock


\bibitem[Shafir et~al\mbox{.}(1993)]%
        {SHAFIR199311}
\bibfield{author}{\bibinfo{person}{Eldar Shafir}, \bibinfo{person}{Itamar
  Simonson}, {and} \bibinfo{person}{Amos Tversky}.}
  \bibinfo{year}{1993}\natexlab{}.
\newblock \showarticletitle{Reason-based choice}.
\newblock \bibinfo{journal}{\emph{Cognition}} \bibinfo{volume}{49},
  \bibinfo{number}{1} (\bibinfo{year}{1993}), \bibinfo{pages}{11--36}.
\newblock
\showISSN{0010-0277}
\urldef\tempurl%
\url{https://doi.org/10.1016/0010-0277(93)90034-S}
\showDOI{\tempurl}


\bibitem[Shi and Yin(2020)]%
        {shi2020reconnecting}
\bibfield{author}{\bibinfo{person}{Haolun Shi} {and} \bibinfo{person}{Guosheng
  Yin}.} \bibinfo{year}{2020}\natexlab{}.
\newblock \showarticletitle{Reconnecting p-value and Posterior Probability
  under One-and Two-sided Tests}.
\newblock \bibinfo{journal}{\emph{The American Statistician}}
  (\bibinfo{year}{2020}), \bibinfo{pages}{1--11}.
\newblock


\bibitem[Starns and Ratcliff(2010)]%
        {starns2010effects}
\bibfield{author}{\bibinfo{person}{Jeffrey~J Starns} {and}
  \bibinfo{person}{Roger Ratcliff}.} \bibinfo{year}{2010}\natexlab{}.
\newblock \showarticletitle{The effects of aging on the speed--accuracy
  compromise: Boundary optimality in the diffusion model.}
\newblock \bibinfo{journal}{\emph{Psychology and Aging}} \bibinfo{volume}{25},
  \bibinfo{number}{2} (\bibinfo{year}{2010}), \bibinfo{pages}{377}.
\newblock


\bibitem[Stewart-Williams and Podd(2004)]%
        {stewart2004placebo}
\bibfield{author}{\bibinfo{person}{Steve Stewart-Williams} {and}
  \bibinfo{person}{John Podd}.} \bibinfo{year}{2004}\natexlab{}.
\newblock \showarticletitle{The placebo effect: dissolving the expectancy
  versus conditioning debate.}
\newblock \bibinfo{journal}{\emph{Psychological Bulletin}}
  \bibinfo{volume}{130}, \bibinfo{number}{2} (\bibinfo{year}{2004}),
  \bibinfo{pages}{324}.
\newblock


\bibitem[Team(2021)]%
        {R-base}
\bibfield{author}{\bibinfo{person}{R~Core Team}.}
  \bibinfo{year}{2021}\natexlab{}.
\newblock \bibinfo{booktitle}{\emph{R: A Language and Environment for
  Statistical Computing}}.
\newblock R Foundation for Statistical Computing, Vienna, Austria.
\newblock
\urldef\tempurl%
\url{https://www.R-project.org/}
\showURL{%
\tempurl}


\bibitem[Thapar et~al\mbox{.}(2003)]%
        {thapar_2003_Letter_discrimination_task}
\bibfield{author}{\bibinfo{person}{Anjali Thapar}, \bibinfo{person}{Roger
  Ratcliff}, {and} \bibinfo{person}{Gail McKoon}.}
  \bibinfo{year}{2003}\natexlab{}.
\newblock \showarticletitle{A diffusion model analysis of the effects of aging
  on letter discrimination}.
\newblock \bibinfo{journal}{\emph{Psychology and Aging}} \bibinfo{volume}{18},
  \bibinfo{number}{3} (\bibinfo{year}{2003}), \bibinfo{pages}{415--429}.
\newblock
\urldef\tempurl%
\url{https://doi.org/10.1037/0882-7974.18.3.415}
\showDOI{\tempurl}


\bibitem[Ueno et~al\mbox{.}(2022)]%
        {ueno2022TrustAI}
\bibfield{author}{\bibinfo{person}{Takane Ueno}, \bibinfo{person}{Yuto Sawa},
  \bibinfo{person}{Yeongdae Kim}, \bibinfo{person}{Jacqueline Urakami},
  \bibinfo{person}{Hiroki Oura}, {and} \bibinfo{person}{Katie Seaborn}.}
  \bibinfo{year}{2022}\natexlab{}.
\newblock \showarticletitle{Trust in Human-AI Interaction: Scoping Out Models,
  Measures, and Methods}. In \bibinfo{booktitle}{\emph{Extended Abstracts of
  the 2022 CHI Conference on Human Factors in Computing Systems}} (New Orleans,
  LA, USA) \emph{(\bibinfo{series}{CHI EA '22})}.
  \bibinfo{publisher}{Association for Computing Machinery},
  \bibinfo{address}{New York, NY, USA}, Article \bibinfo{articleno}{254},
  \bibinfo{numpages}{7}~pages.
\newblock
\showISBNx{9781450391566}
\urldef\tempurl%
\url{https://doi.org/10.1145/3491101.3519772}
\showDOI{\tempurl}


\bibitem[Vaccaro et~al\mbox{.}(2018)]%
        {vaccaro2018illusion}
\bibfield{author}{\bibinfo{person}{Kristen Vaccaro}, \bibinfo{person}{Dylan
  Huang}, \bibinfo{person}{Motahhare Eslami}, \bibinfo{person}{Christian
  Sandvig}, \bibinfo{person}{Kevin Hamilton}, {and} \bibinfo{person}{Karrie
  Karahalios}.} \bibinfo{year}{2018}\natexlab{}.
\newblock \showarticletitle{The Illusion of Control: Placebo Effects of Control
  Settings}. In \bibinfo{booktitle}{\emph{Proceedings of the 2018 CHI
  Conference on Human Factors in Computing Systems}} (Montreal QC, Canada)
  \emph{(\bibinfo{series}{Chi '18})}. \bibinfo{publisher}{Association for
  Computing Machinery}, \bibinfo{address}{New York, NY, USA},
  \bibinfo{pages}{1–13}.
\newblock
\showISBNx{9781450356206}
\urldef\tempurl%
\url{https://doi.org/10.1145/3173574.3173590}
\showDOI{\tempurl}


\bibitem[van Berkel and Hornb\ae{}k(2023)]%
        {berkel2023}
\bibfield{author}{\bibinfo{person}{Niels van Berkel} {and}
  \bibinfo{person}{Kasper Hornb\ae{}k}.} \bibinfo{year}{2023}\natexlab{}.
\newblock \showarticletitle{Implications of Human-Computer Interaction
  Research}.
\newblock \bibinfo{journal}{\emph{Interactions}} \bibinfo{volume}{30},
  \bibinfo{number}{4} (\bibinfo{date}{June} \bibinfo{year}{2023}),
  \bibinfo{pages}{50--55}.
\newblock
\showISSN{1072-5520}
\urldef\tempurl%
\url{https://doi.org/10.1145/3600103}
\showDOI{\tempurl}


\bibitem[van~de Schoot et~al\mbox{.}(2021)]%
        {van2021bayesian}
\bibfield{author}{\bibinfo{person}{Rens van~de Schoot}, \bibinfo{person}{Sarah
  Depaoli}, \bibinfo{person}{Ruth King}, \bibinfo{person}{Bianca Kramer},
  \bibinfo{person}{Kaspar Märtens}, \bibinfo{person}{Mahlet~G. Tadesse},
  \bibinfo{person}{Marina Vannucci}, \bibinfo{person}{Andrew Gelman},
  \bibinfo{person}{Duco Veen}, \bibinfo{person}{Joukje Willemsen}, {and}
  \bibinfo{person}{Christopher Yau}.} \bibinfo{year}{2021}\natexlab{}.
\newblock \showarticletitle{Bayesian statistics and modelling}.
\newblock \bibinfo{journal}{\emph{Nature Reviews Methods Primers}}
  \bibinfo{volume}{1}, \bibinfo{number}{1} (\bibinfo{date}{Jan.}
  \bibinfo{year}{2021}), \bibinfo{pages}{1--26}.
\newblock
\urldef\tempurl%
\url{https://doi.org/10.1038/s43586-020-00001-2}
\showDOI{\tempurl}


\bibitem[van~den Bergh et~al\mbox{.}(2020)]%
        {van2020dstarm}
\bibfield{author}{\bibinfo{person}{Don van~den Bergh}, \bibinfo{person}{Francis
  Tuerlinckx}, {and} \bibinfo{person}{Stijn Verdonck}.}
  \bibinfo{year}{2020}\natexlab{}.
\newblock \showarticletitle{DstarM: an R package for analyzing two-choice
  reaction time data with the D*M method}.
\newblock \bibinfo{journal}{\emph{Behavior Research Methods}}
  \bibinfo{volume}{52} (\bibinfo{date}{May} \bibinfo{year}{2020}),
  \bibinfo{pages}{521--543}.
\newblock
\urldef\tempurl%
\url{https://doi.org/10.3758/s13428-019-01249-7}
\showDOI{\tempurl}


\bibitem[Villa et~al\mbox{.}(2023)]%
        {villa2023placebo}
\bibfield{author}{\bibinfo{person}{Steeven Villa}, \bibinfo{person}{Thomas
  Kosch}, \bibinfo{person}{Felix Grelka}, \bibinfo{person}{Albrecht Schmidt},
  {and} \bibinfo{person}{Robin Welsch}.} \bibinfo{year}{2023}\natexlab{}.
\newblock \showarticletitle{The placebo effect of human augmentation:
  Anticipating cognitive augmentation increases risk-taking behavior}.
\newblock \bibinfo{journal}{\emph{Computers in Human Behavior}}
  \bibinfo{volume}{146} (\bibinfo{date}{Sept.} \bibinfo{year}{2023}),
  \bibinfo{pages}{107787}.
\newblock
\showISSN{0747-5632}
\urldef\tempurl%
\url{https://doi.org/10.1016/j.chb.2023.107787}
\showDOI{\tempurl}


\bibitem[Vodrahalli et~al\mbox{.}(2022)]%
        {vodrahalli2022TrustAI}
\bibfield{author}{\bibinfo{person}{Kailas Vodrahalli}, \bibinfo{person}{Roxana
  Daneshjou}, \bibinfo{person}{Tobias Gerstenberg}, {and}
  \bibinfo{person}{James Zou}.} \bibinfo{year}{2022}\natexlab{}.
\newblock \showarticletitle{Do Humans Trust Advice More If It Comes from AI? An
  Analysis of Human-AI Interactions}. In \bibinfo{booktitle}{\emph{Proceedings
  of the 2022 AAAI/ACM Conference on AI, Ethics, and Society}} (Oxford, United
  Kingdom) \emph{(\bibinfo{series}{AIES '22})}. \bibinfo{publisher}{Association
  for Computing Machinery}, \bibinfo{address}{New York, NY, USA},
  \bibinfo{pages}{763–777}.
\newblock
\showISBNx{9781450392471}
\urldef\tempurl%
\url{https://doi.org/10.1145/3514094.3534150}
\showDOI{\tempurl}


\bibitem[Voss et~al\mbox{.}(2015)]%
        {Voss_2015_ddm_cognitiveprocesses}
\bibfield{author}{\bibinfo{person}{Andreas Voss}, \bibinfo{person}{Jochen
  Voss}, {and} \bibinfo{person}{Veronika Lerche}.}
  \bibinfo{year}{2015}\natexlab{}.
\newblock \showarticletitle{Assessing cognitive processes with diffusion model
  analyses: a tutorial based on fast-dm-30}.
\newblock \bibinfo{journal}{\emph{Frontiers in Psychology}}
  \bibinfo{volume}{6} (\bibinfo{year}{2015}).
\newblock
\showISSN{1664-1078}
\urldef\tempurl%
\url{https://doi.org/10.3389/fpsyg.2015.00336}
\showDOI{\tempurl}


\bibitem[Wager and Atlas(2015)]%
        {wager2015neuroscience}
\bibfield{author}{\bibinfo{person}{Tor~D Wager} {and} \bibinfo{person}{Lauren~Y
  Atlas}.} \bibinfo{year}{2015}\natexlab{}.
\newblock \showarticletitle{The neuroscience of placebo effects: connecting
  context, learning and health}.
\newblock \bibinfo{journal}{\emph{Nature Reviews Neuroscience}}
  \bibinfo{volume}{16}, \bibinfo{number}{7} (\bibinfo{year}{2015}),
  \bibinfo{pages}{403--418}.
\newblock


\bibitem[Weber and Cook(1972)]%
        {weber1972subject}
\bibfield{author}{\bibinfo{person}{Stephen~J Weber} {and}
  \bibinfo{person}{Thomas~D Cook}.} \bibinfo{year}{1972}\natexlab{}.
\newblock \showarticletitle{Subject effects in laboratory research: An
  examination of subject roles, demand characteristics, and valid inference.}
\newblock \bibinfo{journal}{\emph{Psychological bulletin}}
  \bibinfo{volume}{77}, \bibinfo{number}{4} (\bibinfo{year}{1972}),
  \bibinfo{pages}{273}.
\newblock


\bibitem[Wells et~al\mbox{.}(2010)]%
        {https://doi.org/10.1111/j.1540-5915.2010.00292.x}
\bibfield{author}{\bibinfo{person}{John~D. Wells}, \bibinfo{person}{Damon~E.
  Campbell}, \bibinfo{person}{Joseph~S. Valacich}, {and}
  \bibinfo{person}{Mauricio Featherman}.} \bibinfo{year}{2010}\natexlab{}.
\newblock \showarticletitle{The Effect of Perceived Novelty on the Adoption of
  Information Technology Innovations: A Risk/Reward Perspective}.
\newblock \bibinfo{journal}{\emph{Decision Sciences}} \bibinfo{volume}{41},
  \bibinfo{number}{4} (\bibinfo{year}{2010}), \bibinfo{pages}{813--843}.
\newblock
\urldef\tempurl%
\url{https://doi.org/10.1111/j.1540-5915.2010.00292.x}
\showDOI{\tempurl}
\showeprint{https://onlinelibrary.wiley.com/doi/pdf/10.1111/j.1540-5915.2010.00292.x}


\bibitem[Wilson and Rutherford(1989)]%
        {wilson1989mental}
\bibfield{author}{\bibinfo{person}{John~R Wilson} {and} \bibinfo{person}{Andrew
  Rutherford}.} \bibinfo{year}{1989}\natexlab{}.
\newblock \showarticletitle{Mental models: Theory and application in human
  factors}.
\newblock \bibinfo{journal}{\emph{Human Factors}} \bibinfo{volume}{31},
  \bibinfo{number}{6} (\bibinfo{date}{Dec.} \bibinfo{year}{1989}),
  \bibinfo{pages}{617--634}.
\newblock
\urldef\tempurl%
\url{https://doi.org/10.1177/001872088903100601}
\showDOI{\tempurl}


\bibitem[Zgonnikov et~al\mbox{.}(2022)]%
        {zgonnikov2022should}
\bibfield{author}{\bibinfo{person}{Arkady Zgonnikov}, \bibinfo{person}{David
  Abbink}, {and} \bibinfo{person}{Gustav Markkula}.}
  \bibinfo{year}{2022}\natexlab{}.
\newblock \showarticletitle{Should I Stay or Should I Go? Cognitive Modeling of
  Left-Turn Gap Acceptance Decisions in Human Drivers}.
\newblock \bibinfo{journal}{\emph{Human Factors}} (\bibinfo{date}{Dec.}
  \bibinfo{year}{2022}), \bibinfo{numpages}{15}~pages.
\newblock
\urldef\tempurl%
\url{https://doi.org/10.1177/00187208221144561}
\showDOI{\tempurl}


\bibitem[Zhang et~al\mbox{.}(2021)]%
        {zhang_ideal_2021}
\bibfield{author}{\bibinfo{person}{Rui Zhang}, \bibinfo{person}{Nathan~J.
  McNeese}, \bibinfo{person}{Guo Freeman}, {and} \bibinfo{person}{Geoff
  Musick}.} \bibinfo{year}{2021}\natexlab{}.
\newblock \showarticletitle{"An Ideal Human": Expectations of AI Teammates in
  Human-AI Teaming}.
\newblock \bibinfo{journal}{\emph{Proceedings of the ACM on Human-Computer
  Interaction}} \bibinfo{volume}{4}, \bibinfo{number}{Cscw3}
  (\bibinfo{date}{Jan.} \bibinfo{year}{2021}), \bibinfo{pages}{1--25}.
\newblock
\urldef\tempurl%
\url{https://doi.org/10.1145/3432945}
\showDOI{\tempurl}


\end{thebibliography}


\newpage
\appendix





\section{\replaced{Verbal descriptions of the sham-AI system}{Verbal descriptions of the system}}
\label{descriptions}
The participants were presented with the following text as an introduction to the study. Depending on their group assignment (either \textsc{Description}), participants initially read the provided text. Followed by either a paragraph with a \textsc{positive verbal description} or a \textsc{negative verbal description}, both concluding with the same paragraph.

The common paragraph was:
\begin{quote}
\textit{People perform more efficiently when the task difficulty level fits their stress level. Therefore, our team has developed ADAPTIMIND\textsuperscript{TM}, an AI system that adjusts task difficulty in reaction-critical contexts by analyzing the user’s behavior and physiological signals, specifically the electrodermal activity (EDA) measured by medical-grade electrodes using two fingers of your hand.}
\end{quote}

\begin{quote}
\textit{Our AI system dynamically adjusts the task's difficulty by altering the task's pace according to your measured stress level. The algorithm is constantly learning from and adapting to the physiological indicators and your performance during the task. It may take some time to notice the changes in pace.}
\end{quote}

In the \textsc{negative description} condition, the following paragraph was then shown to participants:

\begin{quote}
    \textit{The first users of ADAPTIMIND\textsuperscript{TM} reported that when using the system, it decreased their task performance and increased stress making the task more difficult. As it is a new and untried AI system, it is very unreliable and risky to implement in real-world applications. In this study, we want to test these preliminary findings in a controlled setting.}
\end{quote}

For the \textsc{positive description} condition the following paragraph was shown to the participants:

\begin{quote}
    \textit{The first users of ADAPTIMIND\textsuperscript{TM} reported that when using the system,  it increased their task performance and decreased stress, making the task easier. As it is a cutting-edge AI system, it is very reliable and safe to implement in real-world applications. In this study, we want to test these preliminary findings in a controlled setting.}
\end{quote}

The text concluded in the same way for both groups:

\begin{quote}
\textit{We would like to evaluate your performance using AI and compare it to a condition where the AI is inactive (control condition). We will remind you in which of the two conditions you are in before starting the tasks.}
\end{quote}

\section{\added{Information on the sham-AI system status - Active}}
\label{status:AIActive}

Before the two blocks where participants performed the letter discrimination task with the \sAI{} system active, the following text was displayed:

\begin{quote}
   \textit{AI is now ACTIVE\\ The artificial intelligence system will now monitor your behavior and your physiological signals with the electrodes we have placed on your hand.\\ By monitoring your stress levels, the AI system will adjust the task pace. We will be assessing your performance based on reaction speed and accuracy.}
\end{quote}

The next paragraph differed based on the group allocation to positive/negative \textsc{verbal description}:

\textsc{positive verbal description}:

\begin{quote}
    \textit{The system is expected to increase your task performance and decrease stress, making the task easier.}
\end{quote}

\textsc{negative verbal description}:

\begin{quote}
    \textit{The system is expected to decrease your task performance and increase stress, making the task more difficult.}
\end{quote}

The text was concluded with the following instruction for both groups:

\begin{quote}
    \textit{Please keep your hand with the electrodes on the table with your palm pointed upwards.}
\end{quote}

\section{\added{Information on the sham-AI system status - Inactive}}
\label{status:AIInactive}

Before the two blocks where participants performed the letter discrimination task with the \sAI{} system inactive, the following text was displayed:

\begin{quote}
    \textit{AI is now INACTIVE \\ In this part of the study, we want to measure your performance without the AI system. The pace of the task will be random. We will be assessing your performance based on reaction speed and accuracy. \\Please hold your hand on the table with your palm pointed upwards.}

\end{quote}

\section{\added{MAILS, TiA and SIPAS}}

\begin{table}[h]
\centering
\caption{Mean scores and standard deviation as a function of \textsc{Description} for the questionnaires Meta AI Literacy Scale (MAILS), Checklist for Trust between People and Automation (TiA) and Subjective Information Processing Awareness Scale (SIPAS)}
\label{table:MAILSTiASIPAS}
\begin{tabular}{lcccccc}
\toprule
 & \multicolumn{2}{c}{MAILS} & \multicolumn{2}{c}{TiA} & \multicolumn{2}{c}{SIPAS}\\
\cmidrule(lr){2-3} \cmidrule(lr){4-5} \cmidrule(lr){6-7}
Description & $M$ & $SD$ & $M$ & $SD$ & $M$ & $SD$ \\
\midrule
Negative & 108.61 & 28.28 & 47.97 & 9.15 & 3.47 & 1.05 \\
Positive & 118.71 & 27.89 & 47.39 & 9.93 & 3.58 & 1.03 \\
\bottomrule
\end{tabular}
\end{table}

\section{Hierarchical Drift Diffusion Model with \replaced{\SStatus{}}{status} and Description in BRMS}
\label{BRMS_DDM}
All parameters are modeled on the log scale using the Wiener distribution.

\begin{enumerate}
  \item \textbf{Drift rate (\( \nu \))}:
  \begin{equation}
    \log(\nu_{ijkl}) = \beta_{0\nu} + \beta_{1\nu} \cdot \text{\SStatus}_j + \beta_{2\nu} \cdot \text{Description}_k + \beta_{3\nu} \cdot \text{\SStatus}_j \times \text{Description}_k + \beta_{4\nu} \cdot \text{Order}_l + b_{i\nu}
  \end{equation}
  
  \item \textbf{Boundary separation (\( \alpha \))}:
  \begin{equation}
    \log(\alpha_{ijkl}) = \beta_{0\alpha} + \beta_{1\alpha} \cdot \text{\SStatus}_j + \beta_{2\alpha} \cdot \text{Description}_k + \beta_{3\alpha} \cdot \text{\SStatus}_j \times \text{Description}_k + \beta_{4\alpha} \cdot \text{Order}_l + b_{i\alpha}
  \end{equation}
  
  \item \textbf{Non-decision time (\( \tau \))}:
  \begin{equation}
    \log(\tau_{ijkl}) = \beta_{0\tau} + \beta_{1\tau} \cdot \text{\SStatus}_j + \beta_{2\tau} \cdot \text{Description}_k + \beta_{3\tau} \cdot \text{\SStatus}_j \times \text{Description}_k + \beta_{4\tau} \cdot \text{Order}_l + b_{i\tau}
  \end{equation}
\end{enumerate}

\textbf{Parameters and Priors:}
\begin{itemize}
  \item Intercept priors:
    \begin{align*}
      \beta_{0\nu} &\sim \text{Normal}(0.74, 0.5) \\
      \beta_{0\alpha} &\sim \text{Normal}(0.40, 1), \text{ lb = 0.1} \\
      \beta_{0\tau} &\sim \text{Normal}(-15, 1), \text{ lb = -25, ub = 3}
    \end{align*}
  
  \item Slope priors:
    \begin{align*}
      \beta_{1\nu}, \beta_{2\nu}, \beta_{3\nu}, \beta_{4\nu} &\sim \text{Normal}(0, 0.5) \\
      \beta_{1\alpha}, \beta_{2\alpha}, \beta_{3\alpha}, \beta_{4\alpha} &\sim \text{Normal}(0, 0.1) \\
      \beta_{1\tau}, \beta_{2\tau}, \beta_{3\tau}, \beta_{4\tau} &\sim \text{Normal}(0, 0.01)
    \end{align*}
  
  \item Random effects:
    \[ b_{i\nu}, b_{i\alpha}, b_{i\tau} \sim \text{Normal}(0, \sigma) \]
\end{itemize}

\textbf{Reaction Time Modeling:}
\begin{equation}
  f(RT|\log(\nu_{ijkl}), \log(\alpha_{ijkl}), \log(\tau_{ijkl}), \text{bias}=0.5)
\end{equation}
Where \( RT \) is the observed reaction time.

\section{Empirical and predicted individual reaction time difference for \SStatus{}}
\label{ap:individual-react}
\begin{figure} [!htbp]
    \centering
    \includegraphics[width=1\textwidth]{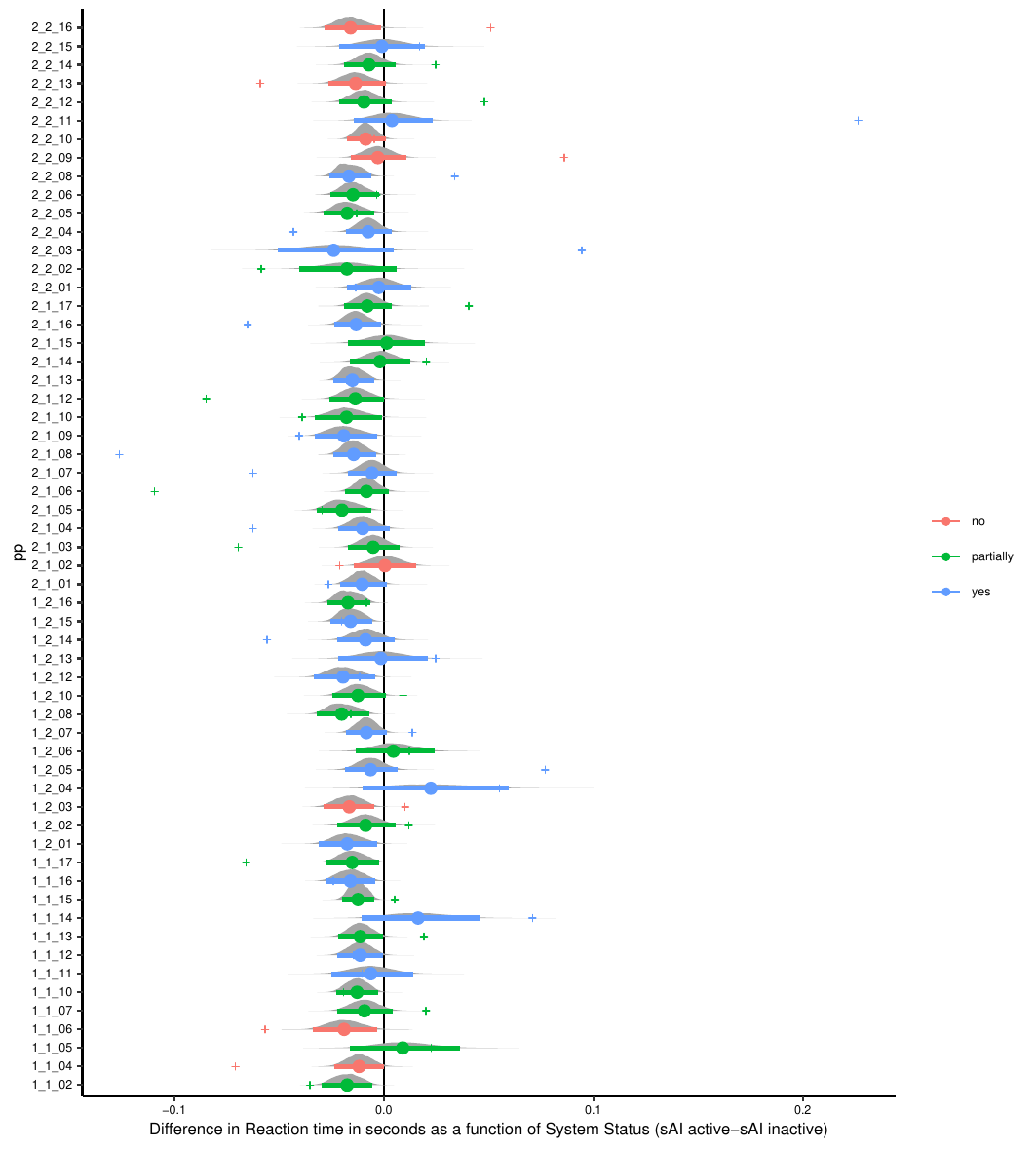}
    \caption{Individual difference (\added{\sAIactive{}} - \added{\sAIinactive{}}) in reaction time predicted by the Drift-Diffusion model with 95$\%$ High-density intervals and the median estimate of the posterior distribution as a function of Manipulation Check (self-reported belief after the debriefing). + indicates the empirical mean difference in reaction time. Distance of empirical RT difference and predicted RT difference shows partial pooling as well as accounting for speed-accuracy trade-offs.}
    \label{fig:enter-label}
\end{figure}
\FloatBarrier

\section{Deviations from the pre-registration}
\begin{table}[h]
\centering
\renewcommand{\arraystretch}{1.5}
\caption{\replaced{Rationale for Deviations from Pre-Registration}{Deviations from the pre-registration}, for the pre-registration see \url{https://aspredicted.org/blind.php?x=SX8_3BG}}
\begin{tabular}{p{3.5cm}|p{10cm}}
\textbf{Section} & \textbf{Deviation} \\
\hline
 Labels & We exchanged the term "nocebo" for the research questions and hypotheses with "negative verbal description" and "placebo" with "positive verbal descriptions." Additionally, the conditions were specified with "sham-AI" (\sAI) in an active or inactive (control condition) state. \\\hline
Participants: Recruiting and testing & 
We deviated from first testing 46 participants for nocebo (negative description), followed by testing 46 for placebo (positive description) due to time constraints. We stopped testing the negative description group after 30 participants were reached and then proceeded with testing the positive description group until we reached 60 participants. After this, we alternated the allocation of the last 6 participants to each group. The last day of testing remained the 18th of August 2023. \\\hline
Reaction time data: Excluding trials & 
We excluded trials with too short responses by filtering RT under 150 ms instead of under 300 ms. This was a necessary deviation as participants were faster in their reactions than anticipated. \\\hline
Reaction time data: Group Analyses & 
Given the AI performance bias, we modeled the data of both groups together instead of separately.\\
\end{tabular}
\label{tab:Deviations}
\end{table}

\section{Model parameters and Diagnostics}
\label{modelsparam}
\subsection{Model found in \Cref{Subjective overall performance}}

\begin{table}[ht]
\setlength{\extrarowheight}{1.5pt}
\caption{Model Formula in Wilkinson notation: $ \textit{Subjective\ overall\ performance\ rating} - 4 \sim 1 + \textit{Description} \times \textit{\SStatus{}} + (1 | \textit{participant}) $}
\centering
\resizebox{\textwidth}{!}{
\begin{tabular}{lrrrrll}
  \hline
Parameter & Median &  $p_b$  & $\hat{R}$ & ESS &  95\% HDI & Prior \\ 
  \hline
Intercept & 0.51 & 0.00\% & 1.00 & 81228.44 & [0.25, 0.77] & student (3, 0.50, 2.50) \\ 
  Description & -0.19 & 6.70\% & 1.00 & 80719.48 & [-0.45, 0.06] & normal (0, 1.39) \\ 
  \SStatus{} & 0.19 & 4.66\% & 1.00 & 130714.43 & [-0.03, 0.42] & normal (0, 1.39) \\ 
  Description $\times$ \SStatus{} & 0.16 & 7.99\% & 1.00 & 136797.53 & [-0.06, 0.38] & normal (0, 1.39) \\ 
  $SD_\text{participant}$ & 0.50 & 0.00\% & 1.00 & 12043.44 & [0.00, 0.85] & student (3, 0, 2.50) \\ 
   \hline
\end{tabular}}
\end{table}

\begin{table}[ht]
\caption{Model Formula in Wilkinson notation: \textit{Subjective estimated task speed rating} - 50 $\sim$ 1 + \textit{Description} $\times$ \textit{\SStatus{}} + (1 | \textit{participant})} 
\centering
\setlength{\extrarowheight}{1.5pt}
\resizebox{\textwidth}{!}{
\begin{tabular}{lrrrrll}
  \hline
Parameter & Median & $p_b$ & $\hat{R}$ & ESS & 95\% HDI & Prior \\ 
  \hline
Intercept & 8.54 & 0.00 & 1.00 & 100738.47 & [5.41, 11.78] & student (3, 7.50, 14.10) \\ 
  Description & -1.31 & 21.09 & 1.00 & 103229.07 & [-4.52, 1.88] & normal (0, 18.04) \\ 
  \SStatus{} & 3.96 & 0.58 & 1.00 & 113865.65 & [0.92, 7.03] & normal (0, 18.04) \\ 
  Description$\times$\SStatus{} & -1.16 & 22.46 & 1.00 & 112055.75 & [-4.17, 1.93] & normal (0, 18.04) \\ 
  $SD_\text{participant}$ & 2.63 & 0.00 & 1.00 & 25022.39 & [0, 6.90] & student (3, 0, 14.10) \\ 
   \hline
\end{tabular}}
\end{table}

\begin{table}[ht]
\caption{Model Formula in Wilkinson notation: $ \textit{Subjective estimated number of correct responses} \sim 1 + \textit{Description} \times \textit{\SStatus{}} \times \textit{Time} + (1 | \textit{participant}) $}
\centering
\setlength{\extrarowheight}{1.5pt}
\resizebox{\textwidth}{!}{
\begin{tabular}{lrrrrll}
  \hline
Parameter & Median & $p_b$ & $\hat{R}$ & ESS & 95\% HDI & Prior \\ 
  \hline
Intercept & 136.30 & 0.00 & 1.00 & 21231.73 & [129.11, 143.47] & student (3, 150, 44.50) \\ 
  Description & -1.38 & 35.22 & 1.00 & 20697.38 & [-8.68, 5.75] & normal (0, 36.71) \\ 
  \SStatus{} & 6.32 & 0.00 & 1.00 & 125110.60 & [3.29, 9.30] & normal (0, 36.71) \\ 
  Time & 6.31 & 0.00 & 1.00 & 130118.28 & [3.36, 9.28] & normal (0, 36.71) \\ 
  Description$\times$\SStatus{} & 0.05 & 48.66 & 1.00 & 125738.08 & [-2.97, 3.07] & normal (0, 36.71) \\ 
  Description$\times$Time & -2.75 & 3.74 & 1.00 & 131880.02 & [-5.76, 0.29] & normal (0, 36.71) \\ 
  \SStatus{}$\times$Time & 2.63 & 4.46 & 1.00 & 131160.58 & [-0.39, 5.62] & normal (0, 36.71) \\ 
  Description$\times$\SStatus{}$\times$Time & -0.07 & 48.21 & 1.00 & 131327.49 & [-3.18, 2.90] & normal (0, 36.71) \\ 
  $SD_\text{participant}$ & 26.56 & 0.00 & 1.00 & 20598.96 & [21.06, 32.61] & student (3, 0, 44.50) \\ 
   \hline
\end{tabular}}

\end{table}

\begin{table}[ht]
\caption{Model Formula in Wilkinson notation: $\textit{Reaction time (s)} * 1000 \sim \textit{\SStatus{}} + \textit{Description} + \textit{Order} + (1 | \textit{participant})$}
\centering
\setlength{\extrarowheight}{1.5pt}
\resizebox{\textwidth}{!}{
\begin{tabular}{lrrrrll}
  \hline
Parameter & Median & $p_b$ & $\hat{R}$ & ESS & 95\% HDI & Prior \\ 
  \hline
Intercept & 606.97 & 0.00 & 1.00 & 2927.02 & [585.57, 628.2] & student (3, 567.10, 123.10) \\ 
  \SStatus{} & -4.17 & 0.00 & 1.00 & 28804.59 & [-6.18, -2.19] & normal (0, 171.61) \\ 
  Description & -5.00 & 32.72 & 1.00 & 2501.51 & [-25.79, 16.94] & normal (0, 171.61) \\ 
  Order& 11.06 & 0.00 & 1.00 & 30137.34 & [9.03, 13.01] & normal (0, 171.61) \\ 
  $SD_\text{participant}$ & 80.75 & 0.00 & 1.00 & 3624.85 & [66.9, 97.36] & student (3, 0, 123.10) \\ 
   \hline
\end{tabular}}
\end{table}

\begin{table}[ht]
\caption{Model Formula in Wilkinson notation: $ \textit{Correctness of responses} \sim \textit{\SStatus{}} + \textit{Description} + \textit{Order} + (1 | \textit{participant})$} 
\centering
\setlength{\extrarowheight}{1.5pt}
\begin{tabular}{lrrrrll}
  \hline
Parameter & Median & $p_b$ & $\hat{R}$ & ESS & 95\% HDI & Prior \\ 
  \hline
Intercept & 2.41 & 0.00 & 1.00 & 6588.72 & [2.19, 2.64] & student (3, 0, 2.50) \\ 
  \SStatus{} & 0.05 & 2.05 & 1.00 & 61408.52 & [0, 0.09] &  student (3, 0, 10)\\ 
  Description & 0.10 & 19.00 & 1.00 & 6137.11 & [-0.12, 0.33] &  student (3, 0, 10)\\ 
  Order& -0.05 & 1.37 & 1.00 & 58438.11 & [-0.09, 0] &  student (3, 0, 10)\\ 
  $SD_\text{participant}$ & 0.83 & 0.00 & 1.00 & 8030.09 & [0.68, 1.01] & student (3, 0, 2.50) \\ 
   \hline
\end{tabular} 
\end{table}


\begin{table}[ht]
\caption{Model Formula in Wilkinson notation: $RTsec | dec(lu) \sim \textit{\SStatus{}} \times \textit{Description} + \textit{Order} + (1 | p | participant)$, Model outputs for the parameters on the log scale. Medians are provided for each parameter, along with their 95\% HDI and $p_b$. Parameters distinguishable from zero are marked with *. We ran the model with two chains and 4000 iterations.}
\centering
\setlength{\extrarowheight}{1.5pt}
\resizebox{\textwidth}{!}{
\begin{tabular}{lrrrrll}
  \hline
Parameter & Median & $p_b$ & $\hat{R}$ & ESS & 95\% HDI & Prior \\ 
  \hline
$\nu$ Intercept & 0.68 & 0.00 & 1.00 & 1097.76 & [0.55, 0.81] & normal (0.74, 0.50) \\ 
$\alpha$ Intercept & 0.37 & 0.00 & 1.00 & 1700.25 & [0.32, 0.43] & normal (0.40, 1) \\ 
$\tau$ Intercept & -1.21 & 0.00 & 1.00 & 1157.78 & [-1.29, -1.13] & normal (-15, 1) \\ 
$\nu$  \SStatus{} & 0.03 & 0.00 & 1.00 & 10716.08 & [0.02, 0.05] & normal (0, 0.50) \\ 
$\nu$  Description & 0.06 & 18.11 & 1.00 & 1308.32 & [-0.06, 0.19] & normal (0, 0.50) \\ 
$\nu$  Order& -0.02 & 0.30 & 1.00 & 10441.23 & [-0.03, 0] & normal (0, 0.50) \\ 
$\nu$  \SStatus{}$\times$Description & 0.01 & 7.68 & 1.00 & 11569.49 & [0, 0.02] & normal (0, 0.50) \\ 
$\alpha$ \SStatus{} & 0.01 & 0.25 & 1.00 & 10301.47 & [0, 0.02] & normal (0, 0.10) \\ 
$\alpha$ Description & 0.02 & 13.90 & 1.00 & 1539.17 & [-0.02, 0.07] & normal (0, 0.10) \\ 
$\alpha$ Order& 0.01 & 14.23 & 1.00 & 9481.10 & [0, 0.01] & normal (0, 0.10) \\ 
$\alpha$ \SStatus{}$\times$Description & 0.01 & 9.55 & 1.00 & 10416.62 & [0, 0.02] & normal (0, 0.10) \\ 
$\tau$ \SStatus{} & -0.02 & 0.00 & 1.00 & 13135.76 & [-0.02, -0.02] & normal (0, 0.01) \\ 
$\tau$ Description & 0.00 & 49.66 & 1.00 & 7734.16 & [-0.02, 0.02] & normal (0, 0.01) \\ 
$\tau$ Order& 0.01 & 0.00 & 1.00 & 13118.79 & [0.01, 0.02] & normal (0, 0.01) \\ 
$\tau$ \SStatus{}$\times$Description & -0.01 & 0.39 & 1.00 & 12630.83 & [-0.01, 0] & normal (0, 0.01) \\ 
  $SD_\text{participant}$ & 0.49 & 0.00 & 1.00 & 2011.08 & [0.40, 0.60] & student (3, 0, 2.50) \\ 
  $SD_\text{participant} \times \alpha$ Intercept & 0.20 & 0.00 & 1.00 & 2108.20 & [0.16, 0.24] & student (3, 0, 2.50) \\ 
  $SD_\text{participant} \times \tau$ Intercept & 0.29 & 0.00 & 1.00 & 1954.95 & [0.23, 0.34] & student (3, 0, 2.50) \\ 
  $cor_\text{participant} \times \nu$ Intercept$\times \alpha$ Intercept & 0.11 & 20.39 & 1.00 & 2067.49 & [-0.15, 0.36] & lkj(2) \\ 
  $cor_\text{participant} \times \nu$ Intercept$\times\tau$ Intercept & 0.31 & 0.98 & 1.00 & 1813.20 & [0.06, 0.54] &  lkj(2) \\ 
  $cor_\text{participant} \times \alpha$ Intercept$\times\tau$ Intercept & -0.56 & 0.00 & 1.00 & 2043.48 & [-0.73, -0.36] & lkj(2)  \\ 
   \hline
\end{tabular}}
\label{tab:model-outputs}

\end{table}

\begin{table}[ht]
\caption{Model Formula in Wilkinson notation: Cognitive Workload ($TLXsum) \sim \textit{Description} \times \textit{\SStatus{}} + \textit{Order} + (1 | participant)$} 
\centering
\setlength{\extrarowheight}{1.5pt}
\resizebox{\textwidth}{!}{
\begin{tabular}{lrrrrll}
  \hline
Parameter & Median & $p_b$ & $\hat{R}$ & ESS & 95\% HDI & Prior \\ 
  \hline
Intercept & 64.75 & 0.00 & 1.00 & 23700.70 & [60.28, 69.33] & student (3,64, 19.30) \\ 
  Description & 0.83 & 35.91 & 1.00 & 22840.38 & [-3.65, 5.50] & normal (0, 20.34) \\ 
  \SStatus{} & -0.08 & 46.82 & 1.00 & 136110.80 & [-2.03, 1.91] & normal (0, 20.34) \\ 
  Order & 1.45 & 7.28 & 1.00 & 137885.40 & [-0.57, 3.37] & normal (0, 20.34) \\ 
  Description$\times$\SStatus{} & 0.43 & 33.28 & 1.00 & 140187.50 & [-1.55, 2.38] & normal (0, 20.34) \\ 
  $SD_\text{participant}$ & 16.69 & 0.00 & 1.00 & 19669.48 & [13.16, 20.54] & student (3, 0, 1) \\ 
   \hline
\end{tabular}}
\end{table}

\begin{table}[ht]
\caption{Model Formula in Wilkinson notation: Physiological Arousal ($SCLmean) \sim \textit{\SStatus{}} \times \textit{Description} + \textit{Order} + (1 | participant)$}
\centering
\setlength{\extrarowheight}{1.5pt}
\resizebox{\textwidth}{!}{
\begin{tabular}{lrrrrll}
  \hline
Parameter & Median & $p_b$ & $\hat{R}$ & ESS & 95\% HDI & Prior \\ 
  \hline
Intercept & -0.96 & 0.00 & 1.00 & 72495.54 & [-1.32, -0.59] & student (3, -0.80, 2.50) \\ 
  \SStatus{} & -0.20 & 11.88 & 1.00 & 48124.82 & [-0.53, 0.13] & normal (0, 0.90) \\ 
  Description & 0.25 & 7.37 & 1.00 & 71529.11 & [-0.10, 0.59] & normal (0, 0.90) \\ 
  Order & 0.20 & 4.20 & 1.00 & 75567.21 & [-0.03, 0.42] & normal (0, 0.90) \\ 
  \SStatus{}$\times$Description & 0.07 & 32.70 & 1.00 & 52413.58 & [-0.24, 0.38] & normal (0, 0.90) \\ 
  $SD_\text{participant}$ & 0.15 & 0.00 & 1.00 & 21592.19 & [0, 0.36] & student (3, 0, 2.50) \\ 
   \hline
\end{tabular}}
\end{table}

\begin{table}[ht]
\caption{Model Formula in Wilkinson notation: $System\ evaluation\ item \ 1 - 4 \sim \textit{Description}$}
\centering
\setlength{\extrarowheight}{1.5pt}
\begin{tabular}{lrrrrll}
  \hline
Parameter & Median & $p_b$ & $\hat{R}$ & ESS & 95\% HDI & Prior \\ 
  \hline
Intercept & -0.60 & 0.45 & 1.00 & 64497.24 & [-1.03, -0.16] & student (3, -1, 2.50) \\ 
  Description & -0.17 & 21.58 & 1.00 & 65782.05 & [-0.63, 0.25] & normal (0, 1.72) \\ 
   \hline
\end{tabular} 
\end{table}

\begin{table}[ht]
\caption{Model Formula in Wilkinson notation: $System\ evaluation\ item \ 2 - 4 \sim \textit{Description}$} 
\centering
\setlength{\extrarowheight}{1.5pt}
\begin{tabular}{lrrrrll}
  \hline
Parameter & Median & $p_b$ & $\hat{R}$ & ESS & 95\% HDI & Prior \\ 
  \hline
Intercept & -0.24 & 12.54 & 1.00 & 64462.65 & [-0.66, 0.16] & student (3, 0, 2.50) \\ 
  Description & -0.41 & 2.48 & 1.00 & 65215.93 & [-0.82, 0] & normal (0, 1.66) \\ 
   \hline
\end{tabular}
\end{table}

\begin{table}[ht]
\caption{Model Formula in Wilkinson notation: $System\ evaluation\ item \ 3 - 4 \sim \textit{Description}$} 
\centering
\setlength{\extrarowheight}{1.5pt}
\begin{tabular}{lrrrrll}
  \hline
Parameter & Median & $p_b$ & $\hat{R}$ & ESS & 95\% HDI & Prior \\ 
  \hline
Intercept & -0.32 & 6.39 & 1.00 & 68372.93 & [-0.74, 0.10] & student (3, 0, 2.50) \\ 
  Description & -0.35 & 5.15 & 1.00 & 65799.87 & [-0.76, 0.07] & normal (0, 1.67) \\ 
   \hline
\end{tabular}
\end{table}

\begin{table}[ht]
\caption{Model Formula in Wilkinson notation: $System\ evaluation\ item \ 4 - 4 \sim \textit{Description}$}
\centering
\setlength{\extrarowheight}{1.5pt}
\begin{tabular}{lrrrrll}
  \hline
Parameter & Median & $p_b$ & $\hat{R}$ & ESS & 95\% HDI & Prior \\ 
  \hline
Intercept & -0.33 & 5.87 & 1.00 & 66744.43 & [-0.74, 0.08] & student (3, 0, 2.50) \\ 
  Description & -0.27 & 9.35 & 1.00 & 67963.57 & [-0.68, 0.14] & normal (0, 1.62) \\ 
   \hline
\end{tabular}
\end{table}

\begin{table}[ht]
\caption{Model Formula in Wilkinson notation: $System\ evaluation\ item \ 5 - 4 \sim \textit{Description}$} 
\centering
\setlength{\extrarowheight}{1.5pt}
\begin{tabular}{lrrrrll}
  \hline
Parameter & Median & $p_b$ & $\hat{R}$ & ESS & 95\% HDI & Prior \\ 
  \hline
Intercept & 0.02 & 44.47 & 1.00 & 59849.21 & [-0.33, 0.39] & student (3, 0, 2.50) \\ 
  Description & -0.12 & 24.80 & 1.00 & 61742.01 & [-0.48, 0.23] & normal (0, 1.42) \\ 
   \hline
\end{tabular}
\end{table}

\begin{table}[ht]
\caption{Model Formula in Wilkinson notation: $System\ evaluation\ item \ 6 - 4 \sim \textit{Description}$} 
\centering
\setlength{\extrarowheight}{1.5pt}
\begin{tabular}{lrrrrll}
  \hline
Parameter & Median & $p_b$ & $\hat{R}$ & ESS & 95\% HDI & Prior \\ 
  \hline
Intercept & 0.09 & 31.37 & 1.00 & 64222.85 & [-0.27, 0.45] & student (3, 0, 2.50) \\ 
  Description & -0.11 & 28.33 & 1.00 & 63489.53 & [-0.46, 0.27] & normal (0, 1.44) \\ 
   \hline
\end{tabular}
\end{table}

\begin{table}[ht]
\caption{Model Formula in Wilkinson notation: $System\ evaluation\ item \ 7 - 4 \sim \textit{Description}$} 
\centering
\setlength{\extrarowheight}{1.5pt}
\begin{tabular}{lrrrrll}
  \hline
Parameter & Median & $p_b$ & $\hat{R}$ & ESS & 95\% HDI & Prior \\ 
  \hline
Intercept & -0.35 & 2.06 & 1.00 & 61105.53 & [-0.68, -0.01] & student (3, 0, 2.50) \\ 
  Description & -0.09 & 29.47 & 1.00 & 59780.62 & [-0.43, 0.25] & normal (0, 1.36) \\ 
   \hline
\end{tabular}
\end{table}

\begin{table}[ht]
\caption{Model Formula in Wilkinson notation: $System\ evaluation\ item \ 8 - 4 \sim \textit{Description}$} 
\centering
\setlength{\extrarowheight}{1.5pt}
\begin{tabular}{lrrrrll}
  \hline
Parameter & Median & $p_b$ & $\hat{R}$ & ESS & 95\% HDI & Prior \\ 
  \hline
Intercept & 1.23 & 0.00 & 1.00 & 66607.93 & [0.93, 1.51] & student (3, 1, 2.50) \\ 
  Description & -0.20 & 8.98 & 1.00 & 67891.85 & [-0.49, 0.09] & normal (0, 1.15) \\ 
   \hline
\end{tabular}
\end{table}

\begin{table}[ht]
\caption{Model Formula in Wilkinson notation: $\text{UEQ-S-pragmatic} - 0 \sim \textit{Description}$}
\centering
\setlength{\extrarowheight}{1.5pt}
\begin{tabular}{lrrrrll}
  \hline
Parameter & Median & $p_b$ & $\hat{R}$ & ESS & 95\% HDI & Prior \\ 
  \hline
Intercept & 0.73 & 0.00 & 1.00 & 57789.76 & [0.46, 0.99] & student (3, 0.80, 2.50) \\ 
  Description & -0.17 & 10.79 & 1.00 & 59970.21 & [-0.43, 0.10] & normal (0, 1.06) \\ 
   \hline
\end{tabular} 
\end{table}

\begin{table}[ht]
\caption{Model Formula in Wilkinson notation: $\text{UEQ-S-hedonic} - 0 \sim \textit{Description}$} 
\centering
\setlength{\extrarowheight}{1.5pt}
\begin{tabular}{lrrrrll}
  \hline
Parameter & Median & $p_b$ & $\hat{R}$ & ESS & 95\% HDI & Prior \\ 
  \hline
  Intercept & 0.78 & 0.00 & 1.00 & 65655.38 & [0.49, 1.08] & student (3, 1, 2.50) \\ 
  Description & -0.04 & 40.80 & 1.00 & 68065.81 & [-0.33, 0.26] & normal (0, 1.17) \\ 
   \hline
\end{tabular}
\end{table}

\begin{table}[ht]
\caption{Model Formula in Wilkinson notation: $\text{SUS-Adapted\ Score} - 68 \sim \textit{Description}$} 
\centering
\setlength{\extrarowheight}{1.5pt}
\begin{tabular}{lrrrrll}
  \hline
Parameter & Median & $p_b$ & $\hat{R}$ & ESS & 95\% HDI & Prior \\ 
  \hline
Intercept & -1.41 & 23.55 & 1.00 & 65479.48 & [-5.39, 2.47] & student (3, -0.50, 14.80) \\ 
  Description & -1.71 & 19.28 & 1.00 & 65245.98 & [-5.69, 2.18] & normal (0, 15.65) \\ 
   \hline
\end{tabular}
\end{table}

\begin{table}[ht]
\caption{Replication Study - Model Formula in Wilkinson notation: $ \textit{Expected\ overall\ performance} - 4 \sim 1 + \textit{Description} $}
\centering
\setlength{\extrarowheight}{1.5pt}
\begin{tabular}{lrrrrll}
  \hline
Parameter & Median & $p_b$ & $\hat{R}$ & ESS & 95\% HDI & Prior \\ 
  \hline
Intercept & 0.85 & 0.00 & 1.00 & 125379.81 & [0.63, 1.08] & student (3, 1, 2.50) \\ 
  Description & -0.26 & 1.05 & 1.00 & 122476.12 & [-0.48, -0.04] & normal (0, 1.08) \\ 
   \hline
\end{tabular} 
\end{table}

\begin{table}[ht]
\caption{Replication Study - Model Formula in Wilkinson notation: $ \textit{Expected\ task\ speed} - 50 \sim 1 + \textit{Description} $}
\centering
\setlength{\extrarowheight}{1.5pt}
\begin{tabular}{lrrrrll}
  \hline
Parameter & Median & $p_b$ & $\hat{R}$ & ESS & 95\% HDI & Prior \\ 
  \hline
Intercept & 15.37 & 0.00 & 1.00 & 65594.30 & [11.73, 19.11] & student (3, 15, 19.30) \\ 
  Description & -4.60 & 0.82 & 1.00 & 61874.58 & [-8.34, -0.86] & normal (0, 18.34) \\ 
   \hline
\end{tabular}
\end{table}

\begin{table}[ht]
\caption{Replication Study - Model Formula in Wilkinson notation: $ \textit{Estimated\ correct} \sim 1 + \textit{Comprehension} \times \textit{\SStatus{}} + (1 | \textit{participant})$}\centering
\setlength{\extrarowheight}{1.5pt}
\resizebox{\textwidth}{!}{
\begin{tabular}{lrrrrll}
  \hline
Parameter & Median & $p_b$ & $\hat{R}$ & ESS & HDI & Prior \\ 
  \hline
Intercept & 147.23 & 0.00 & 1.00 & 35543.08 & [139.7,154.31] & student (3, 150, 44.50) \\ 
  Comprehension & -0.51 & 44.50 & 1.00 & 38871.60 & [-7.80, 6.74] & normal (0, 39.43) \\ 
  \SStatus{} & 5.71 & 0.15 & 1.00 & 151821.25 & [2, 9.39] & normal (0, 39.43) \\ 
  Comprehension$\times$\SStatus{} & -4.36 & 1.03 & 1.00 & 146724.78 & [-8.08, -0.69] & normal (0, 39.43) \\ 
  $SD_\text{participant}$ & 30.22 & 0.00 & 1.00 & 20042.75 & [24.22, 37] & student (3, 0, 44.5) \\ 
   \hline
\end{tabular}}
\end{table}

\end{document}